\documentclass[11pt,a4paper]{article}

\usepackage{hyperref}
\hypersetup{
	colorlinks,
    citecolor=blue,
    linkcolor=blue,
    urlcolor=blue
}
\usepackage{a4wide}
\usepackage[centertags]{amsmath}
\usepackage{amssymb}
\usepackage{cite}
\usepackage{amsmath}
\usepackage{graphicx}

\allowdisplaybreaks
\numberwithin{equation}{section}

\def\surf{\Sigma} % Riemann surface
\def\sdef{\delta}     % Shift of background couplings due to defects
\def\BJ{\mathcal{J}} % Bessel function
\def\MBJ{\mathcal{I}} % Modified Bessel function

\newcommand\blfootnote[1]{%
  \begingroup
  \renewcommand\thefootnote{}\footnote{#1}%
  \addtocounter{footnote}{-1}%
  \endgroup
}

\makeatletter
\def\smallunderbrace#1{\mathop{\vtop{\m@th\ialign{\scriptsize ##\crcr
   $\hfil\displaystyle{#1}\hfil$\crcr
   \noalign{\kern3\p@\nointerlineskip}%
   \tiny\upbracefill\crcr\noalign{\kern3\p@}}}}\limits}
\makeatother

\newcommand\redsout{\bgroup\markoverwith{\textcolor{red}{\rule[0.5ex]{2pt}{0.4pt}}}\ULon}

\begin{document}
\thispagestyle{empty}
\begin{flushright}
\hfill BONN--TH--2021--04\\
\hfill MTIP/21-031
\end{flushright}

\bigskip

\begin{center}
  {\huge \bf Deformations of JT Gravity via\\[1ex]
  Topological Gravity and Applications }  
\end{center}

\vspace*{0.5cm}

\centerline{\large
Stefan F{\"o}rste$^1$,  
Hans Jockers$^2$, 
Joshua Kames-King$^{1,3}$, 
Alexandros Kanargias$^1$%
}

\vspace{0.75cm}\hspace{0.15\textwidth}\begin{minipage}[t]{0.66\textwidth}
	\begin{center}{
	\it
	$^1$Bethe Center for Theoretical Physics\\
	{\footnotesize and}\\
	Physikalisches Institut der Universit\"at Bonn,\\
	Nussallee 12, 53115 Bonn, Germany
	}
	\end{center}
\end{minipage}

\vspace{0.75cm}\hspace{0.15\textwidth}\begin{minipage}[t]{0.66\textwidth}
	\begin{center}{
	\it
	$^2$PRISMA+ Cluster of Excellence and Institute for Physics\\
	Johannes Guttenberg-Universit\"at\\
	Staudinger Weg 7, 55128 Mainz, Germany
	}
	\end{center}
\end{minipage}

\vspace{0.75cm}\hspace{0.15\textwidth}\begin{minipage}[t]{0.66\textwidth}
	\begin{center}{
	\it
	$^3$Kavli Institute for Theoretical Physics\\
	University of California\\
	Santa Barbara, CA93106, USA
	}
	\end{center}
\end{minipage}

%%%%
\vspace*{1cm}
%%%%
\centerline{\bf Abstract}
\vskip .2cm
%%%%%
We study the duality between JT~gravity and the double-scaled matrix model including their respective deformations. For these deformed theories we relate the thermal partition function to the generating function of topological gravity correlators that are determined as solutions to the KdV~hierarchy. We specialise to those deformations of JT gravity coupled to a gas of defects, which conforms with known results in the literature. We express the (asymptotic) thermal partition functions in a low temperature limit, in which non-perturbative corrections are suppressed and the thermal partition function becomes exact. In this limit we demonstrate that there is a Hawking--Page phase transition between connected and disconnected surfaces for this instance of  JT~gravity with a transition temperature affected by the presence of defects. Furthermore, the calculated spectral form factors show the qualitative behaviour expected for a Hawking--Page phase transition. The considered deformations cause the ramp to be shifted along the real time axis. Finally, we comment on recent results related to conical Weil--Petersson volumes and the analytic continuation to two-dimensional de~Sitter space.

\blfootnote{
{\tt \href{mailto:forste@th.physik.uni-bonn.de}{forste@th.physik.uni-bonn.de}}, 
{\tt \href{mailto:jockers@uni-mainz.de}{jockers@uni-mainz.de}},}
%%%
\blfootnote{
{\tt \href{mailto:jvakk@yahoo.com}{jvakk@yahoo.com},}
{\tt \href{mailto:kanargias@physik.uni-bonn.de}{kanargias@physik.uni-bonn.de}}
}

%%%%%%%%%%%%%%%%%%%%%%%%%
\newpage
\pagenumbering{arabic}
\begingroup
\pagestyle{empty}
\hypersetup{linkcolor=black}
\tableofcontents
%\listoffigures
\endgroup
\clearpage
\begingroup
\pagestyle{empty}
\cleardoublepage
\endgroup
\newpage
\renewcommand{\theequation}{\thesection.\arabic{equation}}

%%%%%%%%%%%%%%%%%%%%%%%%%%%%%%%%%%%%
\section{Introduction}
Jackiw--Teitelboim (JT) gravity is a simple model of two-dimensional quantum gravity on backgrounds of constant curvature such as anti-de~Sitter spaces $AdS_2$ \cite{JACKIW,TEITELBOIM,MaldacenaAdS2,AlmheiriPolchinski,Jensen:2016pah,Engelsoy:2016xyb}. It consists of a real scalar field $\phi$ coupled to gravity with the Euclidean action on a Riemann surface $\surf$ being
\begin{equation} \label{eq:JTAction}
  I_\text{JT}=-\frac{S_0}{2}\left(\frac{1}{2}\int_{\surf}\!\!\!d^2x \, \sqrt{g} R   
   +\int_{\partial\surf} \!\!\!\! dx \, \sqrt{h}  K \right)
   -\frac{1}{2}\int_{\surf}\!\!\! d^2x \, \sqrt{g}\phi(R+2)
   +\int_{\partial\surf}\!\!\!\! dx \, \sqrt{h}\phi(K-1)\ ,
\end{equation}
where $R$ is the Ricci scalar,  $g_{\mu\nu}$ the metric, $K$ is the trace of the extrinsic curvature at the boundary $\partial\surf$, and $h_{\mu\nu}$ is the boundary metric induced from $g_{\mu\nu}$. The sum of the first two terms is proportional to the Euler characteristic of the surface $\surf$, which in a black hole context represents the ground-state entropy and for the full gravitational path integral weighs the contribution of geometries in terms of the coupling~$S_0$. The third term sets the constraint of only considering hyperbolic Riemann surfaces
\begin{equation}\label{eq:HyperbolicCondition}
    R(x)+2=0\ ,
\end{equation}
and the last term contains a Gibbons--Hawking--York boundary term together with a counterterm that ensures a finite result when removing the regularisation of the position of the $AdS_2$ boundary. This term captures the Schwarzian dynamics of reparametrisations at the boundary. JT~gravity has been used as a gravitational model in the $AdS_2/CFT_1$ correspondence and in a broader context it encapsulates the low-energy dynamics of near-extremal black holes \cite{Nayak:2018qej,Sarosi:2017ykf}. It can also be linked to the Sachdev–Ye–Kitaev model \cite{SYK1,SYK2} because its low-energy sector is described by the Schwarzian theory and in a certain limit the thermal partition functions agree \cite{RemarksSYK,SYK2}.

In the remarkable work \cite{SSS} Saad, Shenker and Stanford demonstrate that extending the gravitational sector to include geometries consisting of arbitrary number of boundaries and also arbitrary genera furnishes a partition function equivalent to a specific double-scaled Hermitian matrix theory. This duality can be stated as
\begin{align}\label{eq:JTMMDuality}
   Z(\beta_1,\ldots,\beta_n) \mathrel{\widehat{=}}\langle \text{Tr} e^{-\beta_1 H} \ldots \text{Tr}  e^{-\beta_n H}  \rangle_{\text{MM}}\ .
\end{align}
Here the left hand side is the connected thermal partition function~$Z(\beta_1,\ldots,\beta_n)$ of JT~gravity for geometries with $n$ asymptotic boundary components characterised by their inverse temperatures $\beta_i$, $i=1,\ldots,n$. The right hand side is the corresponding correlator of the dual Hermitian matrix integral. Interestingly, these correlators enjoy an interpretation as observables in an ensemble of quantum mechanical systems whose random Hamiltonians~$H$ are given by Hermitian matrices~$H$ of the matrix model \cite{SSS}.\footnote{According to ref.~\cite{BlackHolesandRandomMatrices}, the intriguing appearance of an ensemble of quantum mechanical systems can also be argued for via the relationship of JT~gravity to the Sachdev–Ye–Kitaev model.} This duality is generalised in ref.~\cite{Stanford:2019vob}, where extensions of JT~gravity are associated to other matrix models \cite{Dyson:1962es,Altland:1997zz,Zirnbauer:1996zz}.

The arguments for the proposed duality in ref.~\cite{SSS} rely on two crucial facts: Firstly, as can be seen for the disk, the path integral of the Schwarzian theory localises \cite{WittenStanford}. Secondly, the contributions of Riemann surfaces of higher genera to the JT~gravity path integral reduce to a Schwarzian theory at each boundary component together with an integration over suitable moduli spaces of hyperbolic Riemann surfaces. The latter contributions give rise to Weil--Peterson volumes on the associated moduli spaces of stable curves that --- as proven in ref.~\cite{Eynard_Orantin_2007Volumes} --- obey the same recursion relations as appear in the context of the specific double-scaled Hermitian matrix integral, which in turn suggests the proposed correspondence~\eqref{eq:JTMMDuality}. The duality~\eqref{eq:JTMMDuality} as spelt out above is a priori established perturbatively, i.e.\ on the level of an asymptotic genus expansion. In addition, there are also non-perturbative contributions \cite{SSS}, and hence the matrix model can be viewed as a (non-unique) non-perturbative completion of the genus expansion of JT~gravity. A proposal to deal with potential non-perturbative instabilities is developed in refs.~\cite{CJ1,CJ2,CJ3}. 

In this work we focus on the structure of deformations to JT~gravity and the resulting modifications to the thermal partition functions appearing on the left hand side of the duality~\eqref{eq:JTMMDuality}. A particular deformation to JT~gravity can be incorporated by adding a scalar potential $U(\phi)$ to the Lagrangian of the action~\eqref{eq:JTAction} of the form \cite{Maxfield3gravity,WittenDeformations}
\begin{equation}\label{eq:Dilatonpotential}
    U(\phi) = 2 \epsilon \, e^{-(2\pi-\alpha)\phi} \ , \quad 0 < \alpha < \pi \ .
\end{equation}
This potential does not affect the asymptotic boundary conditions and the gravitational path integral can be evaluated perturbatively in the coupling $\epsilon$ \cite{WittenDeformations}. Carrying out the path integral over the scalar field $\phi$ at the perturbative order $\epsilon^k$ changes the constraint~\eqref{eq:HyperbolicCondition} to \cite{MertensDefects,Maxfield3gravity}
\begin{equation}\label{eq:HyperbolicConditionwithDefect}
  R(x)+2= 2  \sum_{j=1}^{k} %(2\pi-\alpha_{\ell_j}) \ , \delta^{(2)}(x-x_j)\ ,
  (2\pi-\alpha) \, \delta^{(2)}(x-x_j) \ ,
\end{equation}
with a remaining integral of the positions $x_1,\ldots,x_k$ over the Riemann surface $\surf$. Thus the constraint~\eqref{eq:HyperbolicConditionwithDefect} at the given perturbative order $\epsilon^k$ with the two-dimensional $\delta$-distributions introduces on the hyperbolic surfaces $k$~conical singularities at the points $x_1,\ldots,x_k$ with identification angle $\alpha$. As a result, perturbatively the path integral of JT~gravity with the potential~\eqref{eq:Dilatonpotential} can be interpreted as a sum over all possible hyperbolic Riemann surfaces $\Sigma$ with any number of conical singularities with identification angles $\alpha$ at arbitrary positions on $\Sigma$. Furthermore, we can interpret the deformation~\eqref{eq:Dilatonpotential} as coupling JT~gravity to a gas of defects characterized by the coupling constant $\epsilon$ and the idenfication angle $\alpha$ \cite{Maxfield3gravity,WittenDeformations}. The structure can readily be generalised to an arbitrary finite number (possibly even to an infinite number or to a continuous family) of defect species with individual couplings $\epsilon_j$ and identification angles $\alpha_j$ \cite{Maxfield3gravity,WittenDeformations}, such that a more general class of deformations to JT~gravity can be realised.

Instead of directly studying deformations to JT~gravity via scalar potentials of the type \eqref{eq:Dilatonpotential}, we use the connection to two-dimensional topological gravity \cite{WittenIntersection} and the related formulation in terms of moduli spaces of stable curves \cite{KontsevichIntersection,Mirzakhani}. Previously, this approach has been prominently employed in this context, for instance, in refs.~\cite{SSS,OkuyamaSakai1,OkuyamaSakai2,WittenDeformations,Alishahihaetal}. Upon identifying deformations to JT~gravity with solutions to the KdV~hierarchy (which play a central role in topological gravity, see e.g.~ref.~\cite{Itzykson:1992ya}) and using well-established matrix model techniques \cite{Gross1,Gross2,Douglas:1989ve,Brezin:1990rb}, we can study a rather general class of deformations to JT~gravity. From this perspective topological gravity and hence JT~gravity with deformations can be identified with certain minimal string theories and deformations thereof \cite{Douglas:1989ve,GinspargMooreLectures,Belavin:2008kv}. Already in ref.~\cite{SSS} it is observed that JT~gravity can be viewed as the large $p \rightarrow+\infty$ limit of the $(2,2p-1)$ minimal string theory with the associated couplings $t_k$ given by \cite{OkuyamaSakai1,CJ1}
\begin{equation} \label{eq:OSexpansionpoint}
  t_0=t_1=0 \ , \quad  
  t_k=\gamma_k \quad \text{with} \quad 
  \gamma_k=\frac{(-1)^k}{(k-1)!} \quad \text{for} \quad k=2,3,\ldots \ .
\end{equation}
These values for the couplings $t_k$ relate to a specific solution to the above mentioned KdV~hierarchy. In this work we study deformations to JT~gravity by considering more general solutions to the KdV~hierarchy, which on the level of the couplings $t_k$ amounts to deforming them as
\begin{equation}\label{eq:shiftedexpansionpoint}
    t_k=\gamma_k+\sdef_k \quad \text{for} \quad k=0,1,2,\ldots \ .
\end{equation}
For particular choices of $\sdef_k$ --- as established in refs.~\cite{Maxfield3gravity,WittenDeformations} and as discussed in detail in the main text --- this description realises JT~gravity interacting with a gas of defects as described by the scalar potential~\eqref{eq:Dilatonpotential} and generalisations thereof discussed in ref.~\cite{WittenDeformations}. Inspired by the work of Okuyama and Sakai we thoroughly investigate the relationship between general deformations $\sdef_k$ and the specific deformations that are attributed to the interaction of JT~gravity with a gas of defects.

Moreover, we turn to some applications of our general results. First of all, we analyse the low temperature behaviour of the calculated thermal partition functions using techniques developed in refs.~\cite{OkuyamaSakai1,OkuyamaSakai2}. At low temperatures the (asymptotic) genus expansion of the thermal partition function can be given an exact analytic expression \cite{OkuyamaSakai2,Okounkovnpointfunction}, because non-perturbative corrections are suppressed in the performed low temperature double scaling limit. This allows us to study in this low temperature regime Hawking--Page phase transitions and the features of spectral form factors as functions of the deformation parameters with the help of numerical methods. As a second application, we comment on a further instance of JT~gravity, which requires the inclusion of Riemann surfaces with conical singularities, namely the wavefunction of the universe for JT~gravity in de~Sitter space \cite{MaldacenadS,MaloneydS}. This striking connection relies on subtleties of the analytic continuation from sharp to blunt defects or equivalently from small identification angles to large identification angles.

%%%
The structure of the paper is as follows:
%%%
In Section~\ref{JTGravityDeformed JT GravityandTopological Gravity} we first set the stage for the forthcoming analysis and introduce well-established physical and mathematical tools to study correlation functions in topological gravity. Then, applying techniques developed in ref.~\cite{OkuyamaSakai1}, as a genus expansion we calculate for deformed theories of JT~gravity (asymptotic) thermal partition functions (with one or several asymptotic boundary components). The studied class of deformations is suitable to describe interactions of JT~gravity with defects.
%%%
In Section~\ref{Section:LowtemperatureExpansion} we turn to the low temperature expansion of the thermal partition function, which can be computed exactly at leading order in temperature \cite{OkuyamaSakai1,OkuyamaSakai2,Alishahihaetal}. For certain physical applications this analysis is more natural than the previously discussed asymptotic genus expansion because the expansion in temperature naturally sets an energy scale for the accessible states in the computed thermal partition functions. 
%%%
Using the computed low energy limit of the partition functions for JT~gravity coupled to a gas of defects, we show in Section~\ref{Spectral Form Factor} that there is a Hawking--Page phase transition. We numerically compute the associated critical temperature as a function of the deficit coupling constant, and we also analyse the spectral form factor. We find that in the given low temperature approximation the time scale for the onset of the plateau exhibits a simple behaviour in terms of the deficit coupling, which conforms with the observed Hawking--Page phase transition.
%%%
In Section~\ref{ds} we make some basic comments on the connection between the wavefunction of the universe for JT~gravity on de~Sitter space~$dS_2$ and the Weil--Petersson volumes of the associated Riemann surfaces with conical singularities in the light of the recent work \cite{TuriaciBluntDefects}.
%%%
Finally, in Section~\ref{sec:concl} we present our conclusions, where we discuss our results and present some outlook for further investigations.

\smallskip
While completing this work, ref.~\cite{Okuyama:2021ytf} appeared, which has certain overlap with some of our discussions in Section~\ref{JTGravityDeformed JT GravityandTopological Gravity}.

%%%%%%%%%%%%%%%%%
\section{JT Gravity, Deformed JT Gravity and Topological Gravity}\label{JTGravityDeformed JT GravityandTopological Gravity}
%%%%%%%%%%%%%%%%%
In this section we aim to describe JT~gravity together with deformations in terms of two-dimensional topological gravity. The works~\cite{OkuyamaSakai1,OkuyamaSakai2} by Okuyama and Sakai establish a direct link between the partition functions of JT~gravity and correlation functions in topological gravity. Deforming JT~gravity from interactions with defects (as established in ref.~\cite{WittenDeformations,Maxfield3gravity}) yields another instance of two-dimensional topological gravity with modified coupling parameters. While we are indeed interested in JT~gravity coupled to a gas of defects, we study deformations to JT~gravity in a more general setting. By using the results of ref.~\cite{Itzykson:1992ya} we construct thermal partition functions for deformed theories of JT~gravity, which at any intermediate stage of their derivation can be specialised to particular deformed JT~gravity theories (such as JT~gravity interacting with defects). Our approach could offer a starting point towards a dictionary between specific values for the couplings in two-dimensional topological gravity and deformations attributed to scalar potentials added to the JT~gravity action, such as the potential~\eqref{eq:Dilatonpotential} for deformations arising from defect interactions.\footnote{Results in a similar vein of thought are reported in ref.~\cite{TuriaciBluntDefects} as well. See also ref.~\cite{Mertens:2020hbs} for a discussion along these lines from the minimal string theory perspective.}

In part this section uses and reviews some well-established mathematical tools from the intersection theory on the moduli spaces of stable curves to derive the thermal partition functions of deformed JT~gravity. The reader not interested in these derivations should skip these technical details and instead view this section as a collocation of expressions for thermal partition functions and related quantities, which are used in later sections of this work.

%%%%%%%%%%%%%%%%%
\subsection{Weil--Petersson Volumes of Hyperbolic Riemann Surfaces}\label{Weil--Petersson Volumes of Hyperbolic Riemann Surfaces}
To set the stage and to introduce the used notation, we first collect some mathematical preliminaries on the Weil--Petersson volumes of hyperbolic Riemann surfaces with geodesic boundary components and conical singularities from the perspective of intersection theory on the moduli spaces of stable curves. 

Let $\mathcal{M}_{g,n}$ be the moduli space of smooth curves of genus $g$ with $n$ distinct marked points. By construction the moduli space $\mathcal{M}_{g,n}$ is not compact, as it contains neither the limiting curve with a handle degenerating to a nodal point nor the limit as two marked points collide. The Deligne--Mumford compactification $\overline{\mathcal{M}}_{g,n}$ includes the above mentioned limits in terms of stable curves with nodal singularities. The resulting moduli space of stable curves is well-defined to parametrise curves with marked points that do not admit any continuous automorphisms. That is to say $\overline{\mathcal{M}}_{g,n}$ is defined for genus $g\ge 2$ and any number of marked points, for genus one with at least one marked point, and for genus zero with at least three marked points. The complex dimensions of these moduli spaces are given by
\begin{equation} \label{eq:dimM}
  \dim_\mathbb{C} \overline{\mathcal{M}}_{g,n} = 3g - 3 + n \ .
\end{equation}

The moduli space of stable curves $\overline{\mathcal{M}}_{g,n}$ comes equipped with several natural cohomology classes. To each marked point $p_i$, $i=1,\ldots,n$, on the curve $C_g$ one associates at the point $p_i$ the complex cotangent line $T^*_{p_i} C_g$, which patches together to a line bundle~$\mathcal{L}_i$ on $\overline{\mathcal{M}}_{g,n}$. The first Chern class of this line bundle realises a cohomology class on $\overline{\mathcal{M}}_{g,n}$ denoted by
\begin{equation}
  \psi_i = c_1(\mathcal{L}_i) \, \in \, H^2(\overline{\mathcal{M}}_{g,n},\mathbb{Q}) \ .
\end{equation}
The other for us relevant cohomology class is the first Miller--Morita--Mumford class $\kappa_1$, which arises in a similar fashion. Consider the forgetful map $\pi: \overline{\mathcal{M}}_{g,n+1} \to \overline{\mathcal{M}}_{g,n}$ that omits the $(n+1)$-th marked point. Then the cohomology class $\kappa_1$ is given by \cite{MR1486986,MR2482127}
\begin{equation} \label{eq:MMMcl}
  \kappa_1 = \pi_*( c_1(\mathcal{L}_{n+1})^2 ) + \sum_{i=1}^n \psi_i \, \in \, H^2(\overline{\mathcal{M}}_{g,n},\mathbb{Q}) \ ,
\end{equation}  
where the push-forward $\pi_*$ can heuristically be thought of as integrating over the fiber of the map~$\pi$. The class $\kappa_1$ is proportional to the Weil--Petersson K\"ahler form $\omega_\text{WP}$ \cite{MR727702}
\begin{equation} \label{eq:WPKahler}
  \omega_\text{WP} = 2 \pi^2 \kappa_1 \ .
\end{equation}
Upon integrating such cohomology classes over $\overline{\mathcal{M}}_{g,n}$ we obtain (rational) intersection numbers that are collected in correlators. The correlators of particular interest to us are given by 
\begin{equation} \label{topologicalgravitycorrelationfunctions}
  \left\langle \kappa_1^\ell\tau_{d_1} \ldots \tau_{d_n} \right\rangle_{g,n} = \int_{\overline{\mathcal{M}}_{g,n}}\kappa_1^\ell \psi_1^{d_1} \ldots \psi_{n}^{d_n} \ ,
  \quad \ell,d_1,\ldots,d_n \in \mathbb{Z}_{\ge 0} \ ,
\end{equation}  
where the classes $\tau_{d_i}$ are the conventional abbreviations for $\psi_i^{d_i}$ arising from the $i$-th marked point. The defined correlators are only non-vanishing if the integrated class represents a (non-zero) top class of $\overline{\mathcal{M}}_{g,n}$, which together with eq.~\eqref{eq:dimM} amounts to the selection rule
\begin{equation} \label{selectionrules}
    \left\langle \kappa_1^\ell\tau_{d_1} \ldots \tau_{d_n} \right\rangle_{g,n} \ne 0 \quad \Rightarrow \quad
    \ell + d_1 + \ldots + d_n = 3g - 3 + n \ .
\end{equation}    
For these correlators we introduce the generating functions \cite{WittenIntersection}
\begin{equation} \label{generatingfunction}
  F(\{t_k\}) = \sum_{g=0}^{+\infty} g_s^{2g} \left\langle e^{\sum_{d=0}^{\infty} t_d \tau_d} \right\rangle_g 
   =\sum_{g=0}^{+\infty} g_s^{2g} \sum_{\{n_d\}} \left(\prod_{d=0}^{\infty}\frac{t_d^{n_d}}{n_d !} \right) \left\langle \tau_0^{n_0} \tau_1^{n_1}  \ldots \right\rangle_g \ ,
\end{equation}  
and 
\begin{multline}\label{eq:G}
  G(s,\{t_k\}) = \sum_{g=0}^{+\infty} g_s^{2g} \left\langle e^{s \kappa_1 + \sum_{d=0}^{\infty} t_d \tau_d} \right\rangle_g 
   =\sum_{g=0}^{+\infty} \sum_{m=0}^{+\infty}\frac{g_s^{2g} s^m}{m!}\sum_{\{n_d\}} \left(\prod_{d=0}^{\infty} \frac{t_d^{n_d}}{n_d !} \right)
   \left\langle \kappa_1^m \tau_0^{n_0} \tau_1^{n_1}  \ldots \right\rangle_g \ ,
\end{multline}  
in terms of the genus expansion parameter $g_s$ and the couplings $\{t_d\}$. Due to the relation~\eqref{eq:MMMcl} the two generating functions are not independent but instead are related as \cite{MR2379144,MR2482127,Dijkgraaf:2018vnm}
\begin{equation}\label{eq:gammak}
  G(s,\{t_k\}) = F(\{t_k + \gamma_k\}) \ , \qquad \gamma_0 = \gamma_1 = 0 \ , \quad \gamma_k = \frac{(-1)^k}{(k-1)!} s^{k-1} \ .
\end{equation}

As the first Miller--Morita--Mumford class $\kappa_1$ is proportional to the Weil--Petersson K\"ahler form $\omega_\text{WP}$ (cf.\ eq.~\eqref{eq:WPKahler}), the function $G(2\pi^2,\{t_k=0\})$ evaluated at $t_k=0$ readily becomes the generating function of the Weil--Petersson volumes $V_g$ of the moduli space of genus $g$ curves (for $g\ge 2$) without any marked points, i.e.\
\begin{equation}
   G(2\pi^2,\{t_k=0\}) = \sum_{g=2}^{+\infty} g_s^{2g} \int_{\overline{\mathcal{M}}_{g,0}} e^{\omega_\text{WP}} =  \sum_{g=2}^{+\infty} g_s^{2g} \int_{\overline{\mathcal{M}}_{g,0}} \text{vol}_\text{WP} 
   = \sum_{g=2}^{+\infty}g_s^{2g} V_g \ .
\end{equation}
Here $\text{vol}_\text{WP}$ is the Weil--Petersson volume form of the $(3g-3)$-dimensional moduli space $\overline{\mathcal{M}}_{g,0}$. 

As shown in the seminal work~\cite{MR2257394} by Mirzakhani, the Weil--Petersson volume of a hyperbolic Riemann surfaces of genus $g$ with $n$ geodesic boundary components of length $\vec b=(b_1,\ldots,b_n)$ reads in terms of the previously defined cohomology classes on $\overline{\mathcal{M}}_{g,n}$
\begin{equation}\label{WPVolumes}
	V_{g,\vec{b}}=\int_{\overline{\mathcal{M}}_{g,n}}e^{\omega_\text{WP}+\frac{1}{2}\sum_{\ell=1}^n b_\ell^2\psi_\ell}
	=\left\langle{e^{2\pi^2\kappa_1+\frac{1}{2}\sum_{\ell=1}^{n} b_\ell^2\psi_\ell}}\right\rangle_{g,n} \ .
\end{equation}
For hyperbolic Riemann surfaces with geodesic boundary components of uniform length $b$, using eq.~\eqref{topologicalgravitycorrelationfunctions} it is straightforward to verify that the volumes $V_{g,\left(b,\ldots,b\right)}$ are generated by
\begin{equation}
  G(2\pi^2,\{t_k =  \tfrac{b^{2k}}{2^k k!}\delta  \}) = \sum_g g_s^{2g} \sum_{i=0}^{+\infty}
  \frac{\delta^i}{i!} \, V_{g,(\smallunderbrace{b,\ldots,b}_{i\ \text{times}})} \ ,
\end{equation}
or upon rescaling all cohomology classes with a non-zero factor $\lambda$ we obtain with eq.~\eqref{eq:dimM} the generating function
\begin{equation}
  G(2\pi^2\lambda,\{t_k =  \tfrac{\lambda^k b^{2k}}{2^k k!}\delta  \}) = \sum_g \frac{g_s^{2g}}{\lambda^3} \sum_{i=0}^{+\infty}
  \frac{(\lambda\delta)^i}{i!} \, \lambda^{3g} V_{g,(\smallunderbrace{b,\ldots,b}_{i\ \text{times}})} \ .
\end{equation}
For this generating function of Weil--Petersson volumes (and similarly for all other generating functions of Weil--Petersson volumes to be defined in the following), the volumes $V_{g,(b,\ldots,b)}$ that are not in accord with the selection rule~\eqref{selectionrules} are set to zero.\footnote{This in particular implies that the Weil--Petersson volumes are only non-vanishing for stable curves, with the only exception being the Weil--Petersson volume $V_{1}$ for $g=1$ and $n=0$, which is either set to zero or to a constant, see for instance the discussion in ref.~\cite{WittenIntersection}. In this work, however, the volume $V_{1}$ is not relevant as we only consider Riemann surfaces with at least one boundary component.} Furthermore, for boundary components with $p$ distinct geodesic length $b_1,\ldots,b_p$, this generating function readily generalises to
\begin{equation}
  G(2\pi^2\lambda,\{ t_k\! =\!  \sum_{i=1}^p\tfrac{\lambda^kb_i^{2k}}{2^k k!}\delta_j \} )=
  \sum_g \frac{g_s^{2g}}{\lambda^3} \!\!\!\sum_{i_1,\ldots,i_p = 0}^{+\infty} \left(\prod_{s=1}^{p} \frac{(\lambda\delta_s)^{i_s}}{i_s !} \right)
  \lambda^{3g} V_{g,(\smallunderbrace{ b_1,\ldots,b_1}_{i_1\ \text{times}},\ldots ,\smallunderbrace{ b_p,\ldots,b_p}_{i_p\ \text{times}})} \ .
\end{equation}
Finally, a hyperbolic Riemann surface with a conical singularity with identification angle $\alpha$ can simply be obtained by replacing the argument $b$ of a boundary component by $i\alpha$ (for the identification angles in the range $0<\alpha_i<\pi$).\footnote{The identification angle $\alpha$ of a conical singularity corresponds to the deficit angle $2\pi - \alpha$ of the singularity.} Thus, the Weil--Petersson volume $V_{g,\vec b, \vec\alpha}$ of a hyperbolic Riemann surface with boundary components of geodesic lengths $\vec b=(b_1,\ldots,b_p)$ and together with conical singularities $\vec\alpha=(\alpha_1,\ldots,\alpha_q)$ is given by
\begin{equation}\label{eq:conicalWPvolumes}
    V_{g,\vec b,\vec\alpha} = V_{g,(b_1 ,\ldots, b_p,i \alpha_1, \ldots, i \alpha_q)} \ .
\end{equation}
Moreover, the generating function for hyperbolic Riemann surfaces with boundary components of geodesic lengths $b_1,\ldots,b_p$ and conical singularities of with identification angles $\alpha_1,\ldots,\alpha_q$ becomes in terms of the non-zero parameter $\lambda$
\begin{multline} \label{eq:GGeneral}
  G(2\pi^2\lambda,\{ t_k =\sum_{i=1}^p\tfrac{\lambda^kb_i^{2k}}{2^k k!}\delta_i +  \sum_{j=1}^q\tfrac{\lambda^k(-\alpha_j^2)^k}{2^k k!}\epsilon_j \} ) \\
 = \!\sum_g \frac{g_s^{2g}}{\lambda^3} \!\!\!\!\!\sum_{\substack{i_1,\ldots,i_p = 0\\ j_1,\ldots,j_q = 0}}^{+\infty} \!\!
 \! \left(\!\prod_{s=1}^{p} \frac{(\lambda\delta_s)^{i_s}}{i_s !}\!\prod_{t=1}^{q} \frac{(\lambda\epsilon_t)^{j_t}}{j_t !} \!\!\right) 
  \!\lambda^{3g}V_{g,(\smallunderbrace{ b_1,\ldots,b_1}_{i_1\, 
  \text{times}},\ldots ,\smallunderbrace{ b_p,\ldots,b_p}_{i_p\, \text{times}}),
          (\smallunderbrace{\alpha_1,\ldots,\alpha_1}_{j_1\, \text{times}},\ldots ,\smallunderbrace{ \alpha_q,\ldots,\alpha_q}_{j_q\, \text{times}})} 
          %\ 
          .
\end{multline}

%%%%%%%%%%%%%%%%%%%%%%%
\subsection{Deformations of JT~Gravity from Minimal Strings} \label{sec:TopgravityMMstringsandMM}
%%%%%%%%%%%%%%%%%%%%%%%
Before delving into the technical computation of the thermal partition functions of JT~gravity with deformations, in this subsection we briefly spell out the connections among topological gravity, minimal string theories, and JT~gravity. This puts the forthcoming analysis into a broader context.

Saad, Shenker and Stanford already point out that standard JT~gravity relates to the large $p$ limit of the $(2,2p-1)$ minimal string theory \cite{SSS}. Such minimal string theories in turn enjoy a dual matrix model formulation \cite{Douglas:1989ve,KazakovMM,StaudacherMM}, which for finite $p$ comes with a finite number of coupling parameters. In the large $p$ limit, however, an infinite (but countable) number of couplings occur, which for standard JT~gravity are set to specific non-zero values. Furthermore, this infinite number of couplings relate to observables and their correlators in two-dimensional topological gravity, as introduced in the previous subsection.

In the following, as in ref.~\cite{OkuyamaSakai1}, using the connection to topological gravity we want to compute thermal partition functions as a function of this infinite number of couplings in order to describe JT~gravity and deformations thereof. In other words, instead of solely focussing on particular deformation backgrounds --- such as JT~gravity without deformations or JT~gravity interacting with a gas of defects --- we parametrise generic deformations to JT~gravity  in terms of deformations of the $(2,2p-1)$ minimal string theories in the large $p$ limit, using the results of ref.~\cite{Itzykson:1992ya}.

Starting from a JT~gravity action formulation the values of the deformation parameters are ultimately determined from the constraints obtained from integrating out the scalar dilaton field. For instance, JT~gravity coupled to a gas of defects yields the constraint \eqref{eq:Dilatonpotential}, which is dual to specific values of the topological gravity coupling parameters. For a given JT~gravity action functional --- such as JT~gravity interacting with defects --- we refer to coupling values that fulfill these constraints as on-shell couplings and couplings that deviate from this critical condition as off-shell couplings (adapting to a terminology introduced in ref.~\cite{OkuyamaSakai1}). 

Turning this argument around, we can now ask whether specific values for these couplings correspond to a legitimate action functional of a deformed theory of JT~gravity. Intriguingly, as discussed in the following both JT~gravity and JT~gravity coupled to defects give rise to on-shell couplings that are governed by Bessel functions \cite{CJ1,OkuyamaSakai1,WittenDeformations,Maxfield3gravity}. The problem of establishing a dictionary between these deformation spaces raises the question to what extend other transcendental functions for on-shell couplings are linked to action functionals of deformed JT~gravity theories (see, e.g. ref.~\cite{Okuyama:2020qpm} for the realisation of JT~supergravity). For finite $p$ the $(2,2p-1)$ minimal string theories possess a finite dimensional deformation space resulting from finitely many couplings~$t_k$. In the considered limit $p\to\infty$, the deformations $\sdef_k$ in eq.~\eqref{eq:shiftedexpansionpoint} can be characterized by their asymptotic behaviour for large $k$. The values for the couplings $t_k$ for undeformed JT~gravity are suppressed factorially (cf.~eq.~\eqref{eq:OSexpansionpoint}). For deformations arising from a gas of defects (at least for only finitely many types of defect species) the asymptotic behaviour of the couplings~$t_k$ for large $k$ remains the same. On the level of the action functional of JT~gravity such deformations give rise to a scalar potential~\eqref{eq:Dilatonpotential} that is exponentially suppressed for large positive values of the dilaton $\phi$. In general, we expect that the asymptotic behaviour of the scalar potential $U(\phi)$ for large $\phi$ relates to the asymptotic behaviour of the deformations $\sdef_k$ for large $k$.\footnote{Ref.~\cite{Mertens:2020hbs} makes an interesting proposal for a correspondence between a certain limit of Liouville theory coupled to matter and JT~gravity with a $\sinh(\phi)$-dilaton potential with a different asymptotic behaviour for $\phi\to+\infty$ (see also ref.~\cite{Kyono:2017pxs}).} Describing this duality beyond the discussed asymptotic growth behaviours seems a challenging task, which is beyond the scope of this work. Nevertheless, we hope that the description of generic deformations in the context of $(2,2p-1)$ minimal string theories in the large $p$ limit presented here proves useful from the JT~gravity perspective as well.

%%%%%%%%%%%%%%%%%%%%%%%
\subsection{JT Gravity Interacting with a Gas of Defects}\label{JT Partition Function and Specific Coupling Background}
%%%%%%%%%%%%%%%%%%%%%%%
We now study JT gravity interacting with a gas of defects, which is geometrically described in terms of Riemann surfaces with conical singularities \cite{WittenDeformations,Maxfield3gravity}. That is to say, we consider the partition function of JT gravity with contributions from hyperbolic Riemann surfaces with asymptotic boundary conditions together with an arbitrary number of conical singularities and at arbitrary genus. The relevant path integrals localise on the Weil--Petersson volumes of hyperbolic Riemann surfaces with geodesic boundary components and conical defects, folded with the path integral of the Schwarzian theory describing the one-dimensional action at the asymptotic boundaries \cite{SSS}. For a single asymptotic boundary component the resulting partition function reads \cite{WittenDeformations,Maxfield3gravity}
\begin{multline} \label{eq:JTPartitionFunctionDefects}
  Z(\beta) = e^{S_0} Z^{\text{disk}}(\beta) +e^{S_0}\sum_{j=1}^{r}\epsilon_{j}Z^{\text{disk}}(\beta,\alpha_j)\\
  +\sum_{g,n=0}^{\infty}e^{(1-2g)S_0}
   \sum_{j_1,\ldots,j_{n}=1}^{r}\frac{\epsilon_{j_1}\cdots\epsilon_{j_{n}}}{n!}\int_{0}^{\infty}db\,b\,
   Z^{\text{trumpet}}(\beta,b)V_{g,b,(\alpha_{j_1},\ldots,\alpha_{j_{n}})} \ .
\end{multline}
Here the parameters $\epsilon_j$, $j=1,\ldots,r$, are the coupling constants to the $r$ distinct defect types that are characterised by the identification angles $\alpha_j$ of their associated conical singularities on the hyperbolic Riemann surfaces. Furthermore, $\beta$ is the inverse temperature attributed to the configurations of wiggles at the asymptotic boundary of the hyperbolic Riemann surfaces. The distinct topologies of Riemann surfaces are weighted by the action $S_0$ that relates to the gravitational coupling $G_N$ as $G_N \sim 1/S_0$. Hence, the partition function is a non-perturbative expansion in the gravitational coupling $G_N$ of JT gravity \cite{SSS}. The first two terms in this expansion capture the contributions of disks with no conical singularities and a single conical singularity, respectively. The remaining topologies appear in the second line.\footnote{Due to the selection rules~\eqref{selectionrules} for non-vanishing Weil--Petersson volumes $V_{g,b,\vec a}$, the second line of eq.~\eqref{eq:JTPartitionFunctionDefects} does not contain a contribution from disks without any or with a single conical singularity.} The individual terms in this expansion are computed as \cite{WittenStanford,SSS,MertensDefects}
\begin{equation} \label{eq:BuildBlocks}
  Z^\text{disk}(\beta)=\frac{\gamma^{\frac32}e^{\frac{2\pi^2\gamma}{\beta}}}{(2 \pi)^\frac12 \beta^{\frac32}}\ , \quad
  Z^\text{disk}(\beta,\alpha_j)=\frac{\gamma^\frac12e^{\frac{\gamma \alpha_j^2}{2\beta}}}{(2 \pi\beta)^{\frac{1}{2}}} \ , \quad
  Z^{\text{trumpet}}(\beta,b)=\frac{\gamma^\frac12e^{-\frac{\gamma b^2}{2\beta}}}{(2 \pi\beta)^{\frac{1}{2}}}\ ,
\end{equation}
where $\gamma$ is the coupling constant to the one-dimensional Schwarzian action.

First we observe that the summation over defects in eq.~\eqref{eq:JTPartitionFunctionDefects} can be rewritten as
\begin{equation}
     \sum_{n=0}^{+\infty}\sum_{j_1,\ldots,j_{n}=1}^{r}\frac{\epsilon_{j_1}\cdots\epsilon_{j_{n}}}{n!}V_{g,b,(\alpha_{j_1},\ldots,\alpha_{j_{n}})} 
     = \sum_{n_1,\ldots,n_r = 0}^{+\infty} \left(\prod_{j=1}^r \frac{ \epsilon_j^{n_j}}{n_j !}\right)
     V_{g,b,(\smallunderbrace{ \alpha_1,\ldots,\alpha_1}_{n_1\ \text{times}},\,.\,.\,.\,,\smallunderbrace{ \alpha_r,\ldots,\alpha_r}_{n_r\ \text{times}})} \ .
\end{equation}
Summed over all genera $g$ we readily express the volumes $V_{g,b,(\alpha_{j_1},\ldots,\alpha_{j_n})}$ in terms of the generating function~\eqref{eq:GGeneral} as
\begin{align} 
     \sum_{g,n=0}^{+\infty} g_s^{2g} \!\!\!\!\!\!\sum_{j_1,...,j_{n}=1}^{r}\!\!\!\!\!\!
     \frac{\epsilon_{j_1}...\epsilon_{j_{n}}}{n!}\lambda^{3g} V_{g,b,(\alpha_{j_1},...,\alpha_{j_{n}})} &=
      \left. \lambda^2\frac{\partial}{\partial\delta} 
     G(2\pi^2\lambda,\{t_k =\tfrac{\lambda^kb^{2k}}{2^k k!}\delta +  \sum_{j=1}^r\tfrac{\lambda^{k-1}(-\alpha_j^2)^k}{2^k k!}\epsilon_j \} ) \right|_{\delta=0} 
     \nonumber\\
     &
     \hspace*{-2em}= \sum_\ell \frac{b^{2\ell}\lambda^{\ell+2}}{2^\ell \ell!} \frac{\partial}{\partial{t_\ell}}  
     G(2\pi^2\lambda,\{t_k\! =\!\sum_{j=1}^r\tfrac{\lambda^{k-1}(-\alpha_j^2)^k}{2^k k!}\epsilon_j  \} )  \ .
\label{eq:VAsGen1}
\end{align}
We insert this expression into eq.~\eqref{eq:JTPartitionFunctionDefects} with the relation 
\begin{equation} \label{eq:Defgs}
   e^{-S_0} = \lambda^\frac32 g_s \ ,
\end{equation}   
and carry out the integration over the geodesic boundary lengths in eq.~\eqref{eq:JTPartitionFunctionDefects} using
\begin{equation}\label{eq:bintegration}
    \int_{0}^{\infty}db\,b^{2n+1}\,e^{-\frac{\gamma b^2}{2\beta}} = \frac{n!}2 \left(\frac{2\beta}\gamma\right)^{n+1}  \ .
\end{equation}
Then we arrive for the partition function $Z(\beta)$ at
\begin{equation} \label{eq:Zsingle1}
   Z(\beta) = \frac{1}{\sqrt{2\pi} g_s} \!\left( \!\! \frac{\gamma}{\lambda \beta} \right)^\frac32 
    \!\left( e^{\frac{2\pi^2\gamma}\beta} \!\! + \frac\beta\gamma \sum_{j=1}^r \epsilon_j\,e^{\frac{\gamma \alpha_j^2}{2\beta}} 
  \!\! +  \sum_{\ell=0}^{+\infty} \left( \frac{\lambda\beta}{\gamma}\right)^{\ell+2} \!\!\!\frac{\partial}{\partial t_\ell} G(2\pi^2\lambda,\{ t_k = \sdef_k \}) 
  \! \!\right)  ,
\end{equation}
with
\begin{equation} \label{eq:defGtilde}
    \sdef_k = \sum_j \sdef_{k,j} \ , \qquad \sdef_{k,j} = \tfrac{\lambda^{k-1}(-\alpha_j^2)^k}{2^k k!}\epsilon_j  \ .     
\end{equation}
Note that only the last term of the partition function $Z(\beta)$ given in eq.~\eqref{eq:Zsingle1} is mapped to the topological correlators~\eqref{topologicalgravitycorrelationfunctions}, whereas the first two terms associated to disk topologies capture the semi-classical contributions to the partition function in the presence of a gas of defects. 

It is straightforward to generalise the partition function $Z(\beta)$ to geometries with multiple asymptotic boundaries \cite{Maxfield3gravity,WittenDeformations}. For $m$ boundaries we define the partition function of connected hyperbolic Riemann surfaces by $Z(\beta_1,\ldots,\beta_m)$, where the inverse temperatures $\beta_1, \ldots, \beta_m$ describe the thermodynamics of the wiggles at the $m$ distinct asymptotic boundary components. 

Similarly as for the partition function $Z(\beta)$ of a single asymptotic boundary, the partition function $Z(\beta_1,\beta_2)$ with two asymptotic boundaries splits into two pieces
\begin{equation} \label{eq:Zsplit}
    Z(\beta_1,\beta_2) = Z(\beta_1,\beta_2)^\text{non-top.} + Z(\beta_1,\beta_2)^\text{top.} \ .
\end{equation}    
The first term does not relate to topological correlators~\eqref{topologicalgravitycorrelationfunctions}, while the second term arises from an integral transformation of the Weil--Petersson volumes of hyperbolic Riemann surfaces with two geodesic boundary components that are computable in terms of topological correlators, cf. eqs.~\eqref{WPVolumes} and \eqref{eq:conicalWPvolumes}. The non-topological piece $Z(\beta_1,\beta_2)^\text{non-top.}$ receives only a contribution at genus zero from the topology of a cylinder (without any conical singularities). Using eqs.~\eqref{eq:BuildBlocks} and \eqref{eq:bintegration}, this cylindrical contribution is obtained by gluing two trumpets along their geodesic boundary components, as computed in ref.~\cite{SSS}
\begin{equation}
  Z(\beta_1,\beta_2)^\text{non-top.}=\int\displaylimits_{0}^{\infty}db \,b\, Z^{\text{trumpet}}(\beta_1,b)\, Z^{\text{trumpet}}(\beta_2,b)
     =\frac{\sqrt{\beta_1  \beta_2}}{2 \pi  \beta_1 +2 \pi  \beta_2 }\ .
\end{equation}
The selection rule~\eqref{selectionrules} implies that the partition functions $Z(\beta_1,\ldots,\beta_m)$ with $m>2$ receive only contributions of the topological type, i.e.\
\begin{equation}
    Z(\beta_1,\ldots,\beta_m) = Z(\beta_1,\ldots,\beta_m)^\text{top.} \quad \text{for} \quad m>2 \ .
\end{equation}
For any $m\ge 1$ the topological part of the partition function $Z(\beta_1,\ldots,\beta_m)$ reads
\begin{multline} \label{eq:DefZk}
   Z(\beta_1,\ldots,\beta_m)^\text{top.} = 
   \sum_{g,n=0}^{\infty}e^{(2-2g-m)S_0} 
   \sum_{j_1,\ldots,j_{n}=1}^{r}\frac{\epsilon_{j_1}\cdots\epsilon_{j_{n}}}{n!} \\
   \cdot \prod_{i=1}^m\int_{0}^{\infty} \!\!\! db_i\,b_i\,
   Z^{\text{trumpet}}(\beta_i,b_i)V_{g,(b_1,\ldots,b_m),(\alpha_{j_1},\ldots,\alpha_{j_{n}})} \ .
\end{multline}
Analogously to the formula~\eqref{eq:VAsGen1} for a single boundary component, we express the volumes $V_{g,(b_1,\ldots,b_m),(\alpha_{j_1},\ldots,\alpha_{j_n})}$ in terms of the generating function~\eqref{eq:GGeneral} as
\begin{multline}
    \sum_{g,n} g_s^{2g} \!\! \sum_{j_1,\ldots,j_{n}=1}^{r} \lambda^{3g} \frac{\epsilon_{j_1}\cdots\epsilon_{j_{n}}}{n!}
    V_{g,(b_1,\ldots,b_m),(\alpha_{j_1},\ldots,\alpha_{j_{n}})} \\
    = \lambda^{3-m} \prod_{i=1}^m \left( \sum_{\ell=0}^{+\infty} \frac{b_i^{2\ell} \lambda^\ell}{2^\ell \ell!} \frac{\partial}{\partial t_\ell} \right)
    G(2\pi^2\lambda,\{t_k = \sum_{j=1}^r\tfrac{\lambda^{k-1}(-\alpha_j^2)^k}{2^k k!}\epsilon_j  \} )  \ .
\end{multline}  
Inserting this expression into eq.~\eqref{eq:DefZk} and carrying out the integrals~\eqref{eq:bintegration}, we obtain 
\begin{equation} \label{eq:Zmultitop}
   Z(\beta_1,\ldots,\beta_m)^\text{top.} = \frac1{g_s^2} \mathcal{B}(\beta_1) \cdots \mathcal{B}(\beta_m) G(2\pi^2 \lambda,\{ t_k = \sdef_k \} ) 
   \quad\text{for}\quad m \ge 1 \ ,
\end{equation}
with $\delta_k$ as defined in eq.~\eqref{eq:defGtilde} and in terms of the differential operator
\begin{equation}
  \mathcal{B}(\beta) = g_s \sqrt{ \frac{\lambda \beta}{2\pi\gamma} } \, 
  \sum_{\ell=0}^{+\infty} \left( \frac{\lambda\beta}{\gamma} \right)^\ell \frac{\partial}{\partial t_\ell} \ .
\end{equation}
It is shown in ref.~\cite{OkuyamaSakai2} that the differential operator $\mathcal{B}(\beta)$ creates an asymptotic boundary component at temperature $\beta$. It is universal in the sense that without any modifications it also creates asymptotic boundary components in the presence of defects. The operator $\mathcal{B}(\beta)$ as a function of $\beta$ relates to the operator in ref.~\cite{Moore:1991ir}, which in the context of two-dimensional topological gravity creates in a surface a hole of specified boundary length. Therefore, we refer to $\mathcal{B}(\beta)$ as the boundary creation operator.

The obtained simple forms~\eqref{eq:Zsingle1} and \eqref{eq:Zmultitop} of the partition function $Z(\beta)$ and its multi-boundary generalisations $Z(\beta_1,\ldots,\beta_m)$ in the presence of a gas of defects have a nice interpretation from the topological gravity perspective. The Weil--Petersson volumes~\eqref{WPVolumes} are computed with the K\"ahler class $2\pi^2 \kappa_1$ on the moduli spaces $\overline{\mathcal{M}}_{g,n}$ \cite{MR2257394}. The generating function $G(2\pi^2\lambda,\{ t_k \})$ now expresses these volumes (as functions of the scaling and genus expansion parameters $\lambda$ and $g_s$) in terms of the shifted generating function $F(\{ t_k + \gamma_k \})$ of topological gravity according to eq.~\eqref{eq:gammak}. As explained in ref.~\cite{SSS,OkuyamaSakai1}, JT gravity can be interpreted as topological gravity with non-vanishing background parameters~$\{\gamma_k\}$. Including now a gas of defects (characterised by their couplings $\epsilon_j$ and identification angles $\alpha_j$) further deforms the background couplings $\{\gamma_k\}$. The leading order contribution arises from single-defect interactions while the higher order corrections are due to multi-defect interactions. These order-by-order contributions can be viewed as a Taylor expansion about the JT~gravity background parameters $\{\gamma_k\}$, which altogether sum up to the deformation $\{\gamma_k +\sdef_k\}$. Thus, JT~gravity interacting with a gas of defects yields yet other expansion points of the generating function $F(\{t_k\})$. It would be interesting to see if there are special expansion points that are singled out from the topological gravity point of view.

As in ref.~\cite{OkuyamaSakai1}, in the following we set the coupling $\gamma$ and the scaling parameter $\lambda$ to the convenient values
\begin{equation}
  \lambda = \gamma = \frac1{2\pi^2} \ .
\end{equation}
Then the boundary creation operator $\mathcal{B}(\beta)$ and the background parameters $\sdef_k$ simplify to
\begin{equation} \label{eq:dshifts}
    \mathcal{B}(\beta) = g_s \sqrt{ \frac{\beta}{2\pi} } \, \sum_{\ell=0}^{+\infty} \beta^\ell \frac{\partial}{\partial t_\ell} \ , \qquad
    \sdef_k = \sum_j \left(-\frac{\alpha_j^2}{4\pi^2}\right)^k\frac{2\pi^2 \epsilon_j}{k!} \ ,
\end{equation}
and the partition functions become
\begin{equation} \label{eq:Zfuncs}
\begin{aligned}
   Z(\beta) &= \frac{1}{\sqrt{2\pi} g_s \beta^\frac32}  
    \left( e^{\frac{1}\beta} + 2\pi^2\beta \sum_{j=1}^r \epsilon_j\,e^{\frac{\alpha_j^2}{4\pi^2\beta}} \right)
   + \frac1{g_s^2} \mathcal{B}(\beta) G(1,\{ t_k=\sdef_{k} \}) \ , \\
   Z(\beta_1,\beta_2) &= \frac{\sqrt{\beta_1  \beta_2}}{2 \pi  \beta_1 +2 \pi  \beta_2 }
   +\frac1{g_s^2} \mathcal{B}(\beta_1) \mathcal{B}(\beta_2) G(1,\{ t_k = \sdef_k \}) , \\[1.5ex]
   Z(\beta_1,\ldots,\beta_m) &= \frac1{g_s^2} \mathcal{B}(\beta_1) \cdots \mathcal{B}(\beta_m) G(1,\{ t_k = \sdef_k \}) \quad \text{for}\quad m\ge 3\ ,
\end{aligned}
\end{equation}
where the first two partition functions receive both non-topological and topological contributions.
%%%%%%%%%%%%%%%%%%%%%%%%%%%%%%%%%%%%%%%%%%%%%%%%%%%%%%%%%%%%%%

%%%%%%%%%%%%%%%%%%%%%%%%%%%%%%%%%%%%%%%%%%%%%%%%%%%%%%%%%%%%%%
\subsection{KdV Hierarchy and Off-Shell Partition Functions} \label{sec:KdV}
%%%%%%%%%%%%%%%%%%%%%%%%%%%%%%%%%%%%%%%%%%%%%%%%%%%%%%%%%%%%%%
As conjectured by Witten~\cite{WittenIntersection} and proven by Kontsevich~\cite{KontsevichIntersection} the generating function $F(\{t_k\})$ of correlators in topological gravity defined in eq.~\eqref{generatingfunction} arises as a solution to the KdV hierarchy as follows. Let us define
\begin{align}\label{eq:Ftou}
    u(\{t_k\}) = \frac{\partial^2}{\partial t_0^2}F(\{t_k\})  \ . %, \qquad \hbar = \sqrt{2} g_s  \ .
\end{align}
The function $u(\{t_k\})$ is a tau function to the KdV~hierarchy, i.e.\ it solves the system of graded partial differential equations
\begin{equation} \label{eq:GeneralizedKdV}
   \partial_k u = \partial_0 \mathcal{R}_{k+1}(u,\partial_0u,\partial_0^2u,\ldots)\quad  \text{with} \quad
   \partial_k \equiv \frac{\partial}{\partial t_k} \ , \quad k=0,1,2,3,\ldots \ .
\end{equation}
Here $\mathcal{R}_k$, $k=1,2,3,\ldots$, are the Gelfand--Dikii polynomials \cite{Gelfand:1975rn}, which are polynomials in the derivatives $\partial_0^\ell u(\{t_k\})$, $\ell=0,1,2,\ldots$ of $u(\{t_k\})$, and depend on the parameter $g_s$. Together with the condition $\mathcal{R}_k(\{\partial_0^\ell u \equiv 0 \}) = 0$ they are defined with the initial polynomial $\mathcal{R}_1=u$ recursively as \cite{Gelfand:1975rn}
\begin{equation} \label{eq:generalKdvequation}
     \partial_0 \mathcal{R}_{k+1}=\frac{1}{2k+1}\left( 
    2 u \left(\partial_0 \mathcal{R}_{k}\right) +  \left(\partial_0 u\right) \mathcal{R}_{k} + \frac{g_s^2}{4}\partial_0^3 \mathcal{R}_{k} \right) \ .
\end{equation}
The first three Gelfand--Dikii polynomials read
\begin{equation} \label{eq:GD}
    \mathcal{R}_1 = u \ , \quad 
    \mathcal{R}_{2}=\frac{u^2}{2}+\frac{g_s^2 }{12}\partial_0^2 u \ , \quad
    \mathcal{R}_{3}=\frac{u^3}{3!}+\frac{g_s^2}{24}\left(2 u \partial_0^2 u + (\partial_0u)^2 \right) +\frac{g_s^4}{240} \partial_0^4u \ .
\end{equation}
The leading order term of the Gelfand--Dikii polynomials is given by
\begin{equation} \label{eq:Rk_u0}
  \left. \mathcal{R}_k \right|_{g_s=0} = \frac{u^k}{k!} \ ,
\end{equation}
independent of any derivatives $\partial_0 u(\{t_k\})$, $\partial_0^2 u(\{t_k\})$, $\partial_0^3 u(\{t_k\})$, $\ldots$. 

As the KdV hierarchy~\eqref{eq:GeneralizedKdV} depends only implicitly on the couplings $t_k$, the function $v(\{t_k\})= \partial_0^2 F(\{ t_k + \Delta t_k \})$ is a tau function for any set of constants $\{\Delta t_k\}$. In particular, a tau function arises from the generating function~$G(s,\{ t_k \})$ of Weil--Petersson volumes (cf.\ eq.~\eqref{eq:gammak}) and from the generating function $H(\{t_k\})$ of correlators on hyperbolic Riemann surfaces with conical singularities given by
\begin{equation} \label{eq:DefH}
   H(\{ t_k \}) = G(1,\{t_k + \sdef_k \}) = F(\{t_k + \gamma_k + \sdef_k\}) \ ,
\end{equation}
in terms of the constants $\Delta t_k=\gamma_k+\sdef_k$, cf. eqs.~\eqref{eq:gammak} and \eqref{eq:dshifts}.

The particular tau function $u(\{t_k\})$ of topological gravity and hence the tau function $v(\{t_k\})$ with the shifted couplings obey the string equation \cite{Dijkgraaf:1990rs}
\begin{equation} \label{eq:stringeq}
   \partial_0 u = 1 + \sum_{k=1}^{+\infty} t_k \partial_k u \ , \qquad  
   \partial_0 v = 1 + \sum_{k=1}^{+\infty} (t_k + \Delta t_k) \,  \partial_k v \ . 
\end{equation}
The string equation together with the KdV hierarchy determine unambiguously the tau functions $u(\{t_k\})$ and $v(\{t_k\})$ \cite{WittenIntersection}. The string equation can be viewed as the initial condition specifying a unique solution to the KdV hierarchy.

The partition functions $Z(\beta_1,\ldots,\beta_m)$ defined in eq.~\eqref{eq:Zfuncs} do not depend on the coupling parameters $\{t_k\}$ appearing in the definition of $H(\{t_k\})$. Instead the generating function $H(\{ t_k \})$ is evaluated at the specific values $t_k=0$ (corresponding to $t_k =\gamma_k + \sdef_k$ in terms of the generating functions $F(\{t_k\})$). We can define partition functions $Z^F(\{ t_k \}; \beta_1,\ldots,\beta_m)$ based on $F(\{ t_k \})$ or alternatively the partition functions $Z^H(\{ t_k \}; \beta_1,\ldots,\beta_m)$ based on $H(\{ t_k \})$ depending on $\{ t_k \}$ by generalising the topological part in eqs.~\eqref{eq:Zfuncs} to
\begin{equation} \label{eq:offshellPart}
\begin{aligned}
   Z^F(\{ t_k \}; \beta_1,\ldots,\beta_m)^\text{top.} &= \frac{1}{g_s^2} \mathcal{B}(\beta_1) \cdots \mathcal{B}(\beta_m) F(\{ t_k \}) \ , \\
   Z^H(\{ t_k \}; \beta_1,\ldots,\beta_m)^\text{top.} &= \frac{1}{g_s^2} \mathcal{B}(\beta_1) \cdots \mathcal{B}(\beta_m) H(\{ t_k \}) \ .
 \end{aligned}
 \end{equation}
Following ref.~\cite{OkuyamaSakai1} we refer to $Z^F(\{ t_k \}; \beta_1,\ldots,\beta_m)$ and $Z^H(\{ t_k \}; \beta_1,\ldots,\beta_m)$ as the off-shell partition functions, and upon specialising to suitable values for the couplings $\{t_k\}$ --- denoted as on-shell values --- we get back the result $Z(\beta_1,\ldots,\beta_m)$ referred to as the on-shell partition function, i.e.\
\begin{equation} \label{eq:onshellPart}
\begin{aligned}
   Z(\beta_1,\ldots,\beta_m) &= Z^F(\{ t_k=\gamma_k + \delta_k \}; \beta_1,\ldots,\beta_m) \ , \\
   Z(\beta_1,\ldots,\beta_m) &= Z^H(\{ t_k = 0\}; \beta_1,\ldots,\beta_m) \ .
\end{aligned}   
\end{equation}
These two classes of off-shell partition functions enjoy distinct interpretations. Whereas the off-shell partition function $Z^F(\{ t_k \}; \beta_1,\ldots,\beta_m)$ is defined in the setting of topological gravity in the context of intersection theory on the moduli spaces of stable curves \cite{WittenIntersection,KontsevichIntersection}, the partition functions $Z^H(\{ t_k \}; \beta_1,\ldots,\beta_m)$ directly relate to correlators on hyperbolic Riemann surfaces (possibly coupled to a gas of defects as described by the constants $\{ \delta_k \}$) in the context of JT~gravity \cite{SSS,OkuyamaSakai1}. These two classes of off-shell partition functions are related as $Z^F(\{ \gamma_k + \delta_k + t_k \}; \beta_1,\ldots,\beta_m)=Z^H(\{ t_k \}; \beta_1,\ldots,\beta_m)$. 

Let us now determine the introduced off-shell partition functions explicitly. The tau function~\eqref{eq:Ftou} and the generating function~$F(\{t_k\})$ enjoy the genus expansion
\begin{equation}
   u(\{t_k\}) = \sum_{\ell=0}^{+\infty} g_s^{2\ell} \, u_\ell (\{ t_k \}) \ , \qquad
   F(\{t_k\}) = \sum_{\ell=0}^{+\infty} g_s^{2\ell} \, F_\ell(\{t_k \}) \ ,
\end{equation}
such that $F_g = \partial_0^2 u_g$. The KdV hierarchy~\eqref{eq:GeneralizedKdV} with eq.~\eqref{eq:Rk_u0} and the string equation~\eqref{eq:stringeq} imply for the genus zero contribution the partial differential equations
\begin{equation} \label{eq:u0diffeq}
   \partial_k u_0 = \frac{\partial_0 u_0^{k+1}}{(k+1)!}  \ , \qquad \partial_0 u_0 = 1 + \sum_{k=1}^{+\infty} t_k \,\partial_k u_0 \ .
\end{equation}
Defining the series 
\begin{equation} \label{eq:DefI}
  I_n(u_0,\{ t_k \}) = \sum_{k=0}^{+\infty} t_{k+n} \frac{u_0^k}{k!} \quad \text{for} \quad n=0,1,2,\ldots \ , 
\end{equation}
and using the partial differential equations~\eqref{eq:u0diffeq}, Itzykson and Zuber show for the genus zero part $u_0(\{ t_k \})$ of the tau function~$u(\{t_k\})$ the remarkable functional relation \cite{Itzykson:1992ya}
\begin{equation}
   u_0 - I_0(u_0,\{t_k\}) = 0 \ .
\end{equation}
With the ansatz $u_0(\{t_k\}) =  \sum_{N=0}^{+\infty} \sum_{{\sum n_k = N}} u_{0,\{n_k\}} t_0^{1-N+\sum k n_k} \left(t_1^{n_1} t_2^{n_2} \cdots\right)$ summed over non-negative integral sets $\{n_k\}$, one readily determines order-by-order the formal expansion in the coupling parameters $\{t_k\}$
\begin{equation} \label{eq:u0exp}
  u_0(\{t_k\}) = t_0 + t_0 t_1 + \left( t_0 t_1^2  + \frac12 t_0^2 t_2 \right) +\left( t_0 t_1^3 + \frac32 t_0^2 t_1 t_2 + \frac16 t_0^3 t_3 \right) + \ldots \ .
\end{equation}
Imposing the correct boundary conditions, the function $u_0$ integrates to \cite{Itzykson:1992ya}
\begin{equation} \label{eq:F0}
  F_0(u_0,\{t_k\}) =  \frac{u_0^3}{3!} - \sum_{k=0}^{+\infty} t_k \frac{u_0^{k+2}}{(k+2)k!} + \frac12 \sum_{k=0}^{+\infty} \frac{u_0^{k+1}}{k+1}
  \sum_{n=0}^k \frac{t_n t_{k-n}}{n!(k-n)!} \ .
\end{equation}
Furthermore, observing that the functions~\eqref{eq:DefI} obey the differential identities
\begin{equation}
  \partial_0 I_0 = \frac{1}{1-I_1} \ , \qquad \partial_0 I_k = \frac{I_{k+1}}{1-I_1} \quad \text{for} \quad k\ge 1 \ 
\end{equation}
Itzykson and Zuber establish that the KdV~hierarchy implies at higher genus the finite non-trivial expansions \cite{Itzykson:1992ya}
\begin{equation} \label{eq:ug}
  u_g = (1- I_1)^{g-1} \sum_{\sum_{k=2}^{3g} (k-1) \ell_k =3 g - 1}  u_{g,\{\ell_k\}} \left( \frac{I_2}{(1-I_1)^2} \right)^{\ell_2} \cdot \ldots \cdot \left( \frac{I_{3g}}{(1-I_1)^{3g}} \right)^{\ell_{3g}} \ .
\end{equation}
Inserting this ansatz into the KdV hierarchy~\eqref{eq:GeneralizedKdV} (recursively in the genus) determines unambiguously the numerical cofficients $u_{g,\{\ell_k\}}$, for instance up to genus $g=2$ we arrive at
\begin{align} 
  u_1 &=  \frac1{12} \left( \frac{I_2}{(1-I_1)^2} \right)^2 + \frac1{24} \frac{I_3}{(1-I_1)^3} \ , \label{eq:u1} \\
  u_2 &=  (1-I_1) \left(\frac{49 I_2^5}{288 (1-I_1)^{10}}+\frac{11 I_3I_2^3}{36 (1-I_1)^9}
    +\frac{7 I_4 I_2^2}{96(1-I_1)^8}+\frac{109 I_3^2 I_2}{1152 (1-I_1)^8}\right.\nonumber\\
    &\qquad\qquad\qquad\qquad\qquad\qquad+\left.\frac{I_5 I_2}{90 (1-I_1)^7}+\frac{17 I_3 I_4}{960 (1-I_1)^7}+\frac{I_6}{1152 (1-I_1)^6}\right) \ . \label{eq:u2}
\end{align}
At genus one $u_1(\{t_k\})$ integrates to
\begin{equation} \label{eq:F1def}
   F_1  =  -\frac{1}{24}\log(1 - I_1)  \ .
\end{equation}
The generating functions $F_g$ for $g>1$ enjoy yet again an expansion of the form \cite{Itzykson:1992ya}
\begin{equation} \label{eq:Fgstructure}
  F_g = (1- I_1)^{g-1} \sum_{\sum_{k=2}^{3g-2} (k-1) \ell_k =3 g - 3}  f_{g,\{\ell_k\}} \left( \frac{I_2}{(1-I_1)^2} \right)^{\ell_2} \cdot \ldots \cdot \left( \frac{I_{3g-2}}{(1-I_1)^{3g-2}} \right)^{\ell_{3g-2}}  \ ,
\end{equation}
in terms of the finitely many coefficients $f_{g,\{\ell_k\}}$ (with the subscript $\{\ell_k\}=\{\ell_2,\ell_3,\ldots\}$). In particular, with eq.~\eqref{eq:u2} we find for $g=2$ the numerical coefficients
\begin{equation} \label{eq:CoeffF2}
  f_{2,{\{3\}}}=\frac{7}{1440}\ , \quad f_{2,{\{1,1\}}}=\frac{29}{5760}\ , \quad f_{2,{\{0,0,1\}}}=\frac{1}{1152} \ ,
\end{equation}
and we arrive at
\begin{equation}
  F_2 = \frac7{1440} \frac{I_2^3}{(1-I_1)^5}+\frac{29}{5760}\frac{I_2\,I_3}{(1-I_1)^4}+\frac1{1152} \frac{I_4}{(1-I_1)^3} \ . 
\end{equation}
Thus, the method of Itzykson and Zuber --- expressing the tau function $u(\{t_k\})$ and hence the generating function $F(\{t_k\})$ in terms of the functions $I_n(u_0,\{t_k\})$ --- offers a very powerful method to compute the generating function $F(\{t_k\})$ order-by-order as a genus expansion \cite{Itzykson:1992ya}. Upon inserting the expression~\eqref{eq:u0exp} to the desired order, one can readily read off the correlators of topological gravity explicitly. 

Solving the KdV hierarchy in terms of the functions $I_n$ allows us to derive a universal expression for the off-shell partition functions~\eqref{eq:offshellPart} with arbitrary shifts $\{ \Delta t_k \}$ in the coupling parameters $\{ t_k \}$. The defined off-shell partition functions~\eqref{eq:offshellPart} are derived from the generating function $F(F_0,\{ I_n \}) = F_0 + \sum_{g=1}^{+\infty} g_s^{2g} F_g(\{ I_n \})$, which --- if expressed in terms of $F_0(u_0(\{t_k\}),\{t_k\})$ and $I_n(u_0(\{t_k\}),\{t_k\})$, $n=1,2,3,\ldots$ --- only implicitly depend on the couplings $\{t_k\}$. Computing the action of the boundary creation operators~\eqref{eq:dshifts} on the functions $F_0$ yields
\begin{equation} \label{eq:BbetaF0}
\begin{aligned}
  \mathcal{B}(\beta) F_0 &= \frac{g_s}{\sqrt{2\pi}\beta^\frac32} \left( e^{\beta I_0}\left(1 - \beta I_0\right)-1 +\sum_{k,\ell=0}^{+\infty} \frac{I_0^{k+\ell+1}}{k+\ell+1} \frac{\beta^{k+2}}{k!} \frac{t_\ell}{\ell!}\right) \ , \\
  \mathcal{B}(\beta_1)\mathcal{B}(\beta_2) F_0 &= \frac{g_s^2 \sqrt{\beta_1\beta_2}}{2\pi \beta_1 +2 \pi \beta_2} \left( e^{(\beta_1+\beta_2)I_0} - 1 \right) \ ,
\end{aligned}
\end{equation}
whereas for $I_n$ we find
\begin{equation}
     \mathcal{B}(\beta) I_0 = g_s\sqrt{\frac{\beta}{2\pi}} \frac{e^{\beta I_0}}{1-I_1} \ ,\qquad
     \mathcal{B}(\beta) I_k = g_s\sqrt{\frac{\beta}{2\pi}} e^{\beta I_0} \left(\beta^k + \frac{I_{k+1}}{(1-I_1)} \right) \quad \text{for} \quad k\ge 1 \ .
\end{equation}
As a consequence of these derivative rules --- except for the leading genus zero contribution to the partition function with one asympototic boundary --- the off-shell partition functions~\eqref{eq:offshellPart} are universally expressible in terms of the functions $I_n$, i.e.\ 
\begin{equation} \label{eq:Zuni}
\begin{aligned}
  Z(\{\mathcal{B}(\beta) F_0,I_n\}; \beta)^{\text{top.}} & =\frac{1}{g_s^2}  \mathcal{B}(\beta) F(\{ t_k \}) = \frac{1}{g_s^2}  \mathcal{B}(\beta) F_0 +Z^{(g>0)}(\{I_n\}; \beta)^{\text{top.}}  \ , \\
  Z(\{I_n \}; \beta_1,\ldots,\beta_m )^\text{top.} &=  \frac{1}{g_s^2} \mathcal{B}(\beta_1) \cdots \mathcal{B}(\beta_m) F(\{ t_k \}) \quad \text{for} \quad m>1 \ .
\end{aligned}  
\end{equation} 
In particular, the partition function with a single asymptotic boundary component enjoys the genus expansion
\begin{equation} \label{eq:Ztopgeneral}
  Z(\{\mathcal{B}(\beta) F_0,I_n\}; \beta)^{\text{top.}} =
  \frac{1}{g_s^2}  \mathcal{B}(\beta) F_0 + \sqrt{\frac\beta{2\pi}} e^{\beta I_0}
  \sum_{g=1}^{+\infty} g_s^{2g-1} (1-I_1)^{g-1} Z_g(\{ I_n \},\beta) \ ,
\end{equation}
where
\begin{equation} \label{eq:Zg1}
  Z_1 = \frac{1}{24}\left(\frac\beta{1-I_1} + \frac{I_2}{(1-I_1)^2}\right) \ ,
\end{equation}
and for $g>0$
\begin{multline} \label{eq:Zguni}
  Z_g 
  =\!\!\!\!\!\!\!\!\! \sum_{\sum_{k=2}^{3g-2} (k-1) \ell_k =3 g - 3} \!\!\!\!\!\!\!\!\!\!\!\!\! f_{g,\{\ell_k\}} 
   \sum_{s=2}^{3g-2} \ell_s \left(\frac{1+2s}{3(1-I_1)} \left( \beta + \frac{I_2}{1-I_1} \right)
  +\frac{I_{s+1}}{I_s(1-I_1)} + \frac{\beta^s}{I_s} \right) \\
  \cdot\left(\frac{I_2}{(1-I_1)^2} \right)^{\ell_2}\cdot\ldots \cdot \left( \frac{I_{3g-2}}{(1-I_1)^{3g-2}} \right)^{\ell_{3g-2}}  ,
\end{multline}
in terms of the constants $f_{g,\{\ell_k\}}$ defined in eq.~\eqref{eq:Fgstructure}. With eq.~\eqref{eq:Zg1} and inserting~\eqref{eq:CoeffF2} into $Z_2$ we find explicitly up to genus two %\hhh{adjusted}%
\begin{multline} \label{eq:Zgupto2}
  Z^{(g>0)}(\{I_n\}; \beta)^{\text{top.}}  =
  \frac{g_s}{24} \sqrt{\frac{\beta}{2\pi}} e^{\beta I_0} \left( \frac\beta{1-I_1} + \frac{I_2}{(1-I_1)^2} \right) \\
  + \frac{g_s^3}{5760} \sqrt{\frac{\beta}{2\pi}} e^{\beta I_0}
  \left( 
  \frac{5\beta^4}{(1-I_1)^4}  
  +\frac{29\beta^3 I_2+29\beta^2 I_3 + 15 \beta I_4+5I_5}{(1-I_1)^5} \hskip20ex \right.\\
  \left. +\frac{84\beta^2 I_2^2+116\beta I_3 I_2 +44 I_4 I_2+29I_3^2}{(1-I_1)^6} 
  +\frac{20 I_2^2 (7\beta I_2 + 10 I_3)}{(1-I_1)^7}
  +\frac{140 I_2^4}{(1-I_1)^8}
   \right) \\
  + \ldots  \ . 
\end{multline}
Similar formulas can be worked out for the universal partition functions with several asymptotic boundary components, namely \begin{equation}
  Z(\{I_n \}; \beta_1,\ldots,\beta_m )^\text{top.}=
  \prod_{i=1}^{m}\left(e^{\beta_iI_0}\sqrt{\frac{\beta_i}{2\pi}}\right)
  \sum_{g=0}^{\infty}g_s^{2g+m-2}(1-I_1)^{g-1}Z_g(\{I_n\},\beta_1,\ldots,\beta_m) \ ,
\end{equation} 
where
\begin{equation}
  \mathcal{B}(\beta_1)\cdots\mathcal{B}(\beta_m) F_g=\frac{g_s\sqrt{\beta_1\cdots\beta_m}}{(2\pi)^\frac{m}2} e^{(\beta_1+\ldots+\beta_m)I_0}(1-I_1)^{g-1}Z_g(\{I_n\},\beta_1,\ldots,\beta_m) \ .
\end{equation}
In particular for two asymptotic boundary components the leading order contributions are given by
\begin{multline}
  Z(\{I_n \}; \beta_1,\beta_2) = \frac{ \sqrt{\beta_1\beta_2}}{2\pi \beta_1 +2 \pi \beta_2} e^{(\beta_1+\beta_2)I_0} \\
  + 
  \frac{ g_s^2 \sqrt{\beta_1\beta_2}}{48\pi}e^{(\beta_1+\beta_2)I_0} 
  \left( \frac{\beta_1^2+\beta_1\beta_2+\beta_2^2}{(1-I_1)^2}
  +\frac{2(\beta_1+\beta_2)I_2+I_3}{(1-I_1)^3}
  +\frac{2I_2^2}{(1-I_1)^4} \right) + \ldots  \ ,
\end{multline}
including the semi-classical contribution, cf.\ eq.~\eqref{eq:Zfuncs}. Thus, any of the off-shell or on-shell partition functions defined in eqs.~\eqref{eq:offshellPart} and \eqref{eq:onshellPart} can be obtained from the universal partition functions~\eqref{eq:Zuni} upon inserting $I_n(\{ t_k \} , u_0(\{ t_k \}))$ with suitable values for the couplings $\{ t_k \}$. For instance, inserting $I_n(\{ t_k + \gamma_k +\sdef_k \}, u_0( \{ t_k + \gamma_k +\sdef_k \}))$ we obtain the off-shell partition functions $Z^H(\{t_k\}; \beta_1,\ldots,\beta_m)$, whereas for $I_n(\{ \gamma_k +\sdef_k \}, u_0( \{ \gamma_k +\sdef_k \}))$ we arrive at the on-shell partition functions $Z(\beta_1,\ldots,\beta_m)$. In the next section, we focus on the partition functions $Z(t_0,t_1;\beta_1,\ldots,\beta_m)$ studied in refs.~\cite{OkuyamaSakai1,OkuyamaSakai2}, where we assign on-shell values to the couplings $t_k$, $k=2,3,4,\ldots$, while keeping the first two couplings $t_0$ and $t_1$ off-shell \cite{ZografLargeGenusAsymptotics}.

While the presented genus expansion in the coupling $g_s \sim e^{-1/G_N}$ is non-perturbative in the gravitational coupling $G_N$ of JT gravity, it is perturbative in the dual matrix model formulation, where the expansion parameter $g_s$ describes quantum fluctuations about the classical energy density of states \cite{SSS,CJ1}. In fact the discussed partition functions $Z(\{ I_n \}; \beta_1,\ldots,\beta_m)$ are divergent series in $g_s$ due to the factorial growth $(2g)!$ of the contributions at order $g_s^{2g}$ \cite{ZografLargeGenusAsymptotics,SSS}. Therefore, the partition functions $Z(\{ I_n \}; \beta_1,\ldots,\beta_m)$ are asymptotic series that require a non-perturbative completion arising from non-perturbative effects of the order $e^{-1/g_s}$. For further details on this issue and the possible emergence of non-perturbative instabilities, we refer the reader to ref.~\cite{SSS,CJ1} and the solutions proposed in refs.~\cite{CJ1,CJ2,CJ3}.

%%%%%%%%%%%%%%%%%%%%%%%%
\subsection{Partition Functions with Leading Order Off-shell Couplings} \label{sec:TwoOffShell}
%%%%%%%%%%%%%%%%%%%%%%%
In the spirit of refs.~\cite{OkuyamaSakai1,OkuyamaSakai2} let us now consider the partition functions $Z(t_0,t_1;\beta_1,\ldots,\beta_m)$ with only the couplings $t_0$ and $t_1$ taken to be off-shell. Then the partition functions for JT~gravity coupled to a gas of defects are defined as
\begin{equation}
  Z(t_0,t_1;\beta_1,\ldots,\beta_m) \equiv Z(\{t_0,t_1,t_{k\ge2} = \gamma_k + \sdef_k \};\beta_1,\ldots,\beta_m) \ ,
\end{equation}
where setting $t_0=\sdef_0$ and $t_1=\sdef_1$ yields the on-shell partition functions in all couplings. Analogously, we can define the function $u(t_0,t_1)$ and the generating function $F(t_0,t_1)$ obtained by evaluating the couplings $t_{k\ge 2}$ of the tau function $u(\{t_k\})$ and of the generating function $F(\{t_k\})$ at their on-shell values, i.e.\
\begin{equation}
   u(t_0,t_1) = u(\{t_0,t_1,t_{k\ge2} = \gamma_k + \sdef_k \}) \ , \qquad
   F(t_0,t_1) = F(\{t_0,t_1,t_{k\ge2} = \gamma_k + \sdef_k \}) \ , 
\end{equation}
with
\begin{equation}
   u(t_0,t_1)=\partial_0^2F(t_0,t_1) \ .
\end{equation}   
All these functions can respectively be obtained from their universal expressions~\eqref{eq:Zuni}, \eqref{eq:ug}, and \eqref{eq:Fgstructure} by inserting the on-shell values of the couplings $t_{k\ge 2}$ into the functions $I_n$. The function $u(t_0,t_1)$ fulfils the first partial differential equation of the KdV hierarchy~\eqref{eq:GeneralizedKdV}, which is just the non-linear partial differential KdV~equation, i.e.,
\begin{equation} \label{eq:KdVeq}
  \partial_{1}u = u \, \partial_{0}u +\frac{g_s^2}{12} \, \partial_{0}^3 u \ .
\end{equation}

With $t_0$ and $t_1$ off-shell we observe that the function $I_1$ depends only on $t_1$ and $u_0\equiv I_0$, while $I_n$ for $n\ge 2$ are series in $u_0$ without an explicit dependence on $t_0$ and $t_1$. Therefore, it is convenient to introduce new (formal) variables $(y,t)$ given  by \cite{ZografLargeGenusAsymptotics,OkuyamaSakai1}
\begin{equation} \label{eq:yt}
    y=u_0 \ , \qquad t=1-I_1 \ .
\end{equation}
Since $I_n$ for $n\ge 2$ is only a function of $y$, we obtain from the universal tau function~\eqref{eq:ug} and the universal generating function~\eqref{eq:Fgstructure} the asymptotic series
\begin{equation}\label{eq:ugenusexpansion}
    u(y,t)=y+\sum_{g=1}^{\infty}g_s^{2g} u_g(y,t)\ , \qquad
    u_g(y,t)=\sum_{k=2g+1}^{5g-1}u_{g,k}(y)t^{-k} \ ,
\end{equation}
and
\begin{equation}\label{eq:Fgenusexpansion}
    F(y,t)=F_0(y,t) - \frac{g_s^2}{24}\log t+ \sum_{g=2}^{\infty}g_s^{2 g}F_g(y,t) \ , \qquad
    F_g(y,t)=\sum_{k=2g-1}^{5g-5}F_{g,k}(y,t)t^{-k}\ .
\end{equation}    
The coefficient functions $u_g(y,t)$ (for $g\ge 1$) and $F_g(y,t)$ (for $g\ge 2$) are Laurent polynomials in the variable $t$, where the range for the powers of $t$ is a consequence of the restricted sums in eqs.~\eqref{eq:ug} and \eqref{eq:Fgstructure}. The degrees of these Laurent polynomials conform with the structure derived by Zograf for the specific on-shell couplings $t_k = \gamma_k$ for $k\ge 2$ \cite{ZografLargeGenusAsymptotics}. Furthermore, at genus one the logarithmic contribution to $F(y,t)$ arises from eq.~\eqref{eq:F1def}, whereas with eq.~\eqref{eq:F0} the genus zero contribution becomes
\begin{multline}
  F_0(y,t) = \frac16 y^3 t^2  +\frac16 y^2t \sum_{k=2}^{+\infty} \frac{y^k(2k+5)(\gamma_k+\sdef_k)}{(k+2)(k+1)(k-2)!}
  + \frac16y \left( \sum_{k=2}^{+\infty} \frac{y^k(\gamma_k+\sdef_k)}{(k+1)(k-2)!}\right)^2 \\
  +\sum_{k=4}^{+\infty} \frac{y^{k+1}}{3(k+1)(k+2)!} \sum_{n=2}^{k-2} \binom{k+4}{n+2}\binom{n}2\binom{k-n}2(\gamma_n+\sdef_n)(\gamma_{k-n}+\sdef_{k-n})  \ .
\end{multline}

Let us now turn to the partition function $Z(t_0,t_1;\beta)$ with a single asymptotic boundary. Since the couplings $t_{k\ge2}$ are taken on-shell we cannot obtain $Z(t_0,t_1;\beta)$ by acting with the boundary creation operator $\mathcal{B}(\beta)$ on the generating function $F(t_0,t_1)$ because the boundary operator $\mathcal{B}(\beta)$ contains derivatives with respect to those parameters that have been fixed to their on-shell values. Thus, either we compute $Z(t_0,t_1;\beta)$ from the universal partition function~\eqref{eq:Zuni} or we determine a differential equation with $Z(t_0,t_1;\beta)$ as its solution. For the latter approach we follow the authors of ref.~\cite{OkuyamaSakai1}. Note that the partial derivatives $\partial_k$ for $k\ge 2$ appearing in the boundary operator $\mathcal{B}(\beta)$ can be rewritten in terms of derivatives with respect to $\partial_0$ due to the KdV hierarchy~\eqref{eq:GeneralizedKdV}, namely 
\begin{equation} \label{eq:ZWrel}
   \partial_0 Z(t_0,t_1;\beta) = \left.\frac1{g_s^2} \mathcal{B}(\beta) \partial_0 F(\{t_k\}) \right|_{\{t_{k\ge 2} = \gamma_k +\sdef_k \}} 
   = -\frac1{g_s\sqrt{2\pi\beta}}+W(t_0,t_1;\beta) \ ,
\end{equation}
with the definition
\begin{equation}
  W(t_0,t_1;\beta) = \frac{1}{g_s\sqrt{2\pi\beta}}  \sum_{\ell=0}^{+\infty} \beta^\ell \mathcal{R}_{\ell} \ ,
\end{equation}
in terms of the Gelfand--Dikii polynomials~\eqref{eq:GD} and $\mathcal{R}_0=1$. The key observation of ref.~\cite{OkuyamaSakai1} is now that the Gelfand--Dikii polynomials obey the non-trivial relation\footnote{This relation can directly be proven by induction with respect to the index $k$ of the Gelfand--Dikii polynomials $\mathcal{R}_k$. The induction step is performed by applying the recursion relation~\eqref{eq:generalKdvequation} of the Gelfand--Dikii polynomials.}
\begin{equation}
  \partial_1 \mathcal{R}_k = u \, \partial_0\mathcal{R}_k + \frac{g_s^2}{12} \partial_0^3\mathcal{R}_k \ ,
\end{equation}
which immediately implies the differential equation
\begin{equation}
  \partial_1 W(t_0,t_1;\beta)  = u \, \partial_0W(t_0,t_1;\beta)  + \frac{g_s^2}{12} \partial_0^3W(t_0,t_1;\beta) \ .
\end{equation}
The partition function $Z(t_0,t_1;\beta)$ can now be determined from this differential equation for $W(t_0,t_1;\beta)$. The function $W(t_0,t_1;\beta)$ is an interesting quantity by itself, see for instance the discussion in ref.~\cite{OkuyamaSakai1}. 

Upon expressing the couplings $(t_0,t_1)$ in terms of the variables $(y,t)$ defined in eq.~\eqref{eq:yt}, the function $W(y,t;\beta)$ enjoys the asymptotic genus expansion
\begin{equation} \label{eq:Wgenusxpansion}
     W(y,t;\,\beta)=\frac{e^{\beta y}}{\sqrt{2 \pi \beta}}\sum_{g=0}^{+\infty} g_s^{2 g-1}\,W_g(y,t;\beta)\ ,
 \end{equation}
 where --- due to the definition~$\mathcal{R}_0=1$ and due to the leading order behaviour~\eqref{eq:Rk_u0} of the Gelfand--Dikii polynomials --- the genus zero contribution reads
 \begin{equation}  \label{eq:Wg0}
   W_0(y,t;\beta) = 1 \ .
 \end{equation}
By inserting the variables~\eqref{eq:yt} into the $t_0$-derivative of the universal expressions~\eqref{eq:Zguni}, we find that the higher genus contributions $W_g(y,t;\beta)$ are polynomials in $t^{-1}$ with coefficient functions in terms of $y$ and $\beta$ of the form
 \begin{equation}\label{eq:WLaurentpolynomial}
     W_g(y,t;\beta)=\sum_{k= 2 g}^{5g-1}W_{g,k}(y;\beta) \, t^{-k} \quad \text{for} \quad g \ge 1 \ .
 \end{equation}
Inserting the asymptotic expansion~\eqref{eq:Wgenusxpansion} into the partial differential equation yields the recursion differential equation \cite{OkuyamaSakai1}
\begin{equation} \label{eq:DiffRec}
  \partial_t W_g = - \sum_{h=0}^{g-1} u_{g-h} \nabla(\beta) W_h - \frac1{12} \nabla(\beta)^3 W_{g-1} \ ,
\end{equation}
with the linear differential operators 
\begin{equation}
    \nabla(\beta) %= e^{-\beta y} \partial_0 e^{\beta y} 
    = \partial_0 + \frac\beta{t}
    %= \frac1t \left( -I_2 \partial_t + \partial_y + \beta \right)
    = \frac1t \left( -I_2 \partial_t + D_y \right)\ , \qquad
    D_y =\partial_y + \beta \ .
\end{equation}
Furthermore, inserting the expansion~\eqref{eq:WLaurentpolynomial} into the differential recursion relation and carrying out a few steps of algebra yields recursion relations for the Laurent modes $W_{g,k}(y;\beta)$. With the initial genus zero contribution~\eqref{eq:Wg0} we arrive for genus $g=1$ at\footnote{Note that the polynomial structure~\eqref{eq:WLaurentpolynomial} of $W_g(y,t;\beta)$ fixes the constant of integration in the differential recursion relation~\eqref{eq:DiffRec} with respect to $t$ unambiguously.}
\begin{equation} 
    W_{1,k} = \frac{\beta}k u_{1,k} + \frac{\beta^3}{24}\delta_{k,2} + \frac1{36} (3 I_2 \beta^2 + I_3 \beta)\delta_{k,3} + \frac\beta{16} I_2^2\delta_{k,4} \quad
    \text{for} \quad k=2,3,4 \ ,
\end{equation}  
which explicitly becomes with eq.~\eqref{eq:u1}
\begin{equation}
    W_1(y,t;\beta) =  \frac{\beta^3}{24t^2} + \frac\beta{24t^3} \left( 2\beta I_2  + I_3 \right) + \frac\beta{12t^4} I_2^2 \ .
\end{equation}
Furthermore, for $g \ge 2$ and $k=2g+1,\ldots, 5g-1$ we arrive at the lengthy but straightforwardly applicable recursion relation
\begin{multline}
    W_{g,k} =  
    \sum_{h=1}^{g-1} \sum_{n=2h}^{5h-1}\left( \frac{n}{k} I_2 u_{g-h,k-n-1} W_{h,n} + \frac1k u_{g-h,k-n} D_y W_{h,n}\right) 
    +\frac\beta{k} u_{g,k} \\
    +\frac1{12k} \Big[D_y^3 W_{g-1,k-2} 
    +\left(3(k-2) I_2 D_y^2 + (3k-8)I_3 D_y + (k-3) I_4\right) W_{g-1,k-3}\\
    +\left(3(k^2-5k +5) I_2^2 D_y + (k-4)(3k-5) I_2 I_3 \right) W_{g-1,k-4}\\
    +(k-5)(k-3)(k-1) I_2^3 W_{g-1,k-5} \Big] \ ,
\end{multline}
where we set $W_{h,n} \equiv 0$  for $n\not\in\{ 2h,\ldots,5h-1 \}$ and $u_{h,n} \equiv 0$ for $n\not\in\{2h+1,\ldots,5h-1\}$. In particular, for genus two we readily compute
\begin{multline}
  W_2(y,t;\beta) = 
  \frac{\beta}{5760}  \left(
  \frac{5\beta^5}{t^4}
  +\frac{44\beta^4 I_2+58\beta^3 I_3+44\beta^2 I_4+20\beta I_5+5 I_6}{t^5} \right. \\
  +\frac{200\beta^3 I_2^2+400\beta^2 I_2 I_3+145\beta I_3^2+220\beta I_2I_4
  +102 I_3 I_4+64 I_2 I_5}{t^6}\\
  +\frac{5I_2 \left(112\beta^2 I_2^2+240\beta I_3 I_2+84 I_4 I_2+109 I_3^2\right)}{t^7}
  \left.+\frac{20 I_2^3 \left(49\beta I_2+88 I_3\right)}{t^8}
  +\frac{980I_2^5}{t^9} \right)
  \ .
\end{multline}

With the help of these recursion formulas we are now in a position to deduce the partition function $Z(y,t;\beta)$ with one asymptotic boundary component as well. The general structure~\eqref{eq:Zguni} implies for the partition function the asymptotic series\footnote{Note that the newly introduced contributions $\tilde Z_g$ to the partition function differ from the definition of $Z_g$ given in eq.~\eqref{eq:Ztopgeneral} by a normalisation.}
\begin{equation}
  Z(y,t;\beta) = \sum_{g=0}^{+\infty} g_s^{2g-1} \tilde Z_g(y,t;\beta) \ .
\end{equation}
The genus zero part splits into the semi-classical and topological contributions
\begin{equation}
  \tilde Z_0(y,t;\beta) = \tilde Z_0(y,t;\beta)^\text{semi.}
  + \tilde Z_0(y,t;\beta)^\text{top.} \ ,
\end{equation}
where --- using eqs.~\eqref{eq:dshifts} and \eqref{eq:Zfuncs}--- the semi-classical part is given by
\begin{multline}
  \tilde Z_0(y,t;\beta)^\text{semi.} = \frac{t(1+y\beta)}{\sqrt{2\pi} \beta^{\frac32}} 
  + \frac1{\sqrt{2\pi\beta}} \sum_{k=2}^{+\infty} \frac{y^k (\gamma_k +\delta_k)}{k(k-2)!}  \\
  +\frac1{\sqrt{2\pi}\beta^{\frac32}} \sum_{k=2}^{+\infty} \frac{y^{k-1}(\gamma_k +\delta_k)}{(k-1)!} 
  +\frac1{\sqrt{2\pi\beta}} \sum_{k=2}^{+\infty} \frac{\delta_k + \gamma_k}{(-\beta)^k} \ ,
\end{multline}
and where --- according to eq.~\eqref{eq:BbetaF0} --- the topological part reads
\begin{multline}
   \tilde Z_0(y,t;\beta)^\text{top.} = \frac{t}{\sqrt{2 \pi} \beta^{\frac32}} \left(e^{\beta y} - (1+y\beta) \right) - \frac{e^{\beta y}}{\sqrt{2\pi\beta}}\sum_{k=2}^{+\infty} \frac{y^k (\gamma_k +\sdef_k)}{k!} 
   -\frac1{\sqrt{2\pi\beta}} \sum_{k=2}^{+\infty} \frac{y^k (\gamma_k +\sdef_k)}{k(k-2)!} \\
   +\frac{e^{\beta y}-1}{\sqrt{2 \pi} \beta^{\frac32}} \sum_{k=2}^{+\infty} \frac{y^{k-1} (\gamma_k +\sdef_k)}{(k-1)!}
   +\frac1{\sqrt{2 \pi} \beta^{\frac32}} \sum_{k=0}^{+\infty} \sum_{\ell=2}^{+\infty} \frac{y^{k+\ell+1}\beta^{k+2}}{(k+\ell+1)!} \binom{k+\ell}{k} (\gamma_\ell+\sdef_\ell) \ .
\end{multline}
Therefore, the total genus zero contribution becomes
\begin{multline} \label{eq:Z0all}
     \tilde Z_0(y,t;\beta) = \frac{e^{\beta y}}{\sqrt{2 \pi} \beta^{\frac32}} \left(  t   - \beta\sum_{k=2}^{+\infty} \frac{y^k (\gamma_k +\sdef_k)}{k!} 
   +\sum_{k=2}^{+\infty} \frac{y^{k-1} (\gamma_k +\sdef_k)}{(k-1)!} \right) \\
   +\frac1{\sqrt{2\pi} \beta^{\frac32}} 
    \sum_{k=0}^{+\infty} \sum_{\ell=2}^{+\infty} \frac{y^{k+\ell+1}\beta^{k+2}}{(k+\ell+1)!} \binom{k+\ell}{k} (\gamma_\ell+\sdef_\ell)
   +\frac{1}{\sqrt{2\pi\beta}}\sum_{k=2}^{+\infty} \frac{\delta_k + \gamma_k}{(-\beta)^k}    \ .
\end{multline}
For the higher genus contributions we arrive with eq.~\eqref{eq:Zguni} at the polynomials in $t^{-1}$
\begin{equation} \label{eq:ZgPoly}
    \tilde Z_g(y,t;\beta) = \frac{e^{\beta y}}{\sqrt{2 \pi} \beta^{\frac32}} \sum_{k=2g-1}^{5g-3} Z_{g,k}(y;\beta) t^{-k} \quad \text{for} \quad g \ge 1 \ .
\end{equation} 
Thus, employing the derived recursion relations for $W_{g,k}(y;\beta)$ we can determine $Z_g(y,t;\beta)$ recursively upon integrating eq.~\eqref{eq:ZWrel}. Note that the constants of integration at each order in $g_s$ are unambiguously determined by the general structure~\eqref{eq:ZgPoly}. Explicitly, we find for genus one --- in agreement with eq.~\eqref{eq:Zg1} --- the result
\begin{equation} \label{eq:Z1onshell}
  \tilde Z_1(y,t;\beta) =  \frac{e^{\beta y}}{24\sqrt{2\pi\beta}} \left( \frac{\beta^2}t + \frac{\beta I_2}{t^2} \right) \ ,
\end{equation}
whereas for genus two --- in agreement with eq.~\eqref{eq:Zgupto2} --- we obtain
\begin{multline} \label{eq:Z2onshell}
  \tilde Z_2(y,t;\beta) = 
  \frac{\sqrt{\beta} \, e^{\beta y}}{5760\sqrt{2\pi}}  \left( 
  \frac{5\beta^4}{t^3}  
  +\frac{29\beta^3 I_2+29\beta^2 I_3 + 15 \beta I_4+5 I_5}{t^4} \right.\\
  \left. +\frac{84\beta^2 I_2^2+116\beta I_3 I_2 +44  I_4 I_2+29 I_3^2}{t^5} 
  +\frac{20 I_2^2 (7\beta I_2 + 10 I_3)}{t^6}
  +\frac{140 \beta I_2^4}{t^7}
   \right) \ .
\end{multline}

Let us give an alternative perspective on the partition function $Z(y,t; \beta)$ in terms of the associated Schr\"odinger problem \cite{Brezin:1990rb,Douglas:1989ve,Gross1} 
\begin{equation}
   \mathcal{H} \psi_E(t_0,t_1) = E \, \psi_E(t_0,t_1) \quad \text{with} \quad 
   \mathcal{H} = \hbar^2 \partial_0^2 + u(t_0,t_1) \ ,
\end{equation}
with $\hbar=\frac{g_s}{\sqrt{2}}$, Hamilton operator $\mathcal{H}$, and the wavefunctions $\psi_E(t_0,t_1)$, which are eigenfunctions with energy eigenvalue~$E$. Here the partially on-shell tau function $u(t_0,t_1)$ becomes the potential of the Schr\"odinger equation, and the partition function can be written as
\begin{equation}
  Z(y,t;\beta) = \int dE \, e^{-\beta E} \rho(E;y,t) \ ,
\end{equation}
in terms of the spectral density $\rho(E;y,t)$ of the energy eigenvalues of the Hamilton operator $\mathcal{H}$. This formulation offers a framework for a non-perturbative description in the genus expansion~$g_s$. However, since in our context the tau function $u(t_0,t_1)$ itself is only given as an asymptotic series in $g_s$, setting up the appropriate non-perturbatively exact Schr\"odinger problem is nevertheless a difficult task. This question has been discussed and analysed with numerical methods in refs.~\cite{CJ1,CJ2,CJ3}. Here we only focus on the leading order contribution at genus zero, which  predicts the integral representation
\begin{equation}
  \tilde Z_0(y,t;\beta) =  \int dE \,e^{-\beta E} \rho_0(E;y,t) \ ,
\end{equation}
in terms of the genus zero spectral density $\rho_0(E;y,t)$. To verify this prediction explicitly, we first express the genus zero partition function~\eqref{eq:Z0all} as
\begin{equation}\label{eq:Ztilde0}
  \tilde Z_0 = \frac{e^{\beta y}}{\sqrt{2\pi} \beta^{\frac32}} \left( t + J'(y)\right) - \frac{e^{\beta y}}{\sqrt{2\pi\beta}}J(y)
  + \sqrt{\frac{\beta}{2\pi}} \int_{-\infty}^y dv \,e^{v\beta} J(v) \ ,
\end{equation}
in terms of the function
\begin{equation} \label{eq:defJC}
  J(y) = \sum_{k=2}^{+\infty} \frac{y^{k} (\gamma_k +\sdef_k)}{k!} \ .
\end{equation}
Here we assume that the function $J(v)$ is continuously differentiable in the interval $(-\infty,y)$, and that the stated integral (for $\beta > 0$) is finite. Performing an integration by parts and using the integral identities
\begin{equation}
    \sqrt{\frac\pi\beta}e^{\beta z} = \int_{-z}^{+\infty} dE \frac{e^{-\beta E}}{\sqrt{E+z}} \ , \qquad
    \frac{\sqrt{\pi}}{2 \beta^{\frac32}} e^{\beta z} =  \int_{-z}^{+\infty} dE e^{-\beta E} \sqrt{E+z}  \ ,
\end{equation} 
we arrive at the expression
\begin{equation}
  \tilde Z_0(y,t;\beta) =  \int_{-y}^{+\infty} dE \,e^{-\beta E} \rho_0(E;y,t) \ ,
\end{equation}
in terms of the (genus zero) spectral density
\begin{equation} \label{eq:rho0}
  \rho_0(E;y,t) = \frac{\sqrt{2}}{\pi} \sqrt{E+y} \left( t +J'(y) \right) 
  -\frac1{\sqrt{2}\pi} \int_{-y}^E dv \frac{J'(-v)}{\sqrt{E-v}} \ .
\end{equation} 

The obtained result agrees with the expected structure of the partition function obtained from the associated Schr\"odinger problem. Note that the obtained function $\rho_0(E;y,t)$ enjoys only the interpretation as a spectral density, if it is non-negative over the energy range $(-y,+\infty)$. The conditions $J'(y)\ge -t$ and $J'(v) \ge 0$ for $v\in (-y,+\infty)$ are sufficient to ensure a non-negative spectral density function (in the genus zero approximation). For some energy ranges $E$ in the interval~$(-y,+\infty)$ we seemingly arrive at a negative function $\rho_0(E;y,t)$. However, as on the classical level a Hawking--Page like first order phase transition can be observed when varying the potential~\eqref{eq:Dilatonpotential} \cite{Witten:2020ert}, it might be expected that here too, a phase transition occurs preventing the aforementioned negativity of the spectral density function. In ref.~\cite{CJ4} this was only observed to be true for a specific class of models for which $U(0)=0$ with $U(\phi)$ (again referring to eq.~\eqref{eq:Dilatonpotential}), while a larger class of models, namely those  for which $U(0) \neq 0$, are declared both perturbatively and non-perturbatively unstable. 

For energies $E$ close to the negative coupling $-y$ the calculated energy density $\rho_0(E;y,t)$ behaves as 
\begin{equation} \label{eq:GroundState}
  \rho_0(E;y,t) = \frac{\sqrt{2}t}{\pi}\sqrt{E+y}+ \mathcal{O}(|E-E_0|^\frac32)  \ .
\end{equation} 
%\end{align}
Therefore, we can interpret the negative coupling $-y$ as the (semi-classical) ground state energy of the Schr\"odinger problem. In particular, for JT~gravity in the absence of defects the on-shell value of $y$ becomes zero, and hence the ground state energy vanishes. Coupling JT~gravity to a gas of defect, however, yields a non-vanishing on-shell value for $y$ according to eqs.~\eqref{eq:dshifts} and \eqref{eq:yt}, which therefore results in a non-trivial shift of the ground state energy. This observation is in agreement with the results obtained in refs.~\cite{Maxfield3gravity,WittenDeformations}, and we get back to this point in the explicit example below and in Section~\ref{Section:LowtemperatureExpansion}.

\bigskip

Finally, let us illustrate the structure of the partially off-shell partition function $Z(y,t;\beta)$ for JT gravity interacting with a single defect type specified by the coupling~$\epsilon$ and identification angle~$\alpha$. Then --- according to eqs.~\eqref{eq:gammak} and \eqref{eq:dshifts} --- the on-shell couplings $t_k$ for $k\ge 2$ become
\begin{equation} \label{eq:onshelltk}
  t_k = \frac{(-1)^k}{(k-1)!} + \left(-\frac{\alpha^2}{4\pi^2}\right)^k \frac{2\pi^2 \epsilon}{k!} 
  \quad \text{for} \quad k\ge 2 \ ,
\end{equation}
whereas the remaining unfixed couplings $t_0$ and $t_1$ acquire their on-shell values upon setting 
\begin{equation}
  \left.(t_0,t_1)\right|_\text{on-shell}= 2 \pi^2 \epsilon\left(1, -\frac{\alpha^2}{4\pi^2}\right) \ .
\end{equation}  
The on-shell values of the variables $(y,t)$ defined in terms of $(t_0,t_1)$ in eq.~\eqref{eq:yt} are governed by the functional relations
\begin{equation} \label{eq:ytonshell}
\begin{aligned}
  0 &= \left. -\sqrt{y} \,\BJ_{1}(2 \sqrt{y})\right|_\text{on-shell}
  + \left.(2\pi^2 \epsilon) \BJ_0\left( \frac{\alpha\sqrt{y}}\pi \right) \right|_\text{on-shell} \ , \\
  \left. t \right|_\text{on-shell} &= \left. \BJ_{0}(2 \sqrt{y})\right|_\text{on-shell}
  + \left.(2\pi^2 \epsilon) \frac{\alpha}{2\pi\sqrt{y}} \BJ_1\left( \frac{\alpha\sqrt{y}}\pi \right) \right|_\text{on-shell}  \ ,
\end{aligned}  
\end{equation}
in terms of the Bessel functions $\BJ_\nu(x)$ of the first kind
\begin{equation} \label{eq:Bfk}
  \BJ_\nu(x) = \left(\frac{x}2\right)^\nu \sum_{k=0}^{+\infty} \frac{(-1)^k}{\Gamma(\nu+k+1) \, k!} \left(\frac{x^2}4\right)^k \ , \quad
  \BJ_{-n}(x)\equiv(-1)^n \BJ_n(x) \ \text{for integer $n$} \ .
\end{equation}
In the limit of vanishing defect interaction $\epsilon \to 0$ the functional relations~\eqref{eq:ytonshell} have the on-shell solution $\left.(y,t)\right|_\text{on-shell}=(0,1)$ in accord with ref.~\cite{OkuyamaSakai1}. Solving for $\left.(y,t)\right|_\text{on-shell}$ in the vicinity of $(0,1)$ for small $\epsilon$ with the implicit function theorem, we obtain for $(y,t)$ the on-shell expansion in the first few orders
\begin{equation}\label{eq:ytexpansionpoint}
\begin{aligned}
   \left.y\right|_\text{on-shell}=&2\pi^2\epsilon+\pi^2\left(2\pi^2-\alpha^2\right)\epsilon^2+\frac{\pi^2(15\alpha^4-72\pi^2\alpha^2+80\pi^4)}{24} \epsilon^3+\ldots \ , \\
   \left.t\right|_\text{on-shell} =& 1+\frac{\alpha^2-4\pi^2}{2} \epsilon-\frac{\alpha^4-8 \pi^2\alpha^2+8 \pi^4}{8} \epsilon^2 \\
   &\qquad\qquad\qquad+\frac{21\alpha^6-216\pi^2\alpha^4+576\pi^4\alpha^2-448\pi^6}{288}\epsilon^3+\ldots \  . 
\end{aligned}
\end{equation}
According to eq.~\eqref{eq:GroundState} these on-shell values give rise to a non-vanishing ground state energy, which to leading order in $\epsilon$ reads
\begin{equation} \label{eq:GSenergy}
  E_0 = - 2 \pi^2 \epsilon + \mathcal{O}(\epsilon^2) \ .
\end{equation}  
Furthermore, inserting the on-shell couplings~\eqref{eq:onshelltk} into the functions $I_n$ for $n\ge2$ yields in terms of the Bessel function~\eqref{eq:Bfk} the expressions
\begin{equation} \label{eq:DefIOneBdry}
   I_n(y) = \frac{(-1)^n}{(\sqrt{y})^{n-1}} \BJ_{n-1}(2 \sqrt{y}) + (2\pi^2 \epsilon) \left( - \frac{\alpha}{2\pi \sqrt{y}} \right)^n \BJ_n\left(\frac{\alpha \sqrt{y}}\pi\right)
   \quad \text{for} \quad n\ge 2 \ .
\end{equation}
Similarly, the function~$J'(y)$ defined via eq.~\eqref{eq:defJC} becomes
\begin{equation} \label{eq:DefJpOneBdry}
  J'(y) = 1+ 2 \pi^2 \epsilon \frac{\alpha^2}{4\pi^2} - \BJ_0(2\sqrt{y}) 
    - (2\pi^2\epsilon) \frac{\alpha}{2\pi\sqrt{y}} \BJ_1\left(\frac{\alpha\sqrt{y}}\pi\right) \ .
\end{equation}
Thus --- according to eq.~\eqref{eq:rho0} --- the genus zero contribution of the spectral density is given in terms of the Bessel functions $\BJ_0$ and $\BJ_1$ and the modified Bessel functions $\MBJ_0$ and $\MBJ_1$ by
\begin{equation} \label{eq:rho0offshell}
  \rho_0(E;y,t) =
  \frac1{\sqrt{2}\,\pi} \int_{-y}^E dv \frac{\MBJ_0(2\sqrt{v}) + (2\pi^2\epsilon)  \frac{\alpha}{2\pi\sqrt{v}} \MBJ_1\left(\frac{\alpha\sqrt{y}}\pi\right)}{\sqrt{E-v}}
\end{equation}
with the modified Bessel functions defined as $\MBJ_\nu(x) = i^{-\nu} \BJ_{\nu}(i x)$. This result is in agreement with refs.~\cite{WittenDeformations,Maxfield3gravity}. Finally, upon inserting the expressions~\eqref{eq:DefIOneBdry} into the general genus one and genus two results~\eqref{eq:Z1onshell} and \eqref{eq:Z2onshell}, we arrive at $Z_1(y,t;\beta)$ and $Z_2(y,t;\beta)$ in terms of Bessel functions. Expanding these results to leading order in the coupling $\epsilon$ we respectively obtain
\begin{multline}  \label{eq:Z1Z2}
     \left.\tilde Z_{1}(y,t;\beta)\right|_{y=2 \pi^2 \epsilon + \mathcal{O}(\epsilon^2), 
     t=1+\frac{1}{2}(\alpha^2-4\pi^2)\epsilon+\mathcal{O}(\epsilon^2)} 
     =\frac{\beta^{\frac32}e^{2\pi^2\epsilon \beta}}{\sqrt{2\pi}} \left(\frac1{24}-\frac{\alpha ^2 \epsilon }{48}+\frac{\pi ^2 \epsilon }{12}\right) \\
     +\frac{\beta^{\frac12}e^{2\pi^2\epsilon \beta}}{\sqrt{2\pi}} \left(\frac1{24}+\frac{\alpha ^4 \epsilon }{384 \pi ^2}-\frac{\alpha ^2 \epsilon }{24}+\frac{\pi ^2 \epsilon }{8}\right) 
     + \mathcal{O}(\epsilon^2) \ ,
\end{multline}
and
\begin{equation}
\begin{aligned}
   & \left.\tilde Z_{2}(y,t;\beta)\right|_{y=2 \pi^2 \epsilon + \mathcal{O}(\epsilon^2), 
     t=1+\frac{1}{2}(\alpha^2-4\pi^2)\epsilon+\mathcal{O}(\epsilon^2)}\\ 
   &\quad =\frac{\beta^{\frac{9}{2}} e^{2\pi^2\epsilon \beta}}{\sqrt{2\pi}} \left(\frac{1}{1152}-\frac{\alpha ^2 \epsilon }{768}+\frac{\pi ^2 \epsilon }{192}\right)
     +\frac{\beta^{\frac{7}{2}} e^{2\pi^2\epsilon \beta}}{\sqrt{2\pi}}  \left(\frac{29}{5760}+\frac{29 \alpha ^4 \epsilon }{92160 \pi ^2}-\frac{29 \alpha ^2 \epsilon }{2880}
     +\frac{203 \pi ^2 \epsilon }{5760}\right)\\
   &\quad+\frac{\beta^{\frac{5}{2}} e^{2\pi^2\epsilon \beta}}{\sqrt{2\pi}} \left(\frac{139}{11520}-\frac{29 \alpha ^6 \epsilon }{1105920 \pi ^4}
   +\frac{7 \alpha ^4 \epsilon }{3840 \pi ^2}-\frac{181 \alpha ^2 \epsilon }{5760}
   +\frac{1697 \pi ^2 \epsilon }{17280}\right)\\
   &\quad+\frac{\beta^{\frac{3}{2}} e^{2\pi^2\epsilon \beta}}{\sqrt{2\pi}} \left(\frac{449}{11520}+\frac{\alpha ^8 \epsilon }{1179648 \pi ^6}
   -\frac{29 \alpha ^6 \epsilon }{276480 \pi ^4}+\frac{461 \alpha ^4 \epsilon }{46080 \pi ^2}
   -\frac{77 \alpha ^2 \epsilon }{576}+\frac{5269\pi ^2 \epsilon }{13824}\right)\\
   &\quad+\frac{\beta^{\frac{1}{2}} e^{2\pi^2\epsilon \beta}}{\sqrt{2\pi}} \Big(-\frac{137}{9216}-\frac{\alpha ^{10} \epsilon }{70778880 \pi ^8}+\frac{11 \alpha ^8 \epsilon }{4423680 \pi ^6}
   -\frac{19 \alpha ^6 \epsilon }{122880 \pi ^4}-\frac{289 \alpha ^4 \epsilon }{138240 \pi ^2}\\
   &\qquad\qquad\qquad\qquad\qquad
   +\frac{1267 \alpha ^2 \epsilon }{27648}-\frac{3239 \pi ^2 \epsilon }{23040}\Big) + \mathcal{O}(\epsilon^2) \ .
\end{aligned}
\end{equation}
We observe that at every order in the inverse temperature $\beta$, there are contributions from the interaction with the defects already at the linear order in the defect coupling $\epsilon$. Therefore, it is obvious that the dynamics of JT~gravity are strongly influenced by the interaction with defects.

Finally, let us remark that the generalisation to multiple species of defects (with defect couplings $\epsilon_j$ and identification angles $\alpha_j$) is straightforward. Namely, the on-shell values of the couplings $(y,t)$ of eq.~\eqref{eq:ytexpansionpoint} generalise to
\begin{equation} 
\begin{aligned}
  \left. y \right|_\text{on-shell} &= 2 \pi^2 \sum_{j} \epsilon_{j}  
    +\sum_{j,k}\left( 2 \pi^4-\frac12\pi^2\alpha_j^2-\frac12\pi^2\alpha_k^2 \right) \epsilon_{j}\epsilon_{k}
    + \ldots \ , \\
  \left. t \right|_\text{on-shell} &= 1+\sum_{j} \frac{\alpha_j^2 - 4 \pi^2}2 \epsilon_j
    -\sum_{j,k} \frac{ \alpha_{k}^4+\alpha_j^4-8\pi^2 \left(\alpha_k^2+\alpha_j^2\right)+ 16\pi^4 }{16}\epsilon_{j}\epsilon_{k} 
  + \ldots  \ .
\end{aligned}  
\end{equation}
Furthermore, the functions~\eqref{eq:DefIOneBdry} and \eqref{eq:DefJpOneBdry} now become
\begin{equation}
\begin{aligned}\label{eq:InJprimeoneshellmanydefects}
  I_n(y) &= \frac{(-1)^n}{(\sqrt{y})^{n-1}} \BJ_{n-1}(2 \sqrt{y}) 
  + 2\pi^2 \sum_j  \epsilon_j \left( - \frac{\alpha_j}{2\pi \sqrt{y}} \right)^n \BJ_n\left(\frac{\alpha_j \sqrt{y}}\pi\right)
   \quad \text{for} \quad n\ge 2 \ , \\
  J'(y) &= 1+ 2 \pi^2 \sum_j \epsilon_j \frac{\alpha_j^2}{4\pi^2} - \BJ_0(2\sqrt{y}) 
    - 2\pi^2 \sum_j \epsilon_j \frac{\alpha_j}{2\pi\sqrt{y}} \BJ_1\left(\frac{\alpha_j\sqrt{y}}\pi\right) \ . 
\end{aligned}
\end{equation}
With these expressions at hand one can again readily compute order-by-order in the genus expansion parameter $g_s$ the partition function of JT~gravity coupled to several species of defects.

%%%%%%%%%%%%%%%%%%%%%%%%%%%%%%%%%%%
\section{Low Temperature Expansion}\label{Section:LowtemperatureExpansion}
%%%%%%%%%%%%%%%%%%%%%%%%%%%%%%%%%%%
So far the partition function has been organised as a genus expansion. That is to say, for any given genus different powers of the temperature $T$ contribute in combination with different powers of the defect couplings~$\epsilon_j$. The contribution to the thermal partition functions at each genus are multiplied by polynomials in the inverse temperature $\beta=1/T$. Hence the magnitude of these polynomials are bounded for high temperatures, and the genus expansion in $g_s$ is sensible in the high temperature regime. However, this expansion breaks down in the low temperature limit $\beta\to+\infty$ unless we keep $g_s \beta^{3/2}$ fixed. Then the perturbative genus expansion remains finite and can be summed exactly \cite{OkuyamaSakai1,OkuyamaSakai2,Alishahihaetal}. This double scaling limit implies for the genus expansion parameter $g_s \to 0^+$, and as a consequence the non-perturbative corrections of the type $\sim e^{-1/g_s}$ vanish in this limit.

To study in the following in the described low temperature limit the interaction of JT~gravity with a gas of defects, the couplings~$\epsilon_j$ --- which are the characteristic energy scales of the defect, see, e.g., eq.~\eqref{eq:GSenergy} --- must be comparable to the low temperature scale $T$. Therefore, we additionally require that for $\beta \to +\infty$ the products $\beta \epsilon_j$ remain constant as well. This limit also implies that non-perturbative corrections of the type $\sim e^{-1/| \epsilon_j|}$ are exponentially suppressed.

%%%%%%%%%%%%%%%%%%%%%%%%%%%%%%%%%%%
\subsection{Low Temperature Limit}\label{LowTemperatureLimit}
%%%%%%%%%%%%%%%%%%%%%%%%%%%%%%%%%%%
Let us consider the low temperature expansion of JT~gravity coupled to a gas of defects of a single species type characterised by the defect coupling $\epsilon$ and the identification angle $\alpha$. To this end, we want to compute the partition functions $Z(\beta_1,\ldots,\beta_m)$ defined in eq.~\eqref{eq:Zfuncs} in the double scaling limit
\begin{equation} \label{eq:lowtemplimit}
  \beta_i \to +\infty \quad \text{with} \quad  g_s \beta_i^{3/2} = \text{const.} \ , \ \epsilon \beta_i = \text{const.} \quad \text{for all} \quad i=1,\ldots,m \ ,
\end{equation}
with distinct inverse temperatures $\beta_i$ for the individual boundary components.\footnote{In the absence of defects the low temperature limit of the partition function $Z(\beta_1,\beta_2)$ was previously derived in ref.~\cite{OkuyamaSakai2}. For the uniform limit $\beta \to +\infty$ with $\beta = \beta_1 = \ldots =\beta_m$ the low temperature limit of the partition functions together with defects has been first reported in ref.~\cite{Alishahihaetal}.} The inverse temperatures of the boundary components are conveniently described in terms of the universal inverse temperature scale $\beta$ and the dimensionless constants
\begin{equation}
  \mathfrak{b}_i = \frac{\beta_i}{\beta} \ .
\end{equation}
Then the above limit becomes $\beta \to+\infty$ for constant positive values $\mathfrak{b}_i$ while keeping $g_s \beta^{3/2}$ and $\epsilon\beta$ fixed. 

In the limit~\eqref{eq:lowtemplimit} (the topological part of) the partition function of eq.~\eqref{eq:Zfuncs} becomes
\begin{equation} \label{eq:ZmLowTempExp}
\begin{aligned}
  Z(\beta_1,&\ldots,\beta_m)^\text{top.} = 
     \frac1{g_s^2} \mathcal{B}(\beta_1) \cdots \mathcal{B}(\beta_m) G(1,\{ t_k = \sdef_k \}) \\
     &= \sum_{g,n=0}^{+\infty} \frac{(g_s \beta^{\frac32})^{2g-2+m}(\epsilon\beta)^n}{(2\pi)^\frac{m}2} \\
     &\quad \cdot\sum_{\ell_1,\ldots,\ell_m=0}^{+\infty}
      \beta^{\ell_1 + \ldots +\ell_m-m-n-3g+3}
      \left. 
      \mathfrak{b}_1^{\ell_1+\frac12}\cdots\mathfrak{b}_m^{\ell_m+\frac12}
      \partial_{\ell_1}\cdots\partial_{\ell_m} G_{g,m+n}(\{t_k\})  \right|_{t_k =\sdef_k/\epsilon}  \ ,
\end{aligned}
\end{equation}
with the generating function $G(1,\{ t_k\}) = \sum_{g,n} g_s^{2g} G_{g,n}(\{t_k\})$ decomposed into the contributions $G_{g,n}$ indexed by their genus $g$ and their number of marked points $n$. Imposing now the selection rule~\eqref{selectionrules} and inserting $\sdef_0= 2\pi^2 \epsilon$, we arrive at
\begin{multline} \label{eq:Zlowtemp}
  Z(\beta_1,\ldots,\beta_m)^\text{top.}  \\
  = \sum_{g,n=0}^{+\infty} 
  \frac{(g_s \beta^{\frac32})^{2g-2+m}(2\pi^2\epsilon\beta)^n}{(2\pi)^\frac{m}2 \, n!}
  \sum_{\ell_1,\ldots,\ell_m=0}^{+\infty} 
   \mathfrak{b}_1^{\ell_1+\frac12}\cdots\mathfrak{b}_m^{\ell_m+\frac12} 
   \left\langle \tau_0^n \tau_{\ell_1} \cdots \tau_{\ell_m} \right\rangle_g 
   + \mathcal{O}(\beta^{-1})\ ,
\end{multline}
in terms of the non-vanishing correlators~\eqref{topologicalgravitycorrelationfunctions} on the moduli space of stable curves $\overline{\mathcal{M}}_{g,m+n}$ of genus $g$ with $m+n$ marked points.\footnote{The correction terms $\mathcal{O}(\beta^{-1})$ depend on the genus expansion parameter $g_s$ and the coupling $\epsilon$ in such a way that in the double scaling limit~\eqref{eq:lowtemplimit} they approach zero at least with the rate $\sim1/\beta$.} The string equation of topological correlators implies (expect for the genus zero correlator $\left\langle \tau_0 \tau_0 \tau_0 \right\rangle_0 = 1$) \cite{WittenIntersection}
\begin{equation} \label{eq:topstringeq}
   \left\langle \tau_0^n \tau_{\ell_1} \cdots \tau_{\ell_m} \right\rangle_g
   = \sum_{p_1 +\ldots+ p_m = n} \frac{n!}{p_1! \cdots p_m!}
   \left\langle \tau_{\ell_1-p_1} \cdots \tau_{\ell_m-p_m} \right\rangle_g \ ,
\end{equation}
where $\left\langle \tau_{a_1} \cdots \tau_{a_m} \right\rangle_g = 0$ if any $a_i$, $i=1,\ldots,m$, is negative. 

Following ref.~\cite{OkuyamaSakai2}, we express the low temperature limit by applying the results of ref.~\cite{Okounkovnpointfunction}. Namely, let us define the generating function~$\mathcal{F}$ of topological correlators with $m$ marked points as
\begin{equation}
\begin{aligned}
  &\mathcal{F}(x) =\frac1{x^2} 
   +\sum_{\ell=0}^{+\infty} \sum_{g=1}^{+\infty} x^\ell \left\langle \tau_{\ell} \right\rangle_g \ ,\quad
   \mathcal{F}(x_1,x_2) = \frac{1}{x_1+x_2} 
   +\sum_{\ell_1,\ell_2=0}^{+\infty} \sum_{g=1}^{+\infty} x_1^{\ell_1} x_2^{\ell_2} 
    \left\langle \tau_{\ell_1}  \tau_{\ell_2} \right\rangle_g\ , \\
  &\mathcal{F}(x_1,\ldots,x_m) = 
   \sum_{\ell_1,\ldots,\ell_m=0}^{+\infty} \sum_{g=0}^{+\infty} x_1^{\ell_1} \cdots x_m^{\ell_m} 
   \left\langle \tau_{\ell_1} \cdots \tau_{\ell_m} \right\rangle_g \quad \text{for} \quad m\ge 3 \ .   
\end{aligned}   
\end{equation}
Using these expressions with the string equation~\eqref{eq:topstringeq} and formula $\left\langle\tau_{\alpha_1}\cdots\tau_{\alpha_n}\right\rangle_0=\frac{(n-3)!}{\alpha_1!\cdots\alpha_n!}$ we arrive from eq.~\eqref{eq:Zlowtemp} (for any $m\ge1$) at
\begin{equation} \label{eq:lowtemp}
 Z(\beta_1,\ldots,\beta_m)
  =\prod_{i=1}^m  \sqrt{\frac{g_s^{\frac23}\beta_i}{2\pi}} \
   e^{2\pi^2\epsilon\beta_i}
  \mathcal{F}(g_s^{2/3} \beta_1,\ldots,g_s^{2/3} \beta_m)
  + \mathcal{O}(\beta^{-1}) \ ,
\end{equation}
because $Z(\beta_1,\ldots,\beta_m)^\text{top.} =Z(\beta_1,\ldots,\beta_m)$ for $m>2$ while the semi-classical terms of the partition functions $Z(\beta_1)$ and $Z(\beta_1,\beta_2)$ are included in the leading non-polynomial terms in $\mathcal{F}(x)$ and $\mathcal{F}(x_1,x_2)$, respectively.

For this generating functions Okounkov has developed a remarkable formula spelt out in ref.~\cite{Okounkovnpointfunction}, namely
 \begin{equation} \label{eq:defF2}
   \mathcal{F}(x_1,\ldots,x_m) = 
   \frac{(2\pi)^{m/2}}{\sqrt{x_1\cdot\ldots\cdot x_m}} \mathcal{G}(2^{-1/3}x_1,\ldots,2^{-1/3}x_m) \ ,
\end{equation}
where
\begin{equation}
  \mathcal{G}(x_1,\ldots,x_m) = \sum_{\alpha \in \Pi_m} \frac{(-1)^{ \ell(\alpha) +1}}{\ell(\alpha)}
  \sum_{\sigma \in S_{\ell(\alpha)}} \mathcal{E}(\sigma(x_{\alpha})) \ . 
\end{equation}
Here the first sum is taken over the partitions $\Pi_m$ of the set $\{1,\ldots,m\}$ with $m$ elements where the individual partitions $\alpha$ characterised by their length $\ell(\alpha)$. Furthermore, to each partition $\alpha$ of length $\ell(\alpha)$ is assigned a vector $x_\alpha$ of length $\ell(\alpha)$, where the individual entries of $x_\alpha$ are in turn given by sums of the variables $x_i$ indexed by the subsets in $\alpha$. For example the partition $\alpha = \left\{ \{1,3,6\}, \{2\}, \{4,5\} \right\} \in \Pi_6$ of length $\ell(\alpha) = 3$ yields the vector $x_\alpha=(x_1+x_3+x_6,x_2,x_{4}+x_{5})$. The second sum runs over the permutations $\sigma$ in the symmetric group $S_{\ell(\alpha)}$ of size $\ell(\alpha)$, where $\sigma(x_\alpha)$ permutes the entries of the vector $x_\alpha$ of length $\ell(\alpha)$. Finally, the function $\mathcal{E}(x_1,\ldots,x_\ell)$ is defined as
\begin{equation}\label{eq:Definitionepsilon}
   \mathcal{E}\left(x_1,\ldots,x_\ell\right)=\frac{1}{2^{\ell}\pi^{\ell/2}}\frac{
   \text{e}^{\frac{1}{12}\sum_{i=1}^{\ell} x_i^3}}{\sqrt{x_1\cdot\ldots\cdot x_\ell}}
   \int\displaylimits_{y_i\geq 0} d y_1 \cdots d y_\ell \; 
   \text{e}^{ -\sum_{i=1}^{\ell}\frac{\left(y_i-y_{i+1} \right)^2}{4 x_i}-\sum_{i=1}^{\ell}\frac{y_i+y_{i+1}}{2}x_i}\ ,
\end{equation}
with $y_{\ell+1} \equiv y_1$. For further details on the function~$\mathcal{G}(x_1,\ldots,x_m)$ see the original definitions in refs.~\cite{Okounkovnpointfunction}.

Using the integral formulation of the generating functions~$\mathcal{F}$, in the low temperature limit the partition function $Z(\beta)$ is calculated to be
\begin{equation} \label{eq:Z1lowtemp}
   Z(\beta) = \frac{\text{e}^{\frac{g_s^2}{24}\beta^3+2\pi^2 \epsilon \beta}}{\sqrt{2\pi}\, g_s \beta^{\frac32}}
   + \mathcal{O}(\beta^{-1}) \ ,
\end{equation}
while the partition function $Z(\beta_1,\beta_2)$ becomes
\begin{equation}
   Z(\beta_1,\beta_2) = \frac{\text{e}^{\frac{g_s^2}{24} (\beta_1+\beta_2)^3 + 2\pi^2 \epsilon(\beta_1 +\beta_2)}}
   {\sqrt{2\pi} \, g_s (\beta_1 + \beta_2)^{\frac32}} \,
   \operatorname{erf}(2^{-3/2} g_s \sqrt{\beta_1\beta_2(\beta_1 + \beta_2)} )
   + \mathcal{O}(\beta^{-1}) \ ,
\end{equation}
in terms of the error function 
\begin{equation}
   \operatorname{erf}(x)  = \frac{2}{\sqrt{\pi}} \int_0^x du \, e^{-u^2}
   = \frac{2}{\sqrt{\pi}} \left( x - \frac{x^3}{3} + \frac{x^5}{10} - \ldots \, \right) \ . 
\end{equation}

%%%%%%%%
\subsection{Low Temperature Expansion Schemes} \label{OffshellExpansioninT}
%%%%%%%%

The corrections $\mathcal{O}(\beta^{-1})$ to the low temperature limit in eq.~\eqref{eq:lowtemp} are perturbatively included order-by-order by evaluating the subleading terms of eq.~\eqref{eq:ZmLowTempExp}. For explicitness we focus on the partition function $Z(T)$ with a single boundary component with temperature $T \equiv \beta^{-1}$, and we want to study its low temperature corrections
\begin{equation} \label{eq:TExpScheme1}
   Z(\epsilon; T) = \frac{T^{\frac32}\,\text{e}^{\frac{g_s^2}{24T^3}+\frac{2\pi^2 \epsilon}T}}{\sqrt{2\pi}\, g_s} 
   \mathcal{Z}_\epsilon(T) %(g_s\beta^{3/2},\epsilon\beta; T) 
   \quad \text{where} \quad
   \mathcal{Z}_\epsilon(T) %(g_s\beta^{3/2},\epsilon\beta; T) 
   = \sum_{\ell=0}^{+\infty} T^\ell \, z_\ell( g_s\beta^{3/2},\epsilon\beta ) \ .
\end{equation}
The coefficient functions $z_\ell$ do not depend on the temperature $T$ in the applied double scaling limit~\eqref{eq:lowtemplimit}. By including these perturbative temperature corrections to all orders, the partition function becomes an asymptotic series in $T$, i.e., the series does not contain any non-perturbative corrections that vanish in the limit $T\to 0$. 

Compared to the (asymptotic) genus expansion studied in detail in Section~\ref{JTGravityDeformed JT GravityandTopological Gravity}, the low temperature expansion~\eqref{eq:TExpScheme1} is more natural from a physics point of view, as for many physical problems one is interested in the result up to a certain energy scale. In particular, we see that by only taking the leading order contribution~\eqref{eq:Z1lowtemp} we can immediately read off the threshold energy~\eqref{eq:GSenergy}. Note, however, that since the coupling $\epsilon$ approaches zero in the low energy limit $T\to0$, the ground state energy and the subleading temperature corrections in the expansion~\eqref{eq:TExpScheme1} depend on the details of the chosen double scaling limit. The limit~\eqref{eq:lowtemp} is naturally adapted to the defect coupling $\epsilon$ and the genus expansion parameter $g_s$. However, alternatively we can study other low temperature limits, where other ratios between physical parameters and the temperature $T$ are kept constant. In the following, we refer to such different choices for the double scaling limits as distinct low temperature expansion schemes.

In addition to the scheme discussed in the previous subsection, we introduce the low temperature expansion scheme of ref.~\cite{OkuyamaSakai1}, which is naturally adapted to the variables $(y,t)$ defined in eq.~\eqref{eq:yt} by the double scaling limit 
\begin{equation} \label{eq:lowtemplimityt}
  \beta \to +\infty \quad \text{with} \quad  \frac{g_s \beta^{\frac32}}{t} = \text{const.} \ , \ y \beta = \text{const.} \ .
\end{equation}
Solving eq.~\eqref{eq:yt} for small deformations $\sdef_k$, $k=1,2,3,\ldots$, away from pure JT~gravity yields for the coupling parameters $(y,t)$ appearing in the above limit the expansion
\begin{equation}
\begin{aligned}
  y &= \sdef_0 + \frac12 (2 \sdef_0\sdef_1 -\sdef_0^2)+\ldots \ ,\\
  t & = 1 - (\sdef_0 + \sdef_1) + (\sdef_0^2 -\sdef_0\sdef_1 -\sdef_0\sdef_2) + \ldots \ ,
\end{aligned}
\end{equation}
which at leading order for a single defect become $y =2\pi^2\epsilon + \mathcal{O}(\epsilon^2)$ and $t = 1 + \mathcal{O}(\epsilon)$ (cf.\ eq.~\eqref{eq:ytexpansionpoint}). This low temperature expansion scheme agrees at leading order in $\epsilon$ with the scheme~\eqref{eq:lowtemplimit}, and in particular, upon inserting the on-shell values for $(y,t)$ in the absence of defects, i.e., setting $\epsilon=0$ such that $(y,t)=(0,1)$, the two low temperature expansion schemes become the same.

In the latter scheme the (asymptotic) low temperature expansion of the partition function reads \cite{OkuyamaSakai1}
\begin{equation}\label{eq:ZT}
    Z(y,t;T)=\frac{T^{\frac32}\,\text{e}^{\frac{g_s^2}{24t^2T^3}+\frac{y}{T}}}{\sqrt{2\pi}\, g_s} \mathcal{Z}_{y,t}(T) 
     \quad \text{where} \quad
     \mathcal{Z}_{y,t}(T) =\sum_{\ell=0}^{+\infty} T^\ell z_\ell(y,t) \ , 
\end{equation}
where the coefficient functions $z_\ell(y,t)$ now differ from the coefficient functions $z_\ell(g_s\beta^{3/2},\epsilon\beta)$ in eq.~\eqref{eq:TExpScheme1} (even after inserting the functional relations among their respective arguments).\footnote{There is actually a subtlety here. While the coefficient functions $z_\ell(g_s\beta^{3/2},\epsilon\beta)$ are temperature independent in the double scaling limit~\eqref{eq:lowtemplimit}, the functions $z_\ell(y,t)$ are still temperature dependent in the limit~\eqref{eq:lowtemplimityt}. One can obviously define temperature independent coefficients in the latter case as well. However, as discussed in the following the coefficient functions $z_\ell(y,t)$ are conveniently computable and comparable with ref.~\cite{OkuyamaSakai1}. Truncating the infinite sum in $\mathcal{Z}_{y,t}$ at some finite value $\ell=N$ yields unambiguously the low temperature corrections up to order $T^N$ in the discussed expansion scheme (because the temperature dependence only gives rise to corrections at order $\mathcal{O}(T^{N+1})$).} 

For completeness, let us briefly review the strategy of ref.~\cite{OkuyamaSakai1} to compute the coefficients $z_\ell(y,t)$. First of all, the coefficients $z_\ell(y,t)$ are conveniently determined from the low temperature expansion of the function $W(y,t;\beta)$ defined in eq.~\eqref{eq:Wgenusxpansion}. Using the ansatz
\begin{equation}\label{eq:WTansatz} 
    W(y,t;T)=\sqrt{\frac{T}{4 \pi}}\,\text{e}^{\frac{g_s^2}{24 t^2 T^3 }+\frac{y}{T}}\mathcal{W}_{y,t}(T)
     \quad \text{where} \quad
   \mathcal{W}_{y,t}(T)=\sum_{\ell=0}^{\infty}T^\ell w_\ell(y,t) \ ,
\end{equation} 
together with equation~\eqref{eq:DiffRec} yields the differential equation
\begin{equation} \label{eq:WT}
  \partial_t \mathcal{W}_{y,t}
  =\frac{g_s^2}{12 t^3T^3}\mathcal{W}_{y,t}-\sum_{g=1}^{\infty} g_s^{2 g} u_g \nabla(T) \mathcal{W}_{y,t} +\frac{g_s^2}{12}\nabla(T)^3 \mathcal{W}_{y,t} \ ,
\end{equation}
in terms of the differential operator
\begin{equation}
   \nabla(T) % = e^{-\frac{h^2}{12 t^2 }-\frac{y}{T}} \partial_0 e^{\frac{h^2}{12 t^2 }+\frac{y}{T}}
 =\partial_0+\frac{1}{t T}+ \frac{g_s^2 \, I_2}{12 \,t^4 \, T^3}
 = \frac1t \left( -I_2 \partial_t + D_y\right)+ \frac{g_s^2 \, I_2}{12 \,t^4 \, T^3} \ , \qquad
 D_y = \partial_y + \frac1T \ ,
\end{equation}
which then recursively determines the coefficient functions $w_\ell(y,t)$.\footnote{The first few coefficient functions $w_\ell$ are calculated and spelled out explicitly in ref.~\cite{OkuyamaSakai1}.} Finally, the relation~\eqref{eq:ZWrel} translates to
\begin{equation}
    \mathcal{W}_{y,t} = T\, \nabla(T) \mathcal{Z}_{y,t}  \ ,
\end{equation}
leading for the coefficient functions $z_\ell(y,t)$ to the recursion formula \cite{OkuyamaSakai1}
\begin{equation}\label{eq:zlwl}
    z_\ell 
    = t \left(\ell! \, w_\ell-\ell \left(\nabla(T)- \frac{1}{t T} \right) z_{\ell-1}\right) \ .
\end{equation}
The first few coefficient functions $z_\ell$ are calculated to be
\begin{gather} \label{eq:zlsolutions}
   z_0=t\ , \quad 
  z_1 =\left(1+\frac{g_s^4 }{240 t^4 T^{6} } \right) I_2 \ , \nonumber \\
   z_2 =\left(\frac{7 g_s^4}{240 t^5 T^{6}}+\frac{g_s^6}{576 t^7 T^9}+\frac{g_s^8}{57600 t^9 T^{12}} \right) I_2^2
   +\left(-2+\frac{g_s^2}{12 t^2 T^{3}}+\frac{g_s^4}{120 t^4 T^{6}}+\frac{g_s^6}{3360 t^6 T^9} \right) I_3\ .
\end{gather}

Let us point out some physical implications of the low temperature expansion scheme in the variables $(y,t)$. For the on-shell values \eqref{eq:ytexpansionpoint} of $(y,t)$ for JT~gravity coupled to a gas of defect and compared to the expansion scheme~\eqref{eq:TExpScheme1}, the low temperature expansion of the partition function depends on the identification angle $\alpha$ already at leading orders in the temperature~$T$. Namely compared to the result~\eqref{eq:Z1lowtemp} one finds upon inserting eq.~\eqref{eq:ytexpansionpoint} into the expansion~\eqref{eq:ZT}
\begin{equation}
  Z(T) =  \frac{(1+\frac{\alpha^2-4\pi^2}{2} \epsilon+\ldots)\,T^{\frac32}\,\text{e}^{\frac{g_s^2}{24 T^3}+\frac{g_s^2(4\pi^2-\alpha^2)}{24 T^3} \epsilon +\frac{2\pi^2}T \epsilon + \ldots }}{\sqrt{2\pi}\, g_s}
   + \mathcal{O}(T) \ ,
\end{equation}
where the dots `$\ldots$' indicate subleading terms in $\epsilon$ at order $\mathcal{O}(\epsilon^2)$.

The above analysis of the low temperature limit is general in the sense that we can consider other on-shell values for the couplings $(y,t)$ (and also for the couplings $t_k=\gamma_k+\sdef_k$ appearing implicitly in the expansion~\eqref{eq:ZT}). In particular, if we consider small deviations from the on-shell values $(y,t)=(1,0)$ (and small perturbations $\sdef_k$ for $k\ge 2$) of pure JT~gravity, we can study the low temperature expansions of deformations to pure JT~gravity together with their scheme dependence.

A particularly interesting example in this context is discussed in ref.~\cite{WittenDeformations}, which corresponds to coupling JT~gravity to a gas of defects with two types of defect species characterized by their couplings $\epsilon_1 = -\epsilon_2 = \epsilon$, which are aligned with opposite sign, and their respective identification angles $\alpha_1$ and $\alpha_2$. On the one hand, for the low temperature double scaling limit~\eqref{eq:lowtemplimit} we arrive at
\begin{equation}
   Z(T) = \frac{T^{\frac32}\,\text{e}^{\frac{g_s^2}{24 T^3}}}{\sqrt{2\pi}\, g_s}
   + \mathcal{O}(T) \ ,
\end{equation}
which results in an expected vanishing threshold energy, cf.\ eq.~\eqref{eq:GSenergy}. On the other hand the double scaling limit~\eqref{eq:lowtemplimityt} yields
\begin{equation}
  Z(T) =  \frac{(1+\frac{(\alpha_1^2-\alpha_2^2)}{2} \epsilon+\ldots)\,T^{\frac32}\,\text{e}^{\frac{g_s^2}{24 T^3}+\frac{g_s^2(\alpha_2^2-\alpha_1^2)}{24 T^3} \epsilon  + \ldots }}{\sqrt{2\pi}\, g_s}
   + \mathcal{O}(T) \ ,
\end{equation}
where a non-trivial dependence on the identification angles $\alpha_1$ and $\alpha_2$ now enters because the couplings $(y,t)$ govern the physical quantities that are kept constant in the double scaling limit~\eqref{eq:lowtemplimityt}.

%%%%%%%%%%%%%%%%%%%%%%%%%%%%%%%%%%%%%%%%%%%
\subsection{Low Temperature Expansion Schemes for Multiple Boundaries}
%%%%%%%%%%%%%%%%%%%%%%%%%%%%%%%%%%%%%%%%%%%
Finally, let us remark that the low temperature discussion of the previous subsection can be repeated with multiple boundary components in the same way. The low temperature expansion in this case is studied by Okuyama and Sakai in ref.~\cite{OkuyamaSakai2}. 

As a preparation for Section~\ref{Spectral Form Factor}, we just record here the result of the low temperature limit for the partition function $Z(\beta_1,\beta_2)$ with two boundary components with inverse temperatures $\beta_1$ and $\beta_2$. Then the low temperature expansion scheme~\eqref{eq:lowtemplimityt} generalises to the double scaling limit
\begin{equation} 
  \beta_i \to +\infty \quad \text{with} \quad  \frac{g_s \beta_i^{\frac32}}{t} = \text{const.} \ , \ y \beta_i = \text{const.}
  \quad \text{for} \quad i=1,2 \ ,
\end{equation}
which yields for the low temperature limit of the partition function $Z(\beta_1,\beta_2)$ the result \cite{OkuyamaSakai2}
\begin{equation} \label{eq:ZZT}
   Z(y,t;\beta_1,\beta_2) = \frac{t\,\text{e}^{\frac{g_s^2(\beta_1+\beta_2)^3}{24 t^2}+y(\beta_1+\beta_2)}}{\sqrt{2\pi}\, g_s (\beta_1+\beta_2)^{\frac32}} \,
   \operatorname{erf}\left(\frac{g_s}{2\sqrt{2}\,t}\sqrt{\beta_1\beta_2(\beta_1+\beta_2))} \right)
   +\mathcal{O}(\beta_1^{-1},\beta_2^{-1}) \ .
\end{equation}

%%%%%%%%%%%%%%%%%%%%%%%%%%%%%%%%%%%%%%%%%%%%%%%%%%%
\section{Phase Transition and Spectral Form Factor}\label{Spectral Form Factor}
%%%%%%%%%%%%%%%%%%%%%%%%%%%%%%%%%%%%%%%%%%%%%%%%%%%
 Using the low temperature limit of the partition functions $Z(y,t;\beta_1,\beta_2)$ and $Z(y,t;\beta)$ of the previous section and applying numerical methods, we study two well-established and related phenomena, namely the phase transition \cite{doubletrumpet,Engelhardt:2020qpv}, which exchanges the dominance between the connected versus the disconnected geometries in the two boundary partition function, and the spectral form factor,\footnote{The spectral form factor was first introduced in the AdS/CFT context in ref.~\cite{Papadodimas:2015xma}.} which arises as a certain analytic continuation of the two-boundary partition function. In particular, we analyse the dependence of these quantities in the presence of defects.
 
%%%%%%%%%%%%%%%%%%%%
\begin{figure}[t]
	\centering
	\includegraphics[scale=0.5]{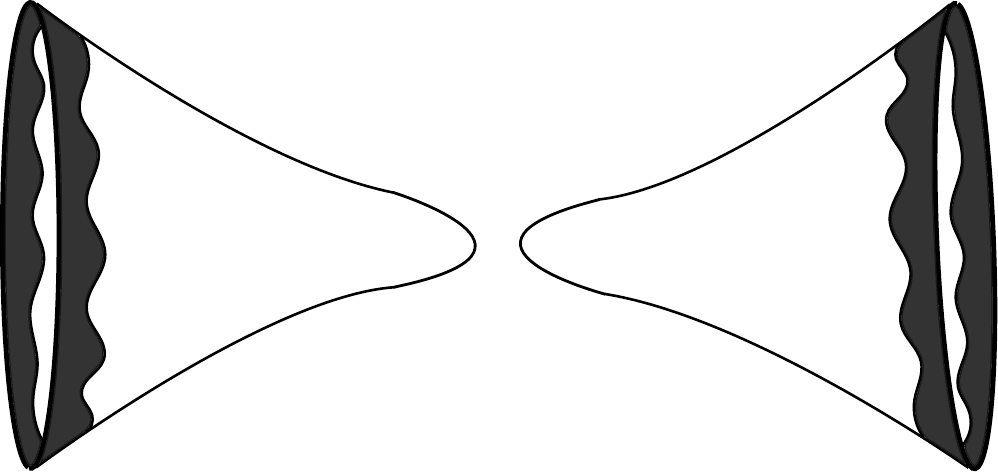}\hspace{20ex}
	\includegraphics[scale=0.4]{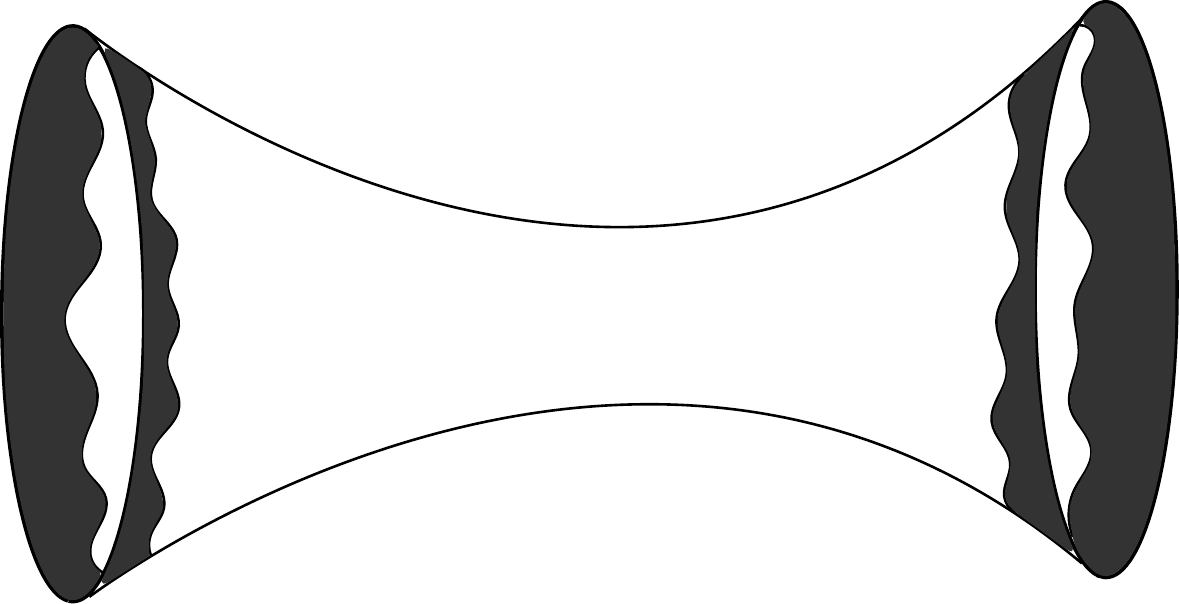}
    \caption{The left figure shows a disconnected geometry --- here illustrated in terms of two $AdS_2$ disks at genus zero --- that dominates the spectral form factor at early times $\tau$, whereas the right figure depicts a connected geometry with two boundaries --- shown is the double trumpet contribution --- that becomes dominant at late times $\tau$. }
	\label{fig:SSFGeometries}
\end{figure}
%%%%%%%%%%%%%%%%%
 
%%%%%%%%%%%%%%%%%%%%%%%%%%%%%%
 \subsection*{Phase Transition}
 There are two types of geometries that contribute to the two-point function. On the one hand there are geometries with two disconnected components, each with a single boundary component, and on the other hand there are connected geometries with two boundary components, as illustrated in fig.~\ref{fig:SSFGeometries} (where only the genus zero contributions are depicted for simplicity). At low temperatures we have according to eqs.~\eqref{eq:ZT} and \eqref{eq:ZZT} (in the chosen low temperatue expansion scheme) the following two quantities
\begin{equation}\label{eq:generaltwopointfunctions}
\begin{aligned}
   Z(y,t;\beta)^2&=\frac{e^{2 y \beta} e^{\frac{g_s^2 \beta^3}{12 t^2}}}{2 \pi g_s^2 \beta^{3}}t^2
   + \mathcal{O}(\beta^{-1}) \ ,\\
   Z(y,t;\beta,\beta) &=\frac{e^{2 y \beta} e^{\frac{\beta^3 g_s^2}{3 t^2}}}{4 \sqrt{ \pi} \beta^{3/2}g_s}\, t\, \operatorname{erf}\left(\frac{\beta^{3/2} g_s}{2 t}\right) 
   + \mathcal{O}(\beta^{-1}) \ .
\end{aligned}
\end{equation}
Independent of the specific choices for the on-shell values of the parameters $(y,t)$, we can make some quite general comments. Taking the ratio of the two-point contributions in eq.~\eqref{eq:generaltwopointfunctions}, the dependence on the shift in energy given by $y$ drops out (at leading order in the temperature). Hence, the phase transition (and as a consequence also the spectral form factor introduced later) is determined by the off-shell parameter $t$. Explicitly analysing the ratio of the two contributions~\eqref{eq:generaltwopointfunctions} in the low temperature regime yields with the dimensionless constant $c:={g_s\beta^{3/2}}/{t}$ the dimensionless (numerical) critical value $c_\text{crit.}$ for the phase transition according to
\begin{equation}
    \frac{Z(y,t;\beta,\beta)}{Z(y,t;\beta)^2}=1 \ \Rightarrow\  \frac{1}{2} \sqrt{\pi } c e^{\frac{c^2}{4}} \text{erf}\left(\frac{c}{2}\right)=1\ \Rightarrow\ 
    c_\text{crit.} \approx \pm 1.24013 \ .
    %\left|c\right|=c_{\text{crit.}}\sim 1.24013
\end{equation}

\begin{figure}[ht]
    \centering
    \includegraphics[scale=1.3]{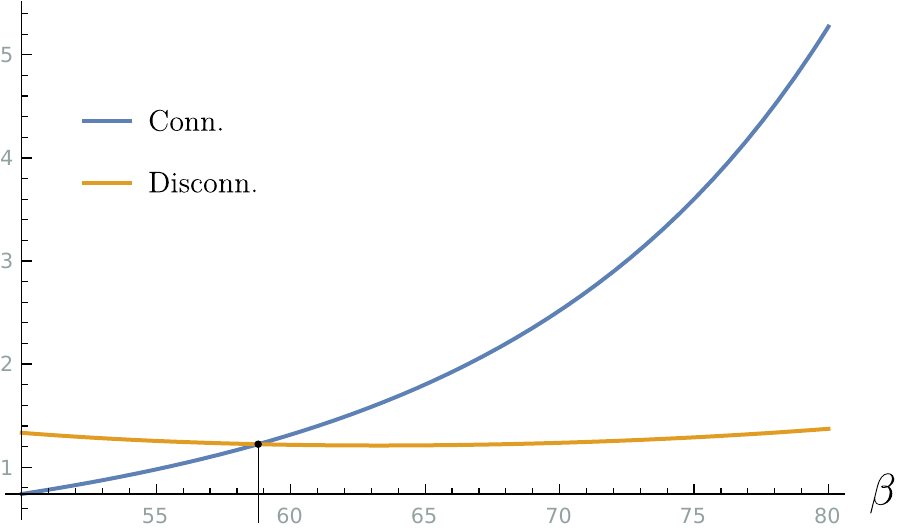}
    \caption{We plot the connected versus the disconnected geometry contributions of eq.~\eqref{eq:twopointfunctions}. The identification angle $\alpha$ is fixed to $\alpha=\frac{\pi}{2}$, the defect amplitude is $\epsilon=0.001$  and $g_s=0.0027$.
    In the range of the plot we have a maximum of $\ \sim 4.2\% $ relative error, which measures the ratio of the terms ignored (order $T^2$) over the terms kept in the low temperature expansion.}
    \label{fig:ConnVSDiscBETA}
\end{figure}

\begin{figure}[ht]
    \centering
    \includegraphics[scale=1.2]{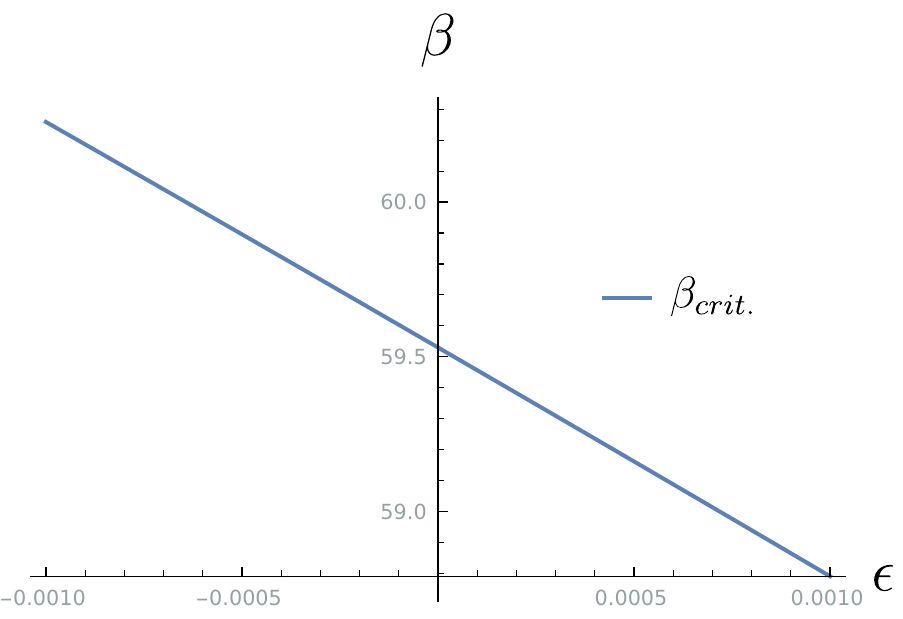}
    \caption{The phase transition temperature (the point for which $Z(\beta,\beta)=Z(\beta)^2$) as a function of the defect amplitude. The identification angle $\alpha$ is fixed to $\alpha=\frac{\pi}{2}$ and $g_s=0.0027$.}
    \label{fig:PhaseTransTempBETA}
\end{figure}

Let us now focus on JT gravity with defects. This means that we take $(y,t)$ to their on-shell values ~\eqref{eq:ytonshell} and that we work with the quantities in eq.~\eqref{eq:twopointfunctions}, where the on-shell values of $(y,t)$ are found numerically for a given set of $\epsilon$ and $\alpha$, i.e.
\begin{equation}\label{eq:twopointfunctions}
    \begin{aligned}
         Z(\beta)^2&=\left.\frac{e^{2 y \beta} e^{\frac{g_s^2 \beta^3}{12 t^2}}}{2 \pi g_s^2 \beta^3}t^2\right|_{y,t \text{ on-shell}}\,,\\
        Z(\beta,\beta) &=\left.\frac{e^{2 y \beta} e^{\frac{\beta^3 g_s^2}{3 t^2}}}{4 \sqrt{ \pi} \beta^{3/2} g_s}\, t\, \text{erf}\left(\frac{\beta^{3/2} g_s}{2 t}\right)\right|_{y,t \text{ on-shell}}.
    \end{aligned}
\end{equation}
\noindent Keeping the above in mind, we plot the connected and disconnected parts of the two-point function in Fig. \ref{fig:ConnVSDiscBETA}.
\noindent We can see the general behaviour of JT~gravity in the absence of defects is reproduced: at high temperatures the disconnected geometry dominates, whereas for low temperatures the connected part constitutes the more dominant contribution \cite{CJ3,OkuyamaSakai2}. This is the two-dimensional instantiation of a Hawking-Page phase transition \cite{WittenHawkingPage,MaldacenaAdS2}. However, we should also notice that, as shown in fig.\ \ref{fig:PhaseTransTempBETA}, for larger $\epsilon$, the phase transition occurs at a smaller value of $\beta$. 
%%%%%%%%%%%%%%%%%%%%%%%%%%%
\begin{figure}[h!]
    \centering
    \includegraphics[scale=1.3]{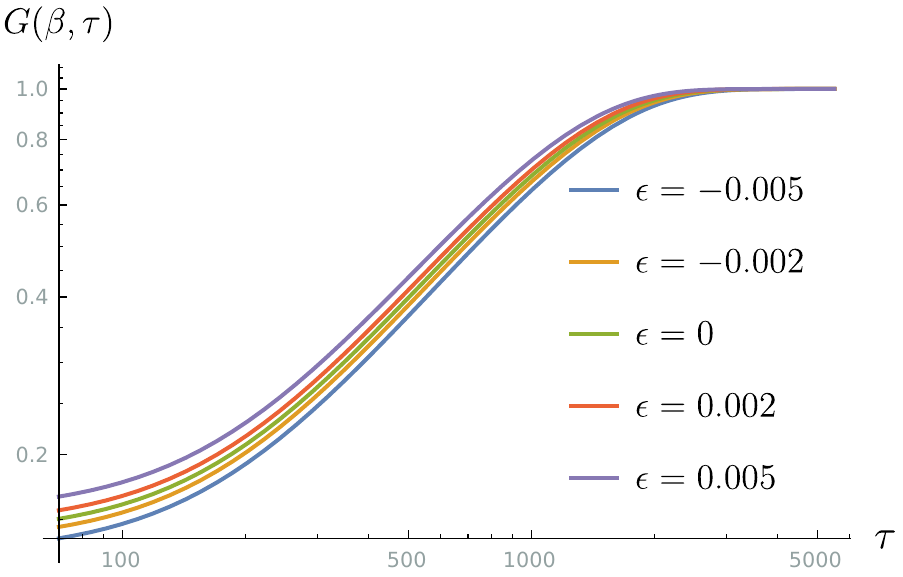}
    \caption{Shown is the spectral form factor for different values of $\epsilon$ with $g_s=\frac{1}{4\cdot180^{3/2}}$, $\beta=180$, $\alpha=\pi/2$.}
    \label{fig:SFFproper}
\end{figure}
%%%%%%%%%%%%%%%%%%%%%%%%%%%

%%%%%%%%%%%%%%%%%%%%%%%%%%%%%%%%%%%%%%
\subsection*{Spectral Form Factor}
Now we come to the analysis of the spectral form factor $Z(\beta+i \tau,\beta-i \tau)$, which is a real function of the time $\tau$ defined via an analytic continuation of the two-point function $Z(\beta_1,\beta_2)$. The spectral form factor is essential in the analysis of quantum chaotic behaviour and plays an increasingly important role in the study of black hole physics \cite{BlackHolesandRandomMatrices}. For the case of JT gravity in the presence of defects the spectral form factor has not yet been analysed. The task is to understand the role of the parameter $\epsilon$.

For large groups of systems obeying quite common assumptions (such as the eigenstate thermalisation hypothesis \cite{PhysRevE.50.888, PhysRevA.43.2046}), one expects the spectral form factor to exhibit certain universal features. Early times are characterised by decay and hence a ``slope", followed by a rise and hence a ``ramp", and lastly at late times we encounter a ``plateau" with fixed value given by the one-point function $Z(2\beta)$.\footnote{The ``plateau'' cannot be obtained if the perturbative series is truncated at some finite $g$. To render an asymptotic series convergent non-perturbative contributions have to be taken into account \cite{doubletrumpet}. In the zero temperature/zero coupling limit considered in ref.~\cite{OkuyamaSakai1} and here the perturbative series converges.} Let us define the normalised spectral form factor in the following manner
\begin{align}\label{eq:nSFF}
    G(\beta,\tau):=\frac{ Z(\beta + i \tau,\beta - i \tau )}{Z(2 \beta)} =\text{erf}\left(\frac{\beta ^{3/2} g_s \sqrt{\frac{\tau ^2}{\beta ^2}+1}}{2t}\right)\,,
\end{align}
where we are normalising with respect to the contribution $Z(2 \beta)$ as this sets the height of the plateau. Due to the low temperature dominance of the connected contribution as outlined above, we would expect late times to be dominated by connected contributions. A closer look at eq.~\eqref{eq:generaltwopointfunctions} shows that this is guaranteed by the functional form of both expressions. We are only considering connected geometries in eq.~\eqref{eq:nSFF} as we are mainly interested in the ramp and plateau behaviour. We want to reiterate some statements of refs.~\cite{SSS,OkuyamaSakai2}, which help in understanding the importance of the corrections outlined in section \ref{LowTemperatureLimit}. The $g=0$ part of the two-boundary correlator only furnishes the ``ramp" behaviour as shown in ref.~\cite{doubletrumpet}. We can see that the approximation \eqref{eq:lowtemplimit} already allows for the creation of the plateau \cite{OkuyamaSakai2}. Furthermore, if we work in the limit \eqref{eq:lowtemplimityt} both the phase transition and spectral form factor become sensitive to the presence of defects.

We note that the transition from ramp to plateau now depends on $\epsilon$. More specifically, for larger values of $\epsilon$ we can move it to earlier times, whereas negative values moves it to later times, which mirrors the behaviour discovered for the phase transition.

We may also consider changes in the identification angle~$\alpha$ while keeping $\epsilon$ fixed for both the phase transition and the spectral form factor. While the dependence on $\alpha$ within the range $0\le\alpha<\pi$ can be studied straightforwardly with the methods presented here, it would be even more interesting to consider changes in $\alpha$ over the full range of identification angles. This could possibly be achieved by implementing the results of ref.~\cite{TuriaciBluntDefects}.

%%%%%%%%%%%%%%%%%%%%%%%%%%%%%%%%%%%%%%%%%
\section{Some Comments on two-dimensional de~Sitter space }\label{ds}
%%%%%%%%%%%%%%%%%%%%%%%%%%%%%%%%%%%%%%%%%
In both refs.~\cite{MaldacenadS,MaloneydS} a proposal is made for the application of the matrix model/JT~duality to two-dimensional de~Sitter space. The logic is the following: As Lorentzian de~Sitter space can be analytically continued to Euclidean Anti-de~Sitter space \cite{Maldacenadsviaads}, in the two-dimensional setting there should exist a map translating the results for the partition function of ref.~\cite{SSS} to the wavefunction of the universe~$\Psi$ at future infinity $\mathcal{I}^{+}$ and past infinity $\mathcal{I}^{-}$ \cite{MaldacenadS,MaloneydS}. For the semi-classical contribution it can be shown that the wavefunction can be mapped to the disk result via the identification
\begin{align}\label{eq:dsbeta}
    \beta \rightarrow \left\{\begin{matrix} -i \ell,\ \text{future}\\
    i \ell, \ \text{past}\end{matrix}\right.
\end{align}
where $\ell$ is the renormalised length of both the future and past circles. For higher genus contributions an approach was outlined in ref.~\cite{MaloneydS}, in which the boundary conditions inherited from de~Sitter space require the analytic continuation of the geodesic length $b\rightarrow i \alpha$ such that this instance of JT gravity requires the inclusion of surfaces with conical singularities. Sticking to the one-point function for the moment, following ref.~\cite{MaloneydS} the wave function on a single future boundary would be given by
\begin{equation}
    \begin{aligned}\label{eq:dssingleboundary}
  \Psi(\ell) &=\frac{(2 \pi^2)^{3/2}}{g_s} Z^{\text{disk}}\left( - i \ell \right)-\sum_{g=1}^{\infty}g_s^{2g-1}\int\displaylimits_{0}^{\infty}d \alpha \alpha \frac{e^{\frac{ i \alpha^2}{4 \pi^2 \ell}}}{2 \pi \sqrt{\pi} \sqrt{- i \ell}}V_{g,(i \alpha)}\\
   &= \frac{(2 \pi^2)^{3/2}}{g_s} Z^{\text{disk}}\left( -i \ell \right)+\frac{1}{g_s^2}B\left(- i \ell\right)F(\{\gamma_k\})\ .
\end{aligned}
\end{equation}
which would indeed correspond to $Z(-i \ell)$. In general this approach implies that the mere analytic continuation \eqref{eq:dsbeta} of the partition function of ref.~\cite{SSS} corresponds to the wave function $\Psi$, i.e.
\begin{equation}
    \begin{aligned}\label{eq:noboundarywavefunction}
  &\Psi_{\text{conn.}} \left(\ell_1,...,\ell_{n_{+}},\ell_{n_{+}+1},\ldots,\ell_{n_{-}}\right)\\ \mathrel{\widehat{=}} &\left\langle\text{tr}\left(e^{i \ell_1 H}\right)\ldots\text{tr}\left(e^{i \ell_{n_+} H}\right)\text{tr}\left(e^{-i \ell_{n_{+}+1} H}\right)\ldots\text{tr}\left(e^{-i \ell_{n_{-}} H}\right) \right\rangle\ ,
\end{aligned}
\end{equation}
However, as clearly stated in ref.~\cite{MaloneydS}, for eqs.~\eqref{eq:dssingleboundary} and \eqref{eq:noboundarywavefunction} to hold in full generality it is necessary that the conical volumes are obtained from a mere analytic continuation as in eq.~\eqref{eq:conicalWPvolumes}. This, however, is only established for $\alpha < \pi$, whereas the results of ref.~\cite{TuriaciBluntDefects} propose for general identification angles~$\alpha$ an implicit definition of Weil--Petersson volumes that goes beyond the analytic continuation prescription of eq.~\eqref{eq:conicalWPvolumes}. Hence, due to the integration range over $\alpha$ in eq.~\eqref{eq:dssingleboundary}, the naive analytic continuation of the individual volumes $V_{g,(b)}$ of the (asymptotic) thermal partition function possibly requires a further modification to the approach of ref.~\cite{MaloneydS} for the computation of the wavefunction $\Psi$.\footnote{Although it is not immediately clear if the path integral may be performed in the same manner as in ref.~\cite{MaloneydS} for the volumes of ref.~\cite{TuriaciBluntDefects}. See comments in ref.~\cite{TuriaciBluntDefects}.} Moreover, the authors of ref.~\cite{MaloneydS} show that eq.~\eqref{eq:noboundarywavefunction} may be derived from the approach of ref.~\cite{Hartle:1983ai}, such that the wavefunction~$\Psi$ is also equivalent to the no-boundary wavefunction. Therefore, further investigation is required in order to understand in how far the correspondence of the Hartle--Hawking construction of ref.~\cite{Hartle:1983ai} and the approach of ref.~\cite{MaloneydS} via continuation to Euclidean Anti-de~Sitter space holds at the non-perturbative level and in how far the validity of eq.~\eqref{eq:noboundarywavefunction} is guaranteed beyond the semi-classical level.\footnote{We would like to thank Joaquin Turiaci for valuable correspondence on these points.}

%%%%%%%%%%%%%%%%%%%%%%%%%%%%%%%%%%%%%%%%%%%%%%%%%%%
\section{Conclusion and Outlook} \label{sec:concl}
%%%%%%%%%%%%%%%%%%%%%%%%%%%%%%%%%%%%%%%%%%%%%%%%%%%
In this work we compute thermal partition functions of deformed JT~gravity theories from solutions to the KdV~hierarchy. These solutions govern the correlation functions of two-dimensional topological gravity, and --- similarly as in ref.~\cite{OkuyamaSakai1} --- we describe both undeformed and a rather general class of deformed theories of JT~gravity in terms of solutions to the KdV~hierarchy. In refs.~\cite{Maxfield3gravity,WittenDeformations} deformations of JT~gravity are described by suitable scalar potentials that do not alter the asymptotic boundaries of the two-dimensional hyperbolic space-time geometries. It would be interesting to relate deformations arising from scalar potentials to solutions of the KdV~hierarchy in the topological gravity description. While we can identify certain classes of deformations in both formulations --- in particular those that arise from a gas of defects with a finite number of defect species --- it would be interesting to investigate whether these two approaches towards deformations of JT~gravity are actually in one-to-one correspondence. As both descriptions yield infinite dimensional deformation spaces, a meaningful comparison of the two approaches to the deformation problem presumably requires a careful treatment using methods of functional analysis.

Interestingly, both standard JT~gravity and JT~gravity interacting with a finite number of defect species are governed by spectral densities given in terms of (modified) Bessel functions, whereas for more general deformations other transcendental functions occur. Therefore, it would be interesting to understand in how far standard JT~gravity and JT~gravity interacting with a gas of defects are singled out from other solutions to the KdV~hierarchy. For instance, the Witten--Kontsevich tau-function relates to the free energy of two-dimensional topological gravity \cite{WittenIntersection, KontsevichIntersection} and the Br\'ezin--Gross--Witten tau-function describes JT~supergravity \cite{Okuyama:2020qpm}. Yet other tau-functions are discussed from the mathematical perspective in ref.~\cite{MR4222602}. As the connection between specific solutions to the KdV~hierarchy and two-dimensional gravitational theories does not seem to be arbitrary, a systematic investigation of tau functions and the associated physical theories is an interesting idea to pursue.

As already addressed in ref.~\cite{SSS}, the discussed solutions to the KdV~hierarchy and the resulting thermal partition functions are asymptotic series in the genus expansion parameter~$g_s$, which are only rendered to analytic functions once non-perturbative effects are taken into account. Therefore, a challenging task is to derive solutions to the KdV~hierarchy that are analytic instead of just being an asymptotic series in the parameter $g_s$. In refs.~\cite{Dalley:1991qg,Dalley:1991vr} a non-perturbative completion of the solutions to the KdV~hierarchy is proposed that has recently been applied to JT~gravity in an interesting series of works \cite{CJ1,CJ3,CJ4}. Both the results of ref.~\cite{OkuyamaSakai1} and our work furnish an easy and systematic access to higher genus contributions, such that modern resurgence techniques could come into play to address non-perturbative effects in this context. Similar considerations in that direction are made in ref.~\cite{finitecutoffresurgence} for JT~gravity with a finite cutoff at the asymptotic space-time boundaries,\footnote{More work on JT gravity restricted to a finite $AdS_2$ subregion can be found in refs.~\cite{Gross:2019ach,Iliesiu:2020zld}. The general paradigm of finite cutoff $AdS/CFT$ was first explored in ref.~\cite{McGough:2016lol}.}
%/$T\Bar{T}$ duality
where a Borel resummation can be performed for the asymptotic series with respect to the cutoff parameter. 

Applying the approach developed by Okuyama and Sakai \cite{OkuyamaSakai1,OkuyamaSakai2}, we compute in a certain low temperature limit the thermal partition functions (with one or more boundary components) for JT~gravity with deformations such as those arising from the presence of a gas of defects. In this limit the studied thermal partition functions become exact because non-perturbative corrections are suppressed. We determine  the critical temperature of the Hawking--Page phase transition as a function of the defect parameters by analysing the two-boundary partition function with numerical methods. Depending on the sign of the defect coupling constant we find that the phase transition either occurs at higher or lower temperatures. The spectral form factor exhibits a similar behaviour, namely the time scale for the onset of the plateau is shifted to earlier or later times depending on the sign of the defect coupling. While we expect that this behaviour of the phase transition and the spectral form factor as a function of the defect parameters does not change upon including further subleading temperature corrections, it is nevertheless desirable to include further terms in the low temperature expansion in order to reliable analyse the Hawking--Page phase transition and the spectral form factor as a function of the defect parameters at higher temperature scales. JT~gravity in the presence of defects is linked to 3d gravity in the near-extremal limit, as reported in ref.~\cite{Maxfield3gravity}. It would be nice to understand and to interpret the changes in both the Hawking--Page phase transition and the spectral form factor more explicitly in that context.

We briefly comment on a possible matrix model/JT~gravity duality for two-dimensional de~Sitter backgrounds. Here we point out an apparent puzzle in light of the recent results of ref.~\cite{TuriaciBluntDefects}, which suggest that the Weil--Petersson volumes in the presence of conical singularities with large identification angles are in general not obtained via analytic continuation from surfaces with conical singularities with small identification angles. As a consequence, computing the wave function of the universe for JT~gravity on two-dimensional de~Sitter by use of analytic continuation techniques may only be an approximation. It is, however, still possible that upon going beyond the study of asymptotic series
the validity of this approach is nevertheless justified. We believe that this issue deserves further study.

\bigskip
%%%%%%%%%%%%%%%%%%%%%%%%%%%%%%%%%%%%%%%%
\section*{Acknowledgements}
We would like to thank Alexander Belavin, Clifford Johnson, Joaquin Turiaci, Kefeng Liu, and Hao Xu for discussions and correspondences.
%%%%
The work of H.J. is supported by the Cluster of Excellence ``Precision Physics, Fundamental Interactions and Structure of Matter'' (PRISMA+ --- EXC 2118/1).
The work of J.K.K. is supported in part by the Heising-Simons Foundation, the Simons Foundation, and National Science Foundation Grant No. NSF PHY-1748958. A.K. is supported by the ``Onassis Foundation" as an ``Onassis Scholar" (Scholarship ID: F ZO 030/2 (2020-2021)). S.F., J.K.K., and A.K.\ acknowledge support by the Bonn Cologne Graduate School of Physics and Astronomy (BCGS).
%%%%%%%%%%%%%%%%%%%%%%%%%%%%%%%%%%%%%%%

\newpage
%\pagenumbering{gobble}
\bibliographystyle{utphys}
\bibliography{references}

\providecommand{\href}[2]{#2}\begingroup\raggedright\begin{thebibliography}{10}

\bibitem{JACKIW}
R.~Jackiw, ``{Lower Dimensional Gravity},''
  \href{http://dx.doi.org/10.1016/0550-3213(85)90448-1}{{\em Nucl. Phys. B}
  {\bfseries 252} (1985) 343--356}.

\bibitem{TEITELBOIM}
C.~Teitelboim, ``{Gravitation and Hamiltonian Structure in Two Space-Time
  Dimensions},'' \href{http://dx.doi.org/10.1016/0370-2693(83)90012-6}{{\em
  Phys. Lett. B} {\bfseries 126} (1983) 41--45}.

\bibitem{MaldacenaAdS2}
J.~Maldacena, D.~Stanford, and Z.~Yang, ``{Conformal symmetry and its breaking
  in two dimensional Nearly Anti-de-Sitter space},''
  \href{http://dx.doi.org/10.1093/ptep/ptw124}{{\em PTEP} {\bfseries 2016}
  no.~12, (2016) 12C104}, \href{http://arxiv.org/abs/1606.01857}{{\ttfamily
  arXiv:1606.01857 [hep-th]}}.

\bibitem{AlmheiriPolchinski}
A.~Almheiri and J.~Polchinski, ``{Models of AdS$_{2}$ backreaction and
  holography},'' \href{http://dx.doi.org/10.1007/JHEP11(2015)014}{{\em JHEP}
  {\bfseries 11} (2015) 014}, \href{http://arxiv.org/abs/1402.6334}{{\ttfamily
  arXiv:1402.6334 [hep-th]}}.

\bibitem{Jensen:2016pah}
K.~Jensen, ``{Chaos in AdS$_2$ Holography},''
  \href{http://dx.doi.org/10.1103/PhysRevLett.117.111601}{{\em Phys. Rev.
  Lett.} {\bfseries 117} no.~11, (2016) 111601},
  \href{http://arxiv.org/abs/1605.06098}{{\ttfamily arXiv:1605.06098
  [hep-th]}}.

\bibitem{Engelsoy:2016xyb}
J.~Engels\"oy, T.~G. Mertens, and H.~Verlinde, ``{An investigation of AdS$_{2}$
  backreaction and holography},''
  \href{http://dx.doi.org/10.1007/JHEP07(2016)139}{{\em JHEP} {\bfseries 07}
  (2016) 139}, \href{http://arxiv.org/abs/1606.03438}{{\ttfamily
  arXiv:1606.03438 [hep-th]}}.

\bibitem{Nayak:2018qej}
P.~Nayak, A.~Shukla, R.~M. Soni, S.~P. Trivedi, and V.~Vishal, ``{On the
  Dynamics of Near-Extremal Black Holes},''
  \href{http://dx.doi.org/10.1007/JHEP09(2018)048}{{\em JHEP} {\bfseries 09}
  (2018) 048}, \href{http://arxiv.org/abs/1802.09547}{{\ttfamily
  arXiv:1802.09547 [hep-th]}}.

\bibitem{Sarosi:2017ykf}
G.~S\'arosi, ``{AdS$_{2}$ holography and the SYK model},''
  \href{http://dx.doi.org/10.22323/1.323.0001}{{\em PoS} {\bfseries
  Modave~2017} (2018) 001}, \href{http://arxiv.org/abs/1711.08482}{{\ttfamily
  arXiv:1711.08482 [hep-th]}}.

\bibitem{SYK1}
S.~Sachdev and J.~Ye, ``{Gapless spin fluid ground state in a random, quantum
  Heisenberg magnet},''
  \href{http://dx.doi.org/10.1103/PhysRevLett.70.3339}{{\em Phys. Rev. Lett.}
  {\bfseries 70} (1993) 3339},
  \href{http://arxiv.org/abs/cond-mat/9212030}{{\ttfamily
  arXiv:cond-mat/9212030}}.

\bibitem{SYK2}
A.~Kitaev and S.~J. Suh, ``{The soft mode in the Sachdev-Ye-Kitaev model and
  its gravity dual},'' \href{http://dx.doi.org/10.1007/JHEP05(2018)183}{{\em
  JHEP} {\bfseries 05} (2018) 183},
  \href{http://arxiv.org/abs/1711.08467}{{\ttfamily arXiv:1711.08467
  [hep-th]}}.

\bibitem{RemarksSYK}
J.~Maldacena and D.~Stanford, ``{Remarks on the Sachdev-Ye-Kitaev model},''
  \href{http://dx.doi.org/10.1103/PhysRevD.94.106002}{{\em Phys. Rev. D}
  {\bfseries 94} no.~10, (2016) 106002},
  \href{http://arxiv.org/abs/1604.07818}{{\ttfamily arXiv:1604.07818
  [hep-th]}}.

\bibitem{SSS}
P.~Saad, S.~H. Shenker, and D.~Stanford, ``{JT gravity as a matrix integral},''
  \href{http://arxiv.org/abs/1903.11115}{{\ttfamily arXiv:1903.11115
  [hep-th]}}.

\bibitem{BlackHolesandRandomMatrices}
J.~S. Cotler, G.~Gur-Ari, M.~Hanada, J.~Polchinski, P.~Saad, S.~H. Shenker,
  D.~Stanford, A.~Streicher, and M.~Tezuka, ``{Black Holes and Random
  Matrices},'' \href{http://dx.doi.org/10.1007/JHEP05(2017)118}{{\em JHEP}
  {\bfseries 05} (2017) 118}, \href{http://arxiv.org/abs/1611.04650}{{\ttfamily
  arXiv:1611.04650 [hep-th]}}. [Erratum: JHEP 09, 002 (2018)].

\bibitem{Stanford:2019vob}
D.~Stanford and E.~Witten, ``{JT Gravity and the Ensembles of Random Matrix
  Theory},'' \href{http://arxiv.org/abs/1907.03363}{{\ttfamily arXiv:1907.03363
  [hep-th]}}.

\bibitem{Dyson:1962es}
F.~J. Dyson, ``{Statistical theory of the energy levels of complex systems.
  I},'' \href{http://dx.doi.org/10.1063/1.1703773}{{\em J. Math. Phys.}
  {\bfseries 3} (1962) 140--156}.

\bibitem{Altland:1997zz}
A.~Altland and M.~R. Zirnbauer, ``{Nonstandard symmetry classes in mesoscopic
  normal-superconducting hybrid structures},''
  \href{http://dx.doi.org/10.1103/PhysRevB.55.1142}{{\em Phys. Rev. B}
  {\bfseries 55} (1997) 1142--1161},
  \href{http://arxiv.org/abs/cond-mat/9602137}{{\ttfamily
  arXiv:cond-mat/9602137}}.

\bibitem{Zirnbauer:1996zz}
M.~R. Zirnbauer, ``{Riemannian symmetric superspaces and their origin in
  random-matrix theory},'' \href{http://dx.doi.org/10.1063/1.531675}{{\em J.
  Math. Phys.} {\bfseries 37} (1996) 4986--5018},
  \href{http://arxiv.org/abs/math-ph/9808012}{{\ttfamily
  arXiv:math-ph/9808012}}.

\bibitem{WittenStanford}
D.~Stanford and E.~Witten, ``{Fermionic Localization of the Schwarzian
  Theory},'' \href{http://dx.doi.org/10.1007/JHEP10(2017)008}{{\em JHEP}
  {\bfseries 10} (2017) 008}, \href{http://arxiv.org/abs/1703.04612}{{\ttfamily
  arXiv:1703.04612 [hep-th]}}.

\bibitem{Eynard_Orantin_2007Volumes}
B.~Eynard and N.~Orantin, ``{Weil-Petersson volume of moduli spaces,
  Mirzakhani's recursion and matrix models},''
  \href{http://arxiv.org/abs/0705.3600}{{\ttfamily arXiv:0705.3600 [math-ph]}}.

\bibitem{CJ1}
C.~V. Johnson, ``{Nonperturbative Jackiw-Teitelboim gravity},''
  \href{http://dx.doi.org/10.1103/PhysRevD.101.106023}{{\em Phys. Rev. D}
  {\bfseries 101} no.~10, (2020) 106023},
  \href{http://arxiv.org/abs/1912.03637}{{\ttfamily arXiv:1912.03637
  [hep-th]}}.

\bibitem{CJ2}
C.~V. Johnson, ``{Jackiw-Teitelboim supergravity, minimal strings, and matrix
  models},'' \href{http://dx.doi.org/10.1103/PhysRevD.103.046012}{{\em Phys.
  Rev. D} {\bfseries 103} no.~4, (2021) 046012},
  \href{http://arxiv.org/abs/2005.01893}{{\ttfamily arXiv:2005.01893
  [hep-th]}}.

\bibitem{CJ3}
C.~V. Johnson, ``{Explorations of nonperturbative Jackiw-Teitelboim gravity and
  supergravity},'' \href{http://dx.doi.org/10.1103/PhysRevD.103.046013}{{\em
  Phys. Rev. D} {\bfseries 103} no.~4, (2021) 046013},
  \href{http://arxiv.org/abs/2006.10959}{{\ttfamily arXiv:2006.10959
  [hep-th]}}.

\bibitem{Maxfield3gravity}
H.~Maxfield and G.~J. Turiaci, ``{The path integral of 3D gravity near
  extremality; or, JT gravity with defects as a matrix integral},''
  \href{http://dx.doi.org/10.1007/JHEP01(2021)118}{{\em JHEP} {\bfseries 01}
  (2021) 118}, \href{http://arxiv.org/abs/2006.11317}{{\ttfamily
  arXiv:2006.11317 [hep-th]}}.

\bibitem{WittenDeformations}
E.~Witten, ``{Matrix Models and Deformations of JT Gravity},''
  \href{http://dx.doi.org/10.1098/rspa.2020.0582}{{\em Proc. Roy. Soc. Lond. A}
  {\bfseries 476} no.~2244, (2020) 20200582},
  \href{http://arxiv.org/abs/2006.13414}{{\ttfamily arXiv:2006.13414
  [hep-th]}}.

\bibitem{MertensDefects}
T.~G. Mertens and G.~J. Turiaci, ``{Defects in Jackiw-Teitelboim Quantum
  Gravity},'' \href{http://dx.doi.org/10.1007/JHEP08(2019)127}{{\em JHEP}
  {\bfseries 08} (2019) 127}, \href{http://arxiv.org/abs/1904.05228}{{\ttfamily
  arXiv:1904.05228 [hep-th]}}.

\bibitem{WittenIntersection}
E.~Witten, ``{Two-dimensional gravity and intersection theory on moduli
  space},'' \href{http://dx.doi.org/10.4310/SDG.1990.v1.n1.a5}{{\em Surveys
  Diff. Geom.} {\bfseries 1} (1991) 243--310}.

\bibitem{KontsevichIntersection}
M.~Kontsevich, ``{Intersection theory on the moduli space of curves and the
  matrix Airy function},''
\href{http://dx.doi.org/10.1007/BF02099526}{{\em Commun. Math. Phys.}
  {\bfseries 147} (1992) 1--23}.
%%CITATION = CMPHA,147,1;%%.

\bibitem{Mirzakhani}
M.~Mirzakhani, ``Simple geodesics and {W}eil-{P}etersson volumes of moduli
  spaces of bordered {R}iemann surfaces,''
  \href{http://dx.doi.org/10.1007/s00222-006-0013-2}{{\em Invent. Math.}
  {\bfseries 167} no.~1, (2007) 179--222}.

\bibitem{OkuyamaSakai1}
K.~Okuyama and K.~Sakai, ``{JT gravity, KdV equations and macroscopic loop
  operators},'' \href{http://dx.doi.org/10.1007/JHEP01(2020)156}{{\em JHEP}
  {\bfseries 01} (2020) 156}, \href{http://arxiv.org/abs/1911.01659}{{\ttfamily
  arXiv:1911.01659 [hep-th]}}.

\bibitem{OkuyamaSakai2}
K.~Okuyama and K.~Sakai, ``{Multi-boundary correlators in JT gravity},''
  \href{http://dx.doi.org/10.1007/JHEP08(2020)126}{{\em JHEP} {\bfseries 08}
  (2020) 126}, \href{http://arxiv.org/abs/2004.07555}{{\ttfamily
  arXiv:2004.07555 [hep-th]}}.

\bibitem{Alishahihaetal}
M.~Alishahiha, A.~Faraji~Astaneh, G.~Jafari, A.~Naseh, and B.~Taghavi, ``{Free
  energy for deformed Jackiw-Teitelboim gravity},''
  \href{http://dx.doi.org/10.1103/PhysRevD.103.046005}{{\em Phys. Rev. D}
  {\bfseries 103} no.~4, (2021) 046005},
  \href{http://arxiv.org/abs/2010.02016}{{\ttfamily arXiv:2010.02016
  [hep-th]}}.

\bibitem{Itzykson:1992ya}
C.~Itzykson and J.~B. Zuber, ``{Combinatorics of the modular group. 2. The
  Kontsevich integrals},''
  \href{http://dx.doi.org/10.1142/S0217751X92002581}{{\em Int. J. Mod. Phys. A}
  {\bfseries 7} (1992) 5661--5705},
  \href{http://arxiv.org/abs/hep-th/9201001}{{\ttfamily arXiv:hep-th/9201001}}.

\bibitem{Gross1}
D.~J. Gross and A.~A. Migdal, ``{Nonperturbative Two-Dimensional Quantum
  Gravity},'' \href{http://dx.doi.org/10.1103/PhysRevLett.64.127}{{\em Phys.
  Rev. Lett.} {\bfseries 64} (1990) 127}.

\bibitem{Gross2}
D.~J. Gross and A.~A. Migdal, ``{A Nonperturbative Treatment of Two-dimensional
  Quantum Gravity},''
  \href{http://dx.doi.org/10.1016/0550-3213(90)90450-R}{{\em Nucl. Phys. B}
  {\bfseries 340} (1990) 333--365}.

\bibitem{Douglas:1989ve}
M.~R. Douglas and S.~H. Shenker, ``{Strings in Less Than One-Dimension},''
  \href{http://dx.doi.org/10.1016/0550-3213(90)90522-F}{{\em Nucl. Phys. B}
  {\bfseries 335} (1990) 635}.

\bibitem{Brezin:1990rb}
E.~Brezin and V.~Kazakov, ``{Exactly Solvable Field Theories of Closed
  Strings},'' \href{http://dx.doi.org/10.1016/0370-2693(90)90818-Q}{{\em Phys.
  Lett. B} {\bfseries 236} (1990) 144--150}.

\bibitem{GinspargMooreLectures}
P.~H. Ginsparg and G.~W. Moore, ``{Lectures on 2-D gravity and 2-D string
  theory},'' in {\em {T}heoretical {A}dvanced {S}tudy {I}nstitute ({TASI 92}):
  {F}rom {B}lack {H}oles and {S}trings to {P}articles}.
\newblock 10, 1993.
\newblock \href{http://arxiv.org/abs/hep-th/9304011}{{\ttfamily
  arXiv:hep-th/9304011}}.

\bibitem{Belavin:2008kv}
A.~A. Belavin and A.~B. Zamolodchikov, ``{On Correlation Numbers in 2D Minimal
  Gravity and Matrix Models},''
  \href{http://dx.doi.org/10.1088/1751-8113/42/30/304004}{{\em J. Phys. A}
  {\bfseries 42} (2009) 304004},
  \href{http://arxiv.org/abs/0811.0450}{{\ttfamily arXiv:0811.0450 [hep-th]}}.

\bibitem{Okounkovnpointfunction}
A.~Okounkov, ``Generating functions for intersection numbers on moduli spaces
  of curves,'' \href{http://dx.doi.org/10.1155/S1073792802110099}{{\em Int.
  Math. Res. Not.} no.~18, (2002) 933--957},
  \href{http://arxiv.org/abs/math/0101201}{{\ttfamily arXiv:math/0101201
  [math.AG]}}.

\bibitem{MaldacenadS}
J.~Maldacena, G.~J. Turiaci, and Z.~Yang, ``{Two dimensional Nearly de Sitter
  gravity},'' \href{http://dx.doi.org/10.1007/JHEP01(2021)139}{{\em JHEP}
  {\bfseries 01} (2021) 139}, \href{http://arxiv.org/abs/1904.01911}{{\ttfamily
  arXiv:1904.01911 [hep-th]}}.

\bibitem{MaloneydS}
J.~Cotler, K.~Jensen, and A.~Maloney, ``{Low-dimensional de Sitter quantum
  gravity},'' \href{http://dx.doi.org/10.1007/JHEP06(2020)048}{{\em JHEP}
  {\bfseries 06} (2020) 048}, \href{http://arxiv.org/abs/1905.03780}{{\ttfamily
  arXiv:1905.03780 [hep-th]}}.

\bibitem{TuriaciBluntDefects}
G.~J. Turiaci, M.~Usatyuk, and W.~W. Weng, ``{Dilaton-gravity, deformations of
  the minimal string, and matrix models},''
  \href{http://arxiv.org/abs/2011.06038}{{\ttfamily arXiv:2011.06038
  [hep-th]}}.

\bibitem{Okuyama:2021ytf}
K.~Okuyama and K.~Sakai, ``{A proof of loop equations in 2d topological
  gravity},'' \href{http://arxiv.org/abs/2106.05643}{{\ttfamily
  arXiv:2106.05643 [hep-th]}}.

\bibitem{Mertens:2020hbs}
T.~G. Mertens and G.~J. Turiaci, ``{Liouville quantum gravity -- holography, JT
  and matrices},'' \href{http://dx.doi.org/10.1007/JHEP01(2021)073}{{\em JHEP}
  {\bfseries 01} (2021) 073}, \href{http://arxiv.org/abs/2006.07072}{{\ttfamily
  arXiv:2006.07072 [hep-th]}}.

\bibitem{MR1486986}
E.~Arbarello and M.~Cornalba, ``Combinatorial and algebro-geometric cohomology
  classes on the moduli spaces of curves,'' {\em J. Algebraic Geom.} {\bfseries
  5} no.~4, (1996) 705--749,
  \href{http://arxiv.org/abs/alg-geom/9406008}{{\ttfamily
  arXiv:alg-geom/9406008 [math.AG]}}.

\bibitem{MR2482127}
K.~Liu and H.~Xu, ``Recursion formulae of higher {W}eil-{P}etersson volumes,''
  \href{http://dx.doi.org/10.1093/imrn/rnn148}{{\em Int. Math. Res. Not. IMRN}
  no.~5, (2009) 835--859}, \href{http://arxiv.org/abs/0708.0565}{{\ttfamily
  arXiv:0708.0565 [math.AG]}}.

\bibitem{MR727702}
S.~Wolpert, ``On the homology of the moduli space of stable curves,''
  \href{http://dx.doi.org/10.2307/2006980}{{\em Ann. of Math. (2)} {\bfseries
  118} no.~3, (1983) 491--523}.

\bibitem{MR2379144}
M.~Mulase and B.~Safnuk, ``Mirzakhani's recursion relations, {V}irasoro
  constraints and the {K}d{V} hierarchy,'' {\em Indian J. Math.} {\bfseries 50}
  no.~1, (2008) 189--218, \href{http://arxiv.org/abs/math/0601194}{{\ttfamily
  arXiv:math/0601194 [math.QA]}}.

\bibitem{Dijkgraaf:2018vnm}
R.~Dijkgraaf and E.~Witten, ``{Developments in Topological Gravity},''
  \href{http://dx.doi.org/10.1142/S0217751X18300296}{{\em Int. J. Mod. Phys. A}
  {\bfseries 33} no.~30, (2018) 1830029},
  \href{http://arxiv.org/abs/1804.03275}{{\ttfamily arXiv:1804.03275
  [hep-th]}}.

\bibitem{MR2257394}
M.~Mirzakhani, ``Weil-{P}etersson volumes and intersection theory on the moduli
  space of curves,''
  \href{http://dx.doi.org/10.1090/S0894-0347-06-00526-1}{{\em J. Amer. Math.
  Soc.} {\bfseries 20} no.~1, (2007) 1--23}.

\bibitem{KazakovMM}
V.~A. Kazakov, ``{The Appearance of Matter Fields from Quantum Fluctuations of
  2D Gravity},'' \href{http://dx.doi.org/10.1142/S0217732389002392}{{\em Mod.
  Phys. Lett. A} {\bfseries 4} (1989) 2125}.

\bibitem{StaudacherMM}
M.~Staudacher, ``{The Yang-lee Edge Singularity on a Dynamical Planar Random
  Surface},'' \href{http://dx.doi.org/10.1016/0550-3213(90)90432-D}{{\em Nucl.
  Phys. B} {\bfseries 336} (1990) 349}.

\bibitem{Okuyama:2020qpm}
K.~Okuyama and K.~Sakai, ``{JT supergravity and Brezin-Gross-Witten
  tau-function},'' \href{http://dx.doi.org/10.1007/JHEP10(2020)160}{{\em JHEP}
  {\bfseries 10} (2020) 160}, \href{http://arxiv.org/abs/2007.09606}{{\ttfamily
  arXiv:2007.09606 [hep-th]}}.

\bibitem{Kyono:2017pxs}
H.~Kyono, S.~Okumura, and K.~Yoshida, ``{Comments on 2D dilaton gravity system
  with a hyperbolic dilaton potential},''
  \href{http://dx.doi.org/10.1016/j.nuclphysb.2017.07.013}{{\em Nucl. Phys. B}
  {\bfseries 923} (2017) 126--143},
  \href{http://arxiv.org/abs/1704.07410}{{\ttfamily arXiv:1704.07410
  [hep-th]}}.

\bibitem{Moore:1991ir}
G.~W. Moore, N.~Seiberg, and M.~Staudacher, ``{From loops to states in 2-D
  quantum gravity},''
  \href{http://dx.doi.org/10.1016/0550-3213(91)90548-C}{{\em Nucl. Phys. B}
  {\bfseries 362} (1991) 665--709}.

\bibitem{Gelfand:1975rn}
I.~Gelfand and L.~Dikii, ``{Asymptotic behavior of the resolvent of
  Sturm-Liouville equations and the algebra of the Korteweg-De Vries
  equations},'' \href{http://dx.doi.org/10.1070/RM1975v030n05ABEH001522}{{\em
  Russ. Math. Surveys} {\bfseries 30} no.~5, (1975) 77--113}.

\bibitem{Dijkgraaf:1990rs}
R.~Dijkgraaf, H.~L. Verlinde, and E.~P. Verlinde, ``{Loop equations and
  Virasoro constraints in nonperturbative 2-D quantum gravity},''
  \href{http://dx.doi.org/10.1016/0550-3213(91)90199-8}{{\em Nucl. Phys. B}
  {\bfseries 348} (1991) 435--456}.

\bibitem{ZografLargeGenusAsymptotics}
P.~Zograf, ``{On the large genus asymptotics of Weil-Petersson volumes},''
  \href{http://arxiv.org/abs/0812.0544}{{\ttfamily arXiv:0812.0544 [math.AG]}}.

\bibitem{Witten:2020ert}
E.~Witten, ``{Deformations of JT Gravity and Phase Transitions},''
  \href{http://arxiv.org/abs/2006.03494}{{\ttfamily arXiv:2006.03494
  [hep-th]}}.

\bibitem{CJ4}
C.~V. Johnson and F.~Rosso, ``{Solving Puzzles in Deformed JT Gravity: Phase
  Transitions and Non-Perturbative Effects},''
  \href{http://dx.doi.org/10.1007/JHEP04(2021)030}{{\em JHEP} {\bfseries 04}
  (2021) 030}, \href{http://arxiv.org/abs/2011.06026}{{\ttfamily
  arXiv:2011.06026 [hep-th]}}.

\bibitem{doubletrumpet}
P.~Saad, S.~H. Shenker, and D.~Stanford, ``{A semiclassical ramp in SYK and in
  gravity},'' \href{http://arxiv.org/abs/1806.06840}{{\ttfamily
  arXiv:1806.06840 [hep-th]}}.

\bibitem{Engelhardt:2020qpv}
N.~Engelhardt, S.~Fischetti, and A.~Maloney, ``{Free energy from replica
  wormholes},'' \href{http://dx.doi.org/10.1103/PhysRevD.103.046021}{{\em Phys.
  Rev. D} {\bfseries 103} no.~4, (2021) 046021},
  \href{http://arxiv.org/abs/2007.07444}{{\ttfamily arXiv:2007.07444
  [hep-th]}}.

\bibitem{Papadodimas:2015xma}
K.~Papadodimas and S.~Raju, ``{Local Operators in the Eternal Black Hole},''
  \href{http://dx.doi.org/10.1103/PhysRevLett.115.211601}{{\em Phys. Rev.
  Lett.} {\bfseries 115} no.~21, (2015) 211601},
  \href{http://arxiv.org/abs/1502.06692}{{\ttfamily arXiv:1502.06692
  [hep-th]}}.

\bibitem{WittenHawkingPage}
E.~Witten, ``{Anti-de Sitter space, thermal phase transition, and confinement
  in gauge theories},''
  \href{http://dx.doi.org/10.4310/ATMP.1998.v2.n3.a3}{{\em Adv. Theor. Math.
  Phys.} {\bfseries 2} (1998) 505--532},
  \href{http://arxiv.org/abs/hep-th/9803131}{{\ttfamily arXiv:hep-th/9803131}}.

\bibitem{PhysRevE.50.888}
M.~Srednicki, ``{Chaos and quantum thermalization},''
  \href{http://dx.doi.org/10.1103/PhysRevE.50.888}{{\em Phys. Rev. E}
  {\bfseries 50} (Aug, 1994) 888--901}.

\bibitem{PhysRevA.43.2046}
J.~M. Deutsch, ``Quantum statistical mechanics in a closed system,''
  \href{http://dx.doi.org/10.1103/PhysRevA.43.2046}{{\em Phys. Rev. A}
  {\bfseries 43} (Feb, 1991) 2046--2049}.

\bibitem{Maldacenadsviaads}
J.~Maldacena, ``{Vacuum decay into Anti de Sitter space},''
  \href{http://arxiv.org/abs/1012.0274}{{\ttfamily arXiv:1012.0274 [hep-th]}}.

\bibitem{Hartle:1983ai}
J.~Hartle and S.~Hawking, ``{Wave Function of the Universe},''
  \href{http://dx.doi.org/10.1103/PhysRevD.28.2960}{{\em Adv. Ser. Astrophys.
  Cosmol.} {\bfseries 3} (1987) 174--189}.

\bibitem{MR4222602}
B.~Dubrovin, D.~Yang, and D.~Zagier, ``On tau-functions for the {K}d{V}
  hierarchy,'' \href{http://dx.doi.org/10.1007/s00029-021-00620-x}{{\em Selecta
  Math. (N.S.)} {\bfseries 27} no.~1, (2021) Paper No. 12, 47},
  \href{http://arxiv.org/abs/1812.08488}{{\ttfamily arXiv:1812.08488
  [math-ph]}}.

\bibitem{Dalley:1991qg}
S.~Dalley, C.~V. Johnson, and T.~R. Morris, ``{Multicritical complex matrix
  models and nonperturbative 2-D quantum gravity},''
  \href{http://dx.doi.org/10.1016/0550-3213(92)90217-Y}{{\em Nucl. Phys. B}
  {\bfseries 368} (1992) 625--654}.

\bibitem{Dalley:1991vr}
S.~Dalley, C.~V. Johnson, and T.~R. Morris, ``{Nonperturbative two-dimensional
  quantum gravity},''
  \href{http://dx.doi.org/10.1016/0550-3213(92)90218-Z}{{\em Nucl. Phys. B}
  {\bfseries 368} (1992) 655--670}.

\bibitem{finitecutoffresurgence}
L.~Griguolo, R.~Panerai, J.~Papalini, and D.~Seminara, ``{Nonperturbative
  effects and resurgence in JT gravity at finite cutoff},''
  \href{http://arxiv.org/abs/2106.01375}{{\ttfamily arXiv:2106.01375
  [hep-th]}}.

\bibitem{Gross:2019ach}
D.~J. Gross, J.~Kruthoff, A.~Rolph, and E.~Shaghoulian, ``{$T\overline{T}$ in
  AdS$_2$ and Quantum Mechanics},''
  \href{http://dx.doi.org/10.1103/PhysRevD.101.026011}{{\em Phys. Rev. D}
  {\bfseries 101} no.~2, (2020) 026011},
  \href{http://arxiv.org/abs/1907.04873}{{\ttfamily arXiv:1907.04873
  [hep-th]}}.

\bibitem{Iliesiu:2020zld}
L.~V. Iliesiu, J.~Kruthoff, G.~J. Turiaci, and H.~Verlinde, ``{JT gravity at
  finite cutoff},'' \href{http://dx.doi.org/10.21468/SciPostPhys.9.2.023}{{\em
  SciPost Phys.} {\bfseries 9} (2020) 023},
  \href{http://arxiv.org/abs/2004.07242}{{\ttfamily arXiv:2004.07242
  [hep-th]}}.

\bibitem{McGough:2016lol}
L.~McGough, M.~Mezei, and H.~Verlinde, ``{Moving the CFT into the bulk with $
  T\overline{T} $},'' \href{http://dx.doi.org/10.1007/JHEP04(2018)010}{{\em
  JHEP} {\bfseries 04} (2018) 010},
  \href{http://arxiv.org/abs/1611.03470}{{\ttfamily arXiv:1611.03470
  [hep-th]}}.

\end{thebibliography}\endgroup
\end{document}